\documentclass[onecolumn]{emulateapj}
\usepackage{apjfonts}
\usepackage{graphicx}
\newcommand{\be}{\begin{equation}}
\newcommand{\ee}{\end{equation} }
\newcommand{\ba}{\begin{eqnarray}}
\newcommand{\ea}{\end{eqnarray}}

\newcommand{\bnabla}{\mbox{\boldmath$\nabla$}}
\newcommand{\bbeta}{\mbox{\boldmath$\beta$}}
\newcommand{\nn}{\mbox{} \nonumber \\ \mbox{} }

\shorttitle{Hot Electromagnetic Outflows II: Jet Breakout}
\shortauthors{Russo \& Thompson}	
\begin{document}
\title{Hot Electromagnetic Outflows II:  Jet Breakout}
\author{Matthew Russo}
\affil{Department of Physics, University of Toronto, 60 St. George St., Toronto, ON M5S 1A7, Canada.}
\author{Christopher Thompson}
\affil{Canadian Institute for Theoretical Astrophysics, 60 St. George St., Toronto, ON M5S 3H8, Canada.}
\slugcomment{Submitted to The Astrophysical Journal}
%%%%%%%%%%%%%%%%%%%%%%%%%%%%%%%%%%%%%%%%%%%%%%%%%%%%%%%%%%%%%%%%%%%%%%%%%%%%%%%%%%%
%%%%%%%%%%%%%%%%%%%%%%%%%%%%%%%%%%%%%%%%%%%%%%%%%%%%%%%%%%%%%%%%%%%%%%%%%%%%%%%%%%%

\begin{abstract}
We consider the interaction between radiation, matter and a magnetic field in a compact, relativistic jet.
The entrained matter accelerates outward as the jet breaks out of a star or other confining medium.
In some circumstances, such as gamma-ray bursts (GRBs), the magnetization of the jet 
is greatly reduced by an advected radiation field while the jet is optically thick to scattering.  
Where magnetic flux surfaces diverge rapidly, a strong outward Lorentz force develops and radiation 
and matter begin to decouple.  The increase in magnetization is coupled to a rapid growth 
in Lorentz factor.   We take two approaches to this problem.  The first examines the flow outside the 
fast magnetosonic critical surface, and calculates the flow speed and the angular distribution
of the radiation field over a range of scattering depths.  The second considers the flow structure on both sides
of the critical surface in the optically thin regime, using a relaxation method.  In both approaches, we find
how the terminal Lorentz factor, and radial profile of the outflow, depend on the radiation intensity and
optical depth at breakout.  The effect of bulk Compton scattering on the radiation spectrum is calculated
by a Monte Carlo method, while neglecting the effects of internal dissipation.  The peak of the scattered
spectrum sits near the seed peak if radiation pressure dominates the acceleration, but is pushed to a higher
frequency if the Lorentz force dominates, and especially if the seed photon cone is broadened by interaction
with a slower component of the outflow. 
\end{abstract}
\keywords{MHD --- plasmas --- radiative transfer --- scattering --- gamma rays: stars}

%%%%%%%%%%%%%%%%%%%%%%%%%%%%%%%%%%%%%%%%%%%%%%%%%%%%%%%%%%%%%%%%%%%%%%%%%%%%%%%%%%%
%%%%%%%%%%%%%%%%%%%%%%%%%%%%%%%%%%%%%%%%%%%%%%%%%%%%%%%%%%%%%%%%%%%%%%%%%%%%%%%%%%%

\section{Introduction}

Gamma-ray bursts involve collimated, relativistic outflows, as deduced from their rapid
variability, extreme apparent energies (which can exceed the binding energy of a neutron star: 
\citealt{kulkarni99,amati02}), and the expected presence of non-relativistic material 
surrounding the engine.  The jet is heated as it works through this denser material,
which may represent a stellar envelope \citep{woosley93,paczynski98}, or neutron-rich debris
from a binary neutron star merger (e.g. \citealt{dessart09}).   As a result, a nearly
blackbody radiation field may carry a significant fraction of the energy flux near
the point of breakout.

We focus here on strongly magnetized jets that are driven outward by a 
combination of the Lorentz force, and the force of radiation scattering
off ionized matter.  The acceleration of such a `hot electromagnetic outflow'
\citep{thompson94, meszaros97, drenkhahn02, thompson06, giannios07, zhang11},
in which radiation pressure
dominates matter pressure, has been treated quantitatively in 
\cite{russo12} (Paper I) in the approximation that the poloidal magnetic field lines threading the outflow
are radial and monopolar.  The radiation field is self-collimating outside the scattering photosphere, but may continue to interact with slower material
that it entrained by the jet.  In Paper I,
the outflow was followed both inside and outside the fast magnetosonic critical point.  The radiation force is
especially important outside the fast point:  even where the kinetic energy flux
of the entrained charged particles is small compared with the magnetic Poynting flux, they provide an efficient
couple between magnetic field and radiation.   The relative influence of the two stresses 
on the asymptotic Lorentz factor depends on the radiation compactness.  Generally, the importance
of radiation pressure is enhanced by bulk relativistic motion at the photosphere.

In this paper we generalize the calculation of Paper I to include non-spherical effects.  
A magnetized outflow experiences a strong Lorentz force where poloidal flux surfaces in the jet
diverge from each other faster than in a monopolar geometry \citep{camenzind87,li92b,begelman94,vlahakis03a,vlahakis03b,beskin06,tchek09}.
In particular, a magnetized jet accelerates rapidly when it breaks out of the confining material
\citep{tchek10}.  The simulations in that paper demonstrated the effect for a cold magnetohydrodynamic
(MHD) outflow, but did not include the effects of radiation pressure and drag.  

The magnetization of a hot electromagnetic outflow remains modest inside its scattering photosphere,
where the radiation is tied to the matter, and the radiation enthalpy contributes to the inertia.
Our first task in this paper is, therefore, to examine how the radiation field begins to decouple from the
matter when the jet material breaks out.  We define a bulk frame in which the radiation force vanishes,
by taking angular moments of the radiation field, and then track the proportions
of the energy flux carried by matter, radiation, and magnetic field, at both large and small optical depths.  

Given the flow profile so obtained, the radiation spectrum is calculated by a Monte Carlo method.
Here we focus on the effects of bulk Compton scattering, which provide a direct probe of the outflow
dynamics.  We neglect the effects of internal dissipation by various process such as MHD wave
damping, magnetic reconnection, or shocks.  

The second principal goal of this paper is to obtain the longitudinal motion along a magnetic
flux surface, taking into account both the radiation force and the singularity in the flow equations which
appears at the fast point.  Our focus here is on the zone near and outside the transparency surface;  previous
efforts to calculate the effect of pressure gradient forces on relativistic outflows (e.g. \citealt{vlahakis03b}) 
have focused on the optically thin regime.  We argued in Paper I that the effect of a magnetic pressure gradient
driven by internal reconnection \citep{drenkhahn02} has been overestimated, because it neglects the addition
to the outflow inertia from particle heating and a strong non-radial magnetic field. 

Coupled wind equations for the matter Lorentz factor and angular momentum are derived in an
arbitrary poloidal field geometry, restricted to the case of small angles near the rotational axis, but
allowing for arbitrary relative flaring of the flux surfaces.  The fast 
point generally sits close to the breakout surface of the jet.  Our main simplification of the
problem is to impose a particular shape for the poloidal flux surfaces, and not to solve self-consistently
for the cross-field force balance.  Two constraints are applied to the imposed magnetic field profile:  
that the rate of flaring is causal, and that the transverse component of the radiation force is at
most a perturbation to the transverse Lorentz force.

The plan of this paper is as follows.  Section 2 reviews the acceleration of a relativistic MHD outflow
driven by the differential flaring of magnetic flux surfaces, and considers the radiation transfer equation
in the limit of small angles.  Equations are derived for the acceleration of a steady MHD outflow outside
its fast point, in combination with the radial evolution of the magnetization, radiation energy flux,
scattering depth, and the frame in which the radiation force vanishes.
These equations are solved in particular cases relevant to GRB jets.  Section 3 presents a simple model
of a spreading thin jet outside its photosphere, and derives the corresponding steady flow equations for arbitrary radiation
force and magnetization.  The effect of radiation pressure on the fast 
point is considered analytically, and numerical solutions for the flow both inside and outside the
fast point are presented.  Section 4 describes Monte Carlo calculations of the emerging radiation spectrum 
in both the causal jet model of Section 2, and the optically thin model of Section 3.  Section 5 summarizes our results.
The Appendix presents a derivation of the radiation force in a thin, transparent jet.

%%%%%%%%%%%%%%%%%%%%%%%%%%%%%%%%%%%%%%%%%%%%%%%%%%%%%%%%%%%%%%%%%%%%%%%%%%%%%%%%%%%
%%%%%%%%%%%%%%%%%%%%%%%%%%%%%%%%%%%%%%%%%%%%%%%%%%%%%%%%%%%%%%%%%%%%%%%%%%%%%%%%%%%

%\section{Hot Electromagnetic Outflow with Decollimating Flux Surfaces}\label{s:causaljet}
\section{Flaring, Hot Magnetized Jet:  Transition to Low Optical Depth (Model I).}\label{s:causaljet}

We consider a stationary, axisymmetric outflow of perfectly conducting material that is tied
to a very strong magnetic field.  The outflow is also a strong source of radiation, which scatters off the
advected light particles (electrons as well as positrons).  Matter pressure gradients are neglected in comparison
with inertial and Lorentz forces as well as the radiation force.

We start by considering the exchange of energy between different components of the outflow.  
Deviations from radial motion are assumed to be small compared with the angular width of the photon beam:  that is,
\begin{figure}[h]
\centerline{\includegraphics[width=0.8\hsize]{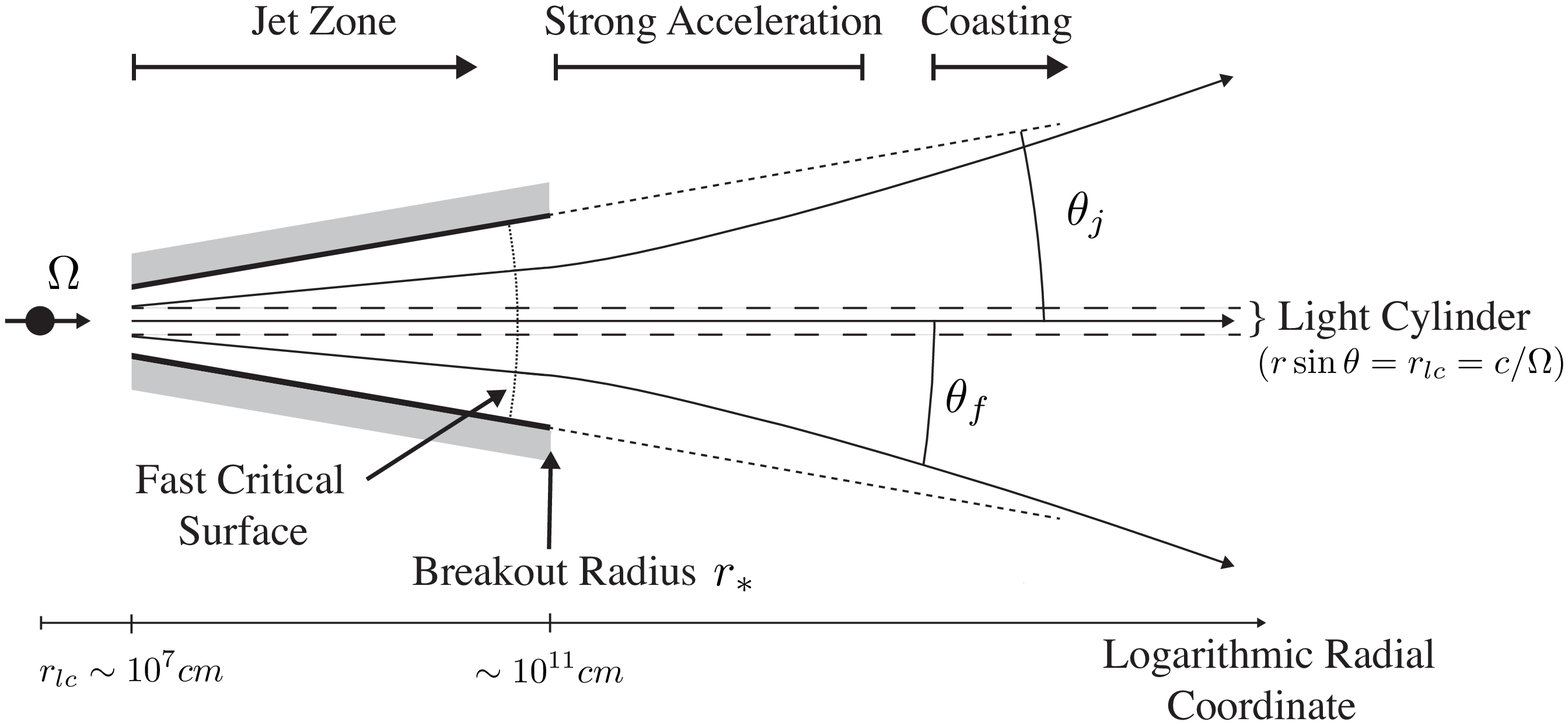}}
%\centerline{\includegraphics[width=1.2\hsize]{JetFigureSec2.eps}}
\caption{Geometry and approximate scale of the flow solutions for jet model I.}
\vskip .1in
\label{fig:JetSec2}
\end{figure}
the interaction between matter and radiation is calculated assuming radial matter motion, but allowance is made
for strong radial Lorentz forces driven by a small amount of magnetic field line flaring.  
The beam angular  
width is set, more or less, by the Lorentz factor of the outflow at its transparency surface.  
Here allowance is made for a finite optical depth of the magnetofluid.  By taking angular moments of the 
radiation field, we track the difference between the Lorentz factor of the matter, and of the frame in which 
the radiation force vanishes, as the matter is accelerated by a strong Lorentz force.   This approach is
suited to a single-component magnetofluid, but also allows for the presence of a second, slower component 
that scatters the radiation field into a broader cone, and plausibly is present in GRBs (Paper I).  

Given the complications introduced by a finite optical depth, we now consider only supermagnetosonic outflows.
In a second approach (Section \ref{s:eom}), we account for non-radial matter motion and follow the flow across the
fast critical surface, but restrict the calculation to low optical depth.
The geometry of the model is shown in Figure \ref{fig:JetSec3}.  After being launched by the central 
engine (with angular frequency $\Omega$) the flow enters the jet zone along the rotation axis.  We ignore the
details of the acceleration while the jet is still very optically thick, and laterally confined.  Our calculation
begins a short distance inside breakout (at radius $r_*$), by which point the flow is assumed to be supermagnetosonic.
Outside breakout, transverse pressure support effectively vanishes and field lines begin 
to diverge differentially.  The outward Lorentz force increases dramatically over a narrow range of
radius, until a loss of causal contact across the jet forces the flow lines to straighten out, and the 
acceleration is cut off.  Although the scattering photosphere could, in principal,
sit anywhere in the outflow, breakout is associated with a large drop in optical depth.  In our calculations,
the photosphere therefore usually sits just outside breakout.  Low optical depth at breakout does produce
an interesting imprint of bulk Compton scattering on the emergent spectrum (Section \ref{s:spectrum}).
%, beginning the calculation just before the flow reaches the breakout radius. 

\subsection{Exchange of Energy between Radiation and Magnetofluid}

We consider the flow along a poloidal magnetic field line $\theta_f(r)$, starting at a large enough radius that 
the streamline sits well outside the light cylinder of the central engine.  Deviations from radial motion are
neglected, except in so far that they influence the radial Lorentz force.  Then the outflow has a fixed total 
luminosity per sterad, including contributions from matter, magnetic field, and radiation,
\be\label{eq:etot0}
{dL\over d\Omega} = {dL_k\over d\Omega} + {dL_P\over d\Omega} + {dL_\gamma\over d\Omega} = {\rm const}.
\ee
Here
\be
{1\over r^2}{dL_k\over d\Omega} = \Gamma c^2\cdot \Gamma \rho v_p = {\Gamma c^2\over r^2}\,{d\dot M\over d\Omega}
\ee
is the kinetic energy flux of material of proper density $\rho$, poloidal (radial) speed $v_p$, and Lorentz factor $\Gamma$.
The poloidal Poynting flux is expressed in terms of the electric and magnetic vectors ${\bf E}$, ${\bf B}$ by
\be\label{eq:lp}
{1\over r^2}{dL_P\over d\Omega} = \hat B_p \cdot {{\bf E}\times{\bf B}\over 4\pi} c.
\ee
Substituting ${\bf E} = -{\bf v}\times{\bf B}/c$ into the induction equation gives
$\partial{\bf B}/\partial t = \bnabla\times({\bf v}\times{\bf B})$, where ${\bf v}$ is the fluid velocity.
The steady solution to this equation involves the pattern angular velocity 
$\Omega_f$ of the magnetic field, which is constant along a poloidal flux surface.  It relates the
toroidal components of ${\bf B}$ and ${\bf v}$ via
\be\label{eq:induction}
B_\phi = {v_\phi - \Omega_f r\sin\theta_f \over v_p}B_p.
\ee
Substituting this into (\ref{eq:lp}) gives
\be
{1\over r^2}{dL_P\over d\Omega} = -\Omega_f r\sin\theta_f {B_pB_\phi\over 4\pi}.
\ee
Far outside the light cylinder, the outflow rotates slowly and the
magnetic field is predominantly toroidal:  $v_\phi \ll v_p \simeq c$ and $|B_\phi| \gg |B_p|$.  Hence
\be
{1\over r^2}{dL_P\over d\Omega} \simeq (\Omega_f r\sin\theta_f)^2 {B_p^2\over 4\pi c}.
\ee

It is useful to normalize all components of the energy flux to the poloidal mass flux, which is conserved 
along a poloidal flux surface in a steady MHD wind, $d\dot M/d\Phi_p = \Gamma\rho v_p/B_p = $ const.
Assuming further that $\Gamma \gg 1$, the magnetization becomes
\be
\sigma = {dL_P/d\Omega\over (d\dot M/d\Omega) c^2} \simeq
         {(\Omega_f r\sin\theta_f)^2 B_p \over 4\pi c^3} {B_p\over \Gamma\rho v_p}\biggr|_*,
\ee
The radius $r_*$ and the label $*$ denote a position in the jet where the confining medium changes rapidly, e.g. the jet
moves beyond the photosphere of a Wolf-Rayet star.  (We will distinguish this from an inner boundary
$r_i$ for the jet integration, which typically is set just interior to the breakout radius.)
Taking $r_* \gg c/\Omega_f$,
\be\label{eq:sigmaf}
{\sigma\over\sigma_*} =  {(r\sin\theta_f)^2\over (r\sin\theta_f)_*^2} {B_p\over B_{p*}}.
\ee
(Note that our definition of $\sigma$ differs by a factor $\Gamma$ from the one used by \citealt{tchek09}
in a similar derivation.)  Defining the normalized photon luminosity by 
\be
{\cal R} = {dL_\gamma/d\Omega\over (d\dot M/d\Omega)c^2},
\ee
the equation of energy conservation (\ref{eq:etot0}) can be written
\be\label{eq:etot}
\Gamma - \Gamma_* = -\left(\sigma - \sigma_* + {\cal R} - {\cal R}_*\right).
\ee
Note that ${\cal R}$ is related to the photon compactness and the scattering depth
measured outward from radius $r$ by
\be\label{eq:chidef}
\chi \equiv \frac{\sigma_{T}}{r\bar{m}c^{3}} \frac{dL_\gamma}{d\Omega}
\sim (2-6) \Gamma^2 \tau_{\rm es} {\cal R},
\ee
where $\sigma_T$ is the Thomson cross section and $\bar m$ is the material inertia per scattering charge.
The coefficient here depends on the acceleration of the outflow, as can be seen from the expression for the 
optical depth of a (radially moving) photon
\be\label{eq:taustar}
\tau_{\rm es}(r) = \alpha_{\rm es}(r)
\int_{r}^\infty\left[1-\beta(r')\right] {r^2 dr'\over \beta(r'){r'}^2};\quad\quad \beta(r) 
               \equiv {v\over c} \simeq {v_r\over c}
\ee
[see equation (\ref{eq:abs}) for notation].  The coefficient is $\sim 2$ when the Lorentz factor 
is constant and reaches $\sim 6$ for a linear growth, $\Gamma \propto r$.

%%%%%%%%%%%%%%%%%%%%%%%%%%%%%%%%%%%%%%%%%%%%%%%%%%%%%%%%%%%%%%%%%%%%%%%%%%%%%%%%%%%

\begin{figure}[h]
\centerline{\includegraphics[width=0.5\hsize]{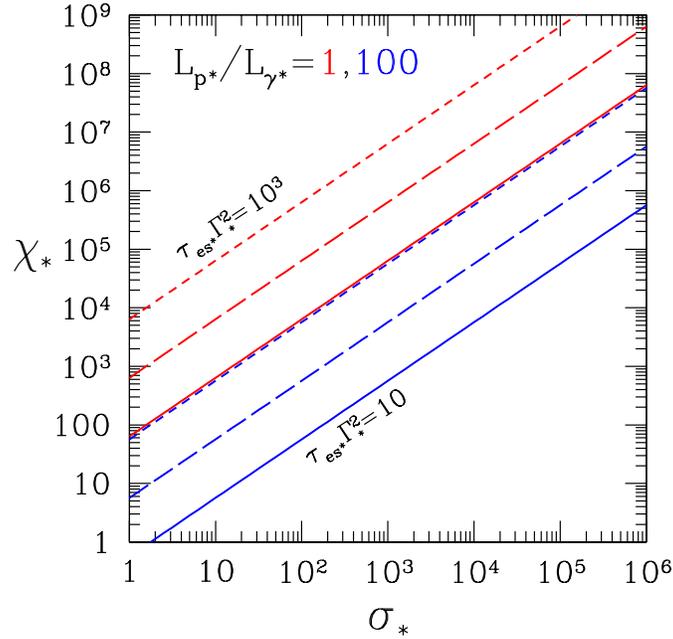}}
%\centerline{\includegraphics[width=0.4\hsize]{smchisigma.eps}}
\caption{Radiation compactness at breakout for different ratios of Poynting to radiation flux: $1$ (red), $100$ (blue). 
Rescaled scattering optical depth $\tau_{\rm es*}\Gamma^2_*=10,10^2,10^3$ is plotted as solid, long-dashed and short-dashed respectively.
The compactness drops below these curves beyond the breakout point.   Strong radiative driving increases
the Lorentz factor above the photospheric value $\Gamma(r_\tau)$ if $\chi_* \gtrsim \Gamma(r_\tau)^3$.}
\label{fig:chisigma}
\end{figure}

\subsection{Importance of Radiative Driving in Outflows with a Relativistically Moving Photosphere}

Jets of a high magnetization can encounter a scattering photosphere not too far outside breakout from
a confining medium such as a Wolf-Rayet envelope or or neutron-rich debris cloud.   Let us take
a fiducial luminosity $4\pi dL_\gamma/d\Omega = 10^{51}\,L_{51}$ erg s$^{-1}$ and a deconfinement radius $r_*
= 10^{10}r_{*,10}$ cm.  The corresponding compactness (\ref{eq:chidef}) is $\chi_* = 1\times 10^8
L_{51}r_{*,10}^{-1}(m_p/\bar m)$.  Since the Lorentz factor increases rapidly after
breakout due to MHD stresses, we fix $r$ and then consider the condition for a photosphere to emerge
at a radius $r_\tau \gtrsim r_*$.
This corresponds to $\Gamma^2(r_\tau) {\cal R}(r_\tau) \sim 10^7 (m_p/\bar m)$.  For example, if 
the jet is hot and strongly magnetized, ${\cal R} \sim \sigma \sim 10^5$, and pairs have largely annihilated
within the bulk of the jet material, then the photosphere emerges at $\Gamma(r_\tau) \sim 10$.

%A collimated flow of radiation defines a
%relativistically moving frame in which a free charge experiences no net radiation force.  The corresponding
%Lorentz factor increases linearly with radius, $\Gamma_{\rm eq} \simeq \Gamma(r_\tau) (r/r_\tau)$ in a luminous
%jet with a photosphere situated at a radius $r_\tau$.  
The radiation field is capable of driving a light baryonic outflow to a terminal Lorentz factor (see Section 2 of Paper I),
\be\label{eq:gaminf}
\Gamma_\infty \sim [\Gamma(r_\tau) \chi(r_\tau)]^{1/4}.
\ee
Moderately relativistic motion at the photosphere enhances $\Gamma_\infty$ and, as we now motivate, a larger radiation
compactness.  When the Lorentz force is taken into account self-consistently, $\Gamma_\infty$ can be greater or smaller
than (\ref{eq:gaminf}), as we detail in this paper.

An upper limit on the photon compactness is derived by demanding that the fluid be optically thin at breakout,
$\tau_{\rm es}(r_*) \leq 1$.  Then the photosphere sits at $r_\tau < r_*$, and the compactness
(\ref{eq:chidef}) at breakout is 
\be
\chi_* = {r_\tau\over r_*}\chi(r_\tau) \simeq 6\frac{r_\tau}{r_*}\Gamma^2(r_\tau){dL_{\gamma *}/d\Omega\over (d\dot M/d\Omega)c^2}. 
\ee
Hence $\Gamma_\infty \propto \Gamma(r_\tau)^{3/4}$ in a jet of a fixed ${\cal R}$.
The radiation flux at breakout is below the Poynting flux if
\be\label{eq:chimax}
\chi_* \lesssim 6\frac{r_\tau}{r_*}\Gamma^2(r_\tau)\sigma_*.
\ee
In Figure \ref{fig:chisigma} we relate the compactness, magnetization and optical depth at breakout for different ratios of Poynting to radiation flux. 

The aforementioned hot jet moving at $\Gamma(r_\tau) \sim 10$ at its photosphere, with a magnetization 
$\sigma \sim {\cal R} \sim 10^5$, can be pushed by radiation pressure up to a terminal Lorentz factor
$\Gamma_\infty \sim (1\times 10^9)^{0.25} \sim 200$.   More relativistic material
accelerated by the Lorentz force will, alternatively, feel a retarding force from the radiation field.

%%%%%%%%%%%%%%%%%%%%%%%%%%%%%%%%%%%%%%%%%%%%%%%%%%%%%%%%%%%%%%%%%%%%%%%%%%%%%%%%%%%

\subsection{Cold MHD Flow without Radiation Pressure}

To begin, we review the case where photons are absent, and assume a thin jet in which the magnetic field
lines have poloidal angle $\theta_f \ll 1$.
Only a small differential bending of the field lines is needed to push a cold magnetofluid to large $\Gamma$:
their polar angle must deviate from conical geometry by $\delta\theta_f/\theta_f \sim \Gamma/\sigma$.  
A basic constraint on the rate of bending is obtained if the transverse component of the Lorentz force
in the matter rest frame is limited to 
\be
{1\over c} \biggl|{\bf J}'\times{\bf B}' - \bbeta\cdot({\bf J}'\times{\bf B}'){\bbeta\over\beta^2}\biggr| 
\lesssim {B_\phi'^2\over 4\pi r/\Gamma}.
\ee
The prime $'$ denotes the rest frame, in which $r/\Gamma$ is a characteristic causal distance, and 
$\bbeta \equiv {\bf v}/c$.  Then
\be\label{eq:causalcons}
r{d(\delta\theta_f)\over dr} \lesssim {1\over \Gamma},
\ee
so that typically $\delta\theta_f \sim 1/\Gamma$.  This is seen in the dynamic cold MHD calculations
of \cite{tchek09}, where strong Lorentz forces are generated in a narrow fan near the jet edge.

Conservation of magnetic flux implies that $B_p r^2\theta_f d\theta_f$ = const.  Hence, writing
$\theta_f = \theta_f(r_*) + \delta\theta_f \equiv \theta_{f*} + \delta\theta_f$, equation (\ref{eq:sigmaf}) becomes 
\be\label{eq:sigev}
\sigma = \sigma_*\left[1 - {d(\delta\theta_f/\theta_{f*})\over d\ln\theta_{f*}}\right].
\ee
The change in the ratio of Poynting and mass fluxes can then be written
\be\label{eq:sigmaevol}
{1\over \sigma_*}{d\sigma\over dr} = - \theta_{f*}{d\over d\theta_{f*}}\left({K\over \Gamma \theta_{f*} r}\right).
\ee
The envelope function $K(\theta_f) \sim 1$ away from the jet axis,
and vanishes on the axis given the assumption of axisymmetry.

%%%%%%%%%%%%%%%%%%%%%%%%%%%%%%%%%%%%%%%%%%%%%%%%%%%%%%%%%%%%%%%%%%%%%%%%%%%%%%%%%%%

\subsection{Radiation Force}

Given the relativistic motion of the matter, the radiation field can be assumed to interact with it via Thomson scattering.
In a frame where the matter moves with velocity $\bbeta c$, and a photon has wave vector ${\bf k} = k\hat k$, 
a scattering charge feels a force
\be\label{eq:Flab2}
{\bf F}^{\rm rad} =
\frac{\sigma_{T}I}{c}\int\left(1-\bbeta\cdot\hat{k}\right)\left[\hat{k}
-\bbeta\Gamma^{2}\left(1 - \bbeta\cdot\hat{k}\right)\right]d\Omega.
\ee 
Here $I(\mu) = \int d\nu I_\nu$ is the spectral intensity integrated over frequency, and
$\mu$ is the direction cosine $\mu = {\rm cos}(\theta) = \hat k\cdot\hat r$.

It is useful to define angular moments of $I$,
\be\label{eq:fn}
F_n \equiv 2\pi \int d\mu (1-\mu)^n I(\mu) \equiv 2\pi \int (\Delta\mu)^n I(\Delta\mu),
\ee
so that for a narrow beam, $\Delta\mu \simeq {1\over 2}\theta^2 \ll 1$, the radiation energy flux is approximately equal to 
$F_0 = {\cal R}\,\Gamma\rho c^3$.  
We may define a frame moving at Lorentz factor $\Gamma_{\rm eq}$ (speed $\beta_{\rm eq}$) in which the radiation field is nearly 
isotropic and the radiation force vanishes.  Defining the bulk frame radiation energy density by $u'$, one has
\be
I(\Delta\mu) \simeq {cu'/4\pi\over [\Gamma_{\rm eq}(1-\beta_{\rm eq}\mu)]^4}
             = {I(0)\over(1 + 2\Gamma_{\rm eq}^2\Delta\mu)^4}.
\ee
Substituting this into (\ref{eq:fn}) yields the relations
\be\label{eq:moment}
F_1 = {1\over 4\Gamma_{\rm eq}^2} F_0; \quad\quad F_2 = {1\over \Gamma_{\rm eq}^2} F_1.
\ee
Expanding the radiation force (\ref{eq:Flab2}) in $\Delta\mu$ gives
\be\label{eq:Flab3}
F_r^{\rm rad} = {\sigma_T \over 4\Gamma^2 c}\left[F_0 - 4\Gamma^4\,F_2\right] 
              = {\sigma_T F_0\over 4\Gamma^2 c}\left[1 - \left({\Gamma\over\Gamma_{\rm eq}}\right)^4\right] .
\ee

The main approximation here is that each field line experiences a small differential bending, so that
the bending angle is small compared with $(2\Delta\mu)^{1/2}$.  
This is consistent with rapid acceleration by the Lorentz force near the jet edge \citep{tchek10},
e.g. $\delta\theta_f \sim \theta_j-\theta_f \ll \theta_f$.
In the context of GRBs, we can also assume that the flow has propagated far
outside the light cylinder, so that $\beta_\phi \ll 1$ and the toroidal radiation force can be neglected.

%%%%%%%%%%%%%%%%%%%%%%%%%%%%%%%%%%%%%%%%%%%%%%%%%%%%%%%%%%%%%%%%%%%%%%%%%%%%%%%%%%%

\subsection{Transfer of a Narrow Photon Beam Near a Relativistic Photosphere}

We work with the transfer equation in the inertial frame into which the outflow is expanding;
a prime denotes the matter rest frame.  The radiation transfer equation is written (e.g. Mihalas 1978)
\be
{dI_\nu\over ds} = \alpha_{\rm es} (S_\nu-I_\nu),
\ee
where
\be\label{eq:abs}
\alpha_{\rm es} = \Gamma(1-\beta\mu)\alpha_{\rm es}' = {\Gamma\rho\sigma\over\bar m} (1-\beta\mu) \equiv 
{\alpha_{{\rm es}*} \over \beta(r/r_*)^2} (1-\beta\mu)
\ee
is the grey scattering opacity, and 
\be
S_\nu = {1\over [\Gamma(1-\beta\mu)]^3} S_\nu' = {1\over 2[\Gamma(1-\beta\mu)]^3}\,\int d\mu' I_\nu'
= {1\over 2\Gamma(1-\beta\mu)]^3}\int d\widetilde\mu (1-\beta\widetilde\mu) I_{\widetilde\nu}
\ee
is the source function in the isotropic scattering approximation.  The Doppler relation between unscattered and scattered 
photon frequencies is $\widetilde\nu(1-\beta\widetilde\mu) = \nu(1-\beta\mu)$, and
\be
I_{\nu}' = [\Gamma(1-\beta\mu)]^3 I_\nu; \quad\quad d\Omega' = 2\pi d\mu' = {2\pi d\mu\over [\Gamma(1-\beta\mu)]^2}
\ee
are the usual transformations.  Integrating over frequency gives
\be
{dI\over ds} = \alpha_{\rm es}(S-I);\quad\quad
S = \int S_\nu d\nu = {1\over 2\Gamma^2(1-\beta\mu)^4}\,\int I(\widetilde\mu) (1-\beta\widetilde\mu)^2 d\widetilde\mu.
\ee
Setting $\Delta\mu \rightarrow 0$, the path length is $ds = dr/\mu \simeq dr$, and one has
$d\Delta\mu/dr \simeq -2\Delta\mu/r$.  Making use of 
\be
{d\over dr}F_n = -{2(n+1)\over r}\,{1\over 2}\int d\mu (\Delta\mu)^n I  + {1\over 2}\int d\mu (\Delta\mu)^n {dI\over dr}, 
\ee
gives
\be\label{eq:f0}
{1\over r^2}{d\over dr}\left(r^2 F_0\right) = -{\alpha_{{\rm es}*}\over 4\Gamma^2(r/r_*)^2}\left(1-4\Gamma^4{F_2\over F_0}\right)F_0
= -{\alpha_{{\rm es}*}\over 4\Gamma^2(r/r_*)^2}\left(1-{\Gamma^4\over \Gamma_{\rm eq}^4}\right)F_0,
\ee
and
\be
{1\over r^4}{d\over dr}\left(r^4F_1\right) = 
{\alpha_{{\rm es}*}\over 8\Gamma^4(r/r_*)^2}\left(1-{\Gamma^4\over \Gamma_{\rm eq}^4}\right)F_0,
\ee
where we have set $\beta \rightarrow 1$ in the coefficient.
These two equations, in combination with (\ref{eq:moment}), allow us to evolve $\Gamma_{\rm eq}$ near the scattering photosphere.

The radial evolution of $\Gamma$ and ${\cal R}$ is obtained by differentiating (\ref{eq:etot}), substituting
(\ref{eq:sigmaevol}) and (\ref{eq:f0}), and expressing the radiation energy flux in terms of ${\cal R}$,  
\be\label{eq:gammaevol}
r{d\Gamma\over dr} = \sigma_* {d\over d\theta_f}\left({K\over \Gamma}\right) - r{d{\cal R}\over dr};
\quad\quad
{d{\cal R}\over dr} = -{\alpha_{{\rm es}*}\over 4\Gamma^2(r/r_*)^2}\left(1-{\Gamma^4\over\Gamma_{\rm eq}^4}\right){\cal R}.
\ee

%%%%%%%%%%%%%%%%%%%%%%%%%%%%%%%%%%%%%%%%%%%%%%%%%%%%%%%%%%%%%%%%%%%%%%%%%%%%%%%%%%%

\subsection{Numerical Results}
Profiles are obtained for $\Gamma(x)$, $\Gamma_{\rm eq}(x)$, $R(x)$ and $\theta_f(x)$ by integrating 
in the radial dimension.  We have made the substitutions $d/d\theta_f\rightarrow \delta\theta_{\rm gradient}^{-1}$ 
and $K = \theta_{f*}/\theta_j$ in (\ref{eq:gammaevol}).  This choice forces the field-line bending 
to zero near the center of the jet.  The gradient angle $\delta\theta_{\rm gradient}$ is of the order of 
$\theta_j$, but may be smaller near the edge of the jet as it emerges from a confining 
medium.  For example, the strong acceleration seen near the jet edge in the simulations of \cite{tchek10} 
is consistent with $\delta\theta_{\rm gradient} \sim \gamma^{-1} \sim 0.1 \theta_j$;  it would presumably
be reduced if the jet did not have a sharp edge.  In our fiducial model we consider a field line anchored at 
$\theta_{f*}=0.1$, and take the jet opening half-angle to be $\theta_j=0.2$. To illustrate the effect of jet breakout 
on the flow parameters, we begin the integration just inside the breakout radius, $r_i=0.8r_*$. 
\begin{figure}[h]
\epsscale{0.92}
\plottwo{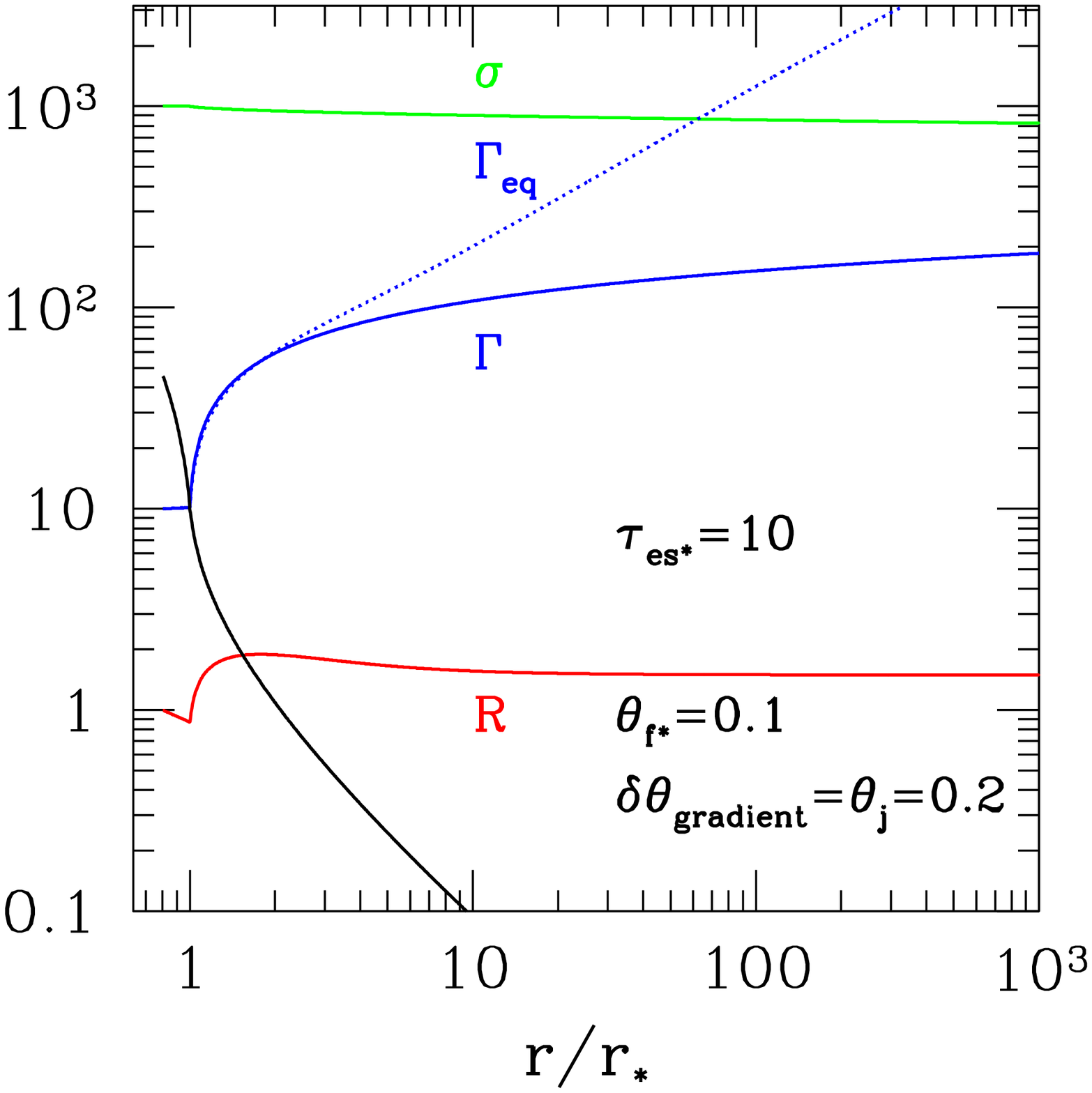}{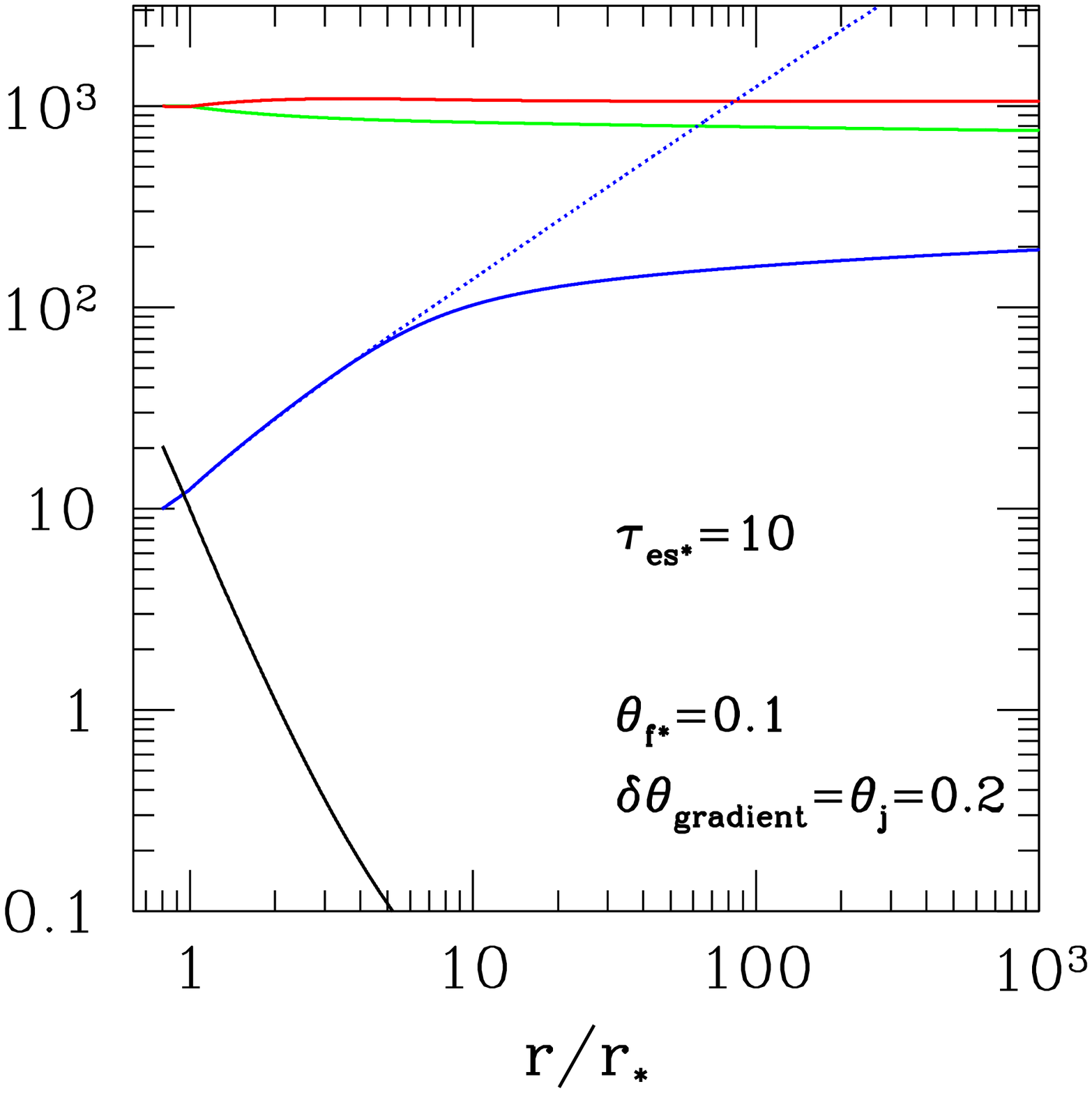}
%\plottwo{smGamCausalExtAtaust10Rs1.eps}{smGamCausalExtAtaust10Rs1000.eps}
\caption{Profiles of Lorentz factor $\Gamma$ (blue), magnetization $\sigma$ (green), and radiation energy flux ${\cal R}$ (red),
in a flaring MHD outflow ($\sigma_* = 10^3$) with radiative driving.  The radiation force vanishes in a frame moving with Lorentz factor
$\Gamma_{\rm eq}$ (dotted blue).  Scattering depth integrated from radius $r$ to infinity (black curve) drops rapidly 
from $\tau_{\rm es*} = 10$ as the flow accelerates. The gradient scale $\delta\theta_{\rm gradient}$ sets the relative degree of flaring between neighboring field lines. \textit{Left panel:} ${\cal R}_i=1$, \textit{right panel:} ${\cal R}_i=1000$.
Other parameters:  $r_i=0.8r_*$, $\Gamma_i = \Gamma_{eq,i} = 10$.}
\label{fig:GamCausal1}
\end{figure}
\begin{figure}[h]
\epsscale{0.92}
\plottwo{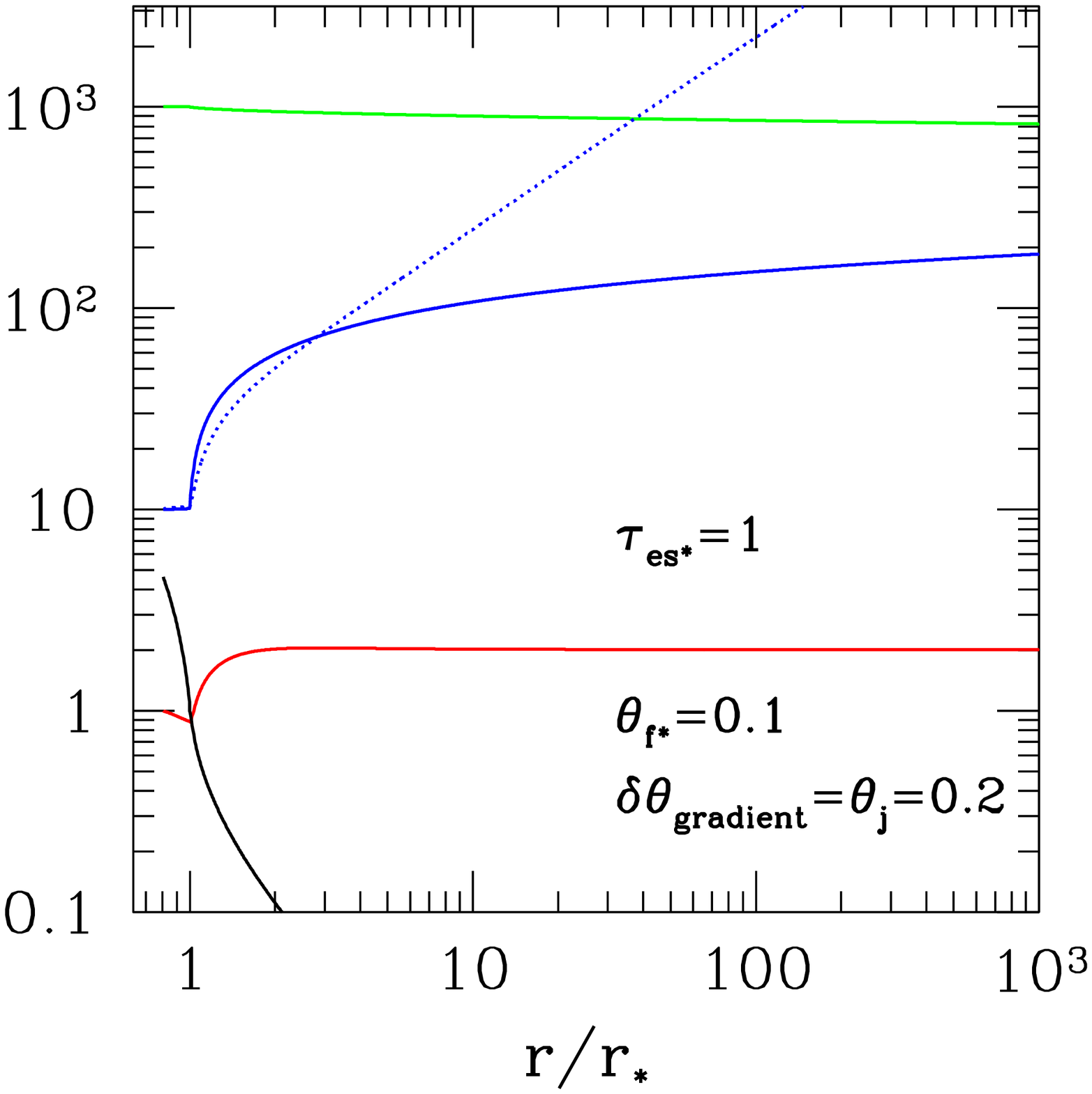}{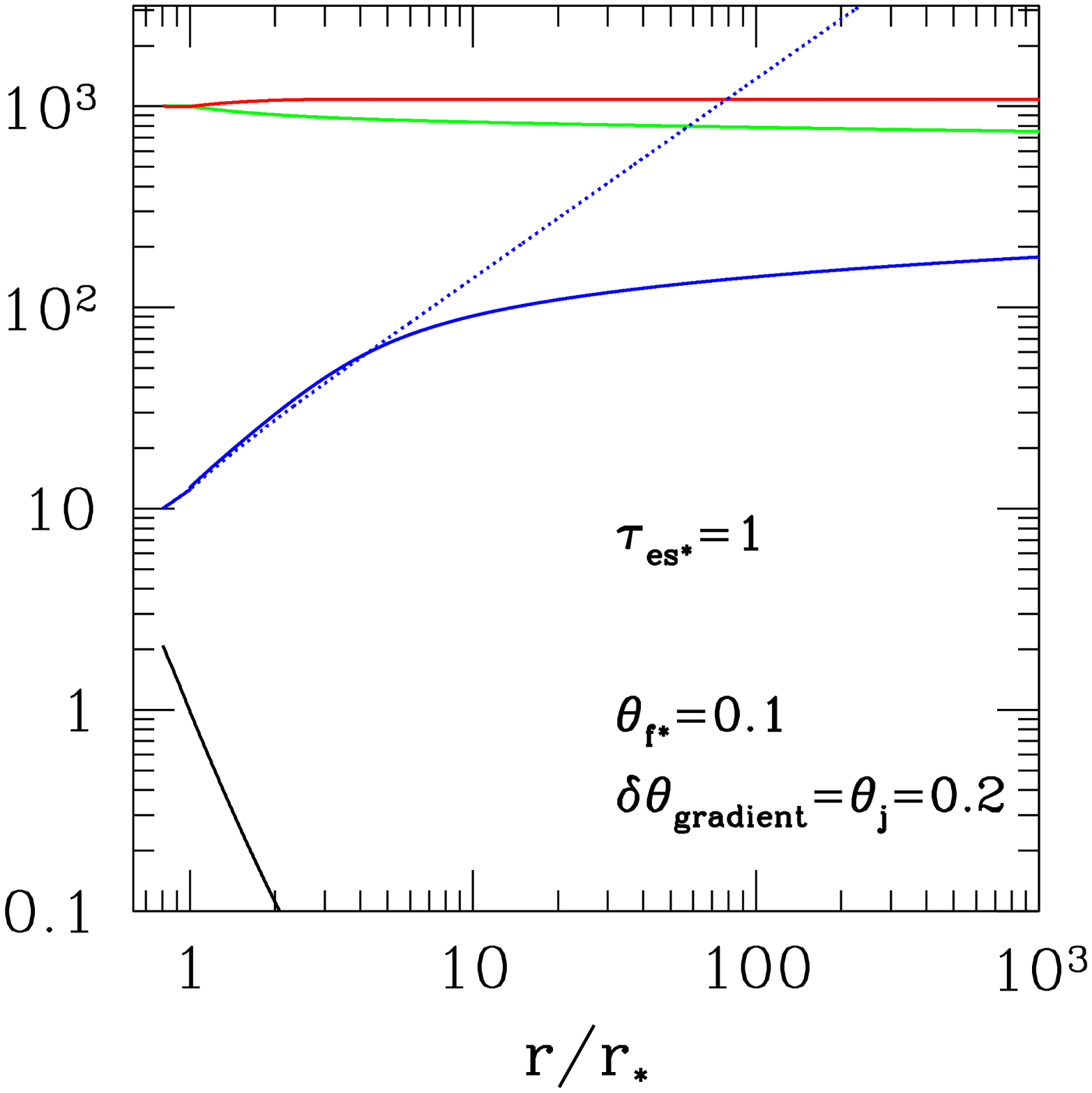}
\plottwo{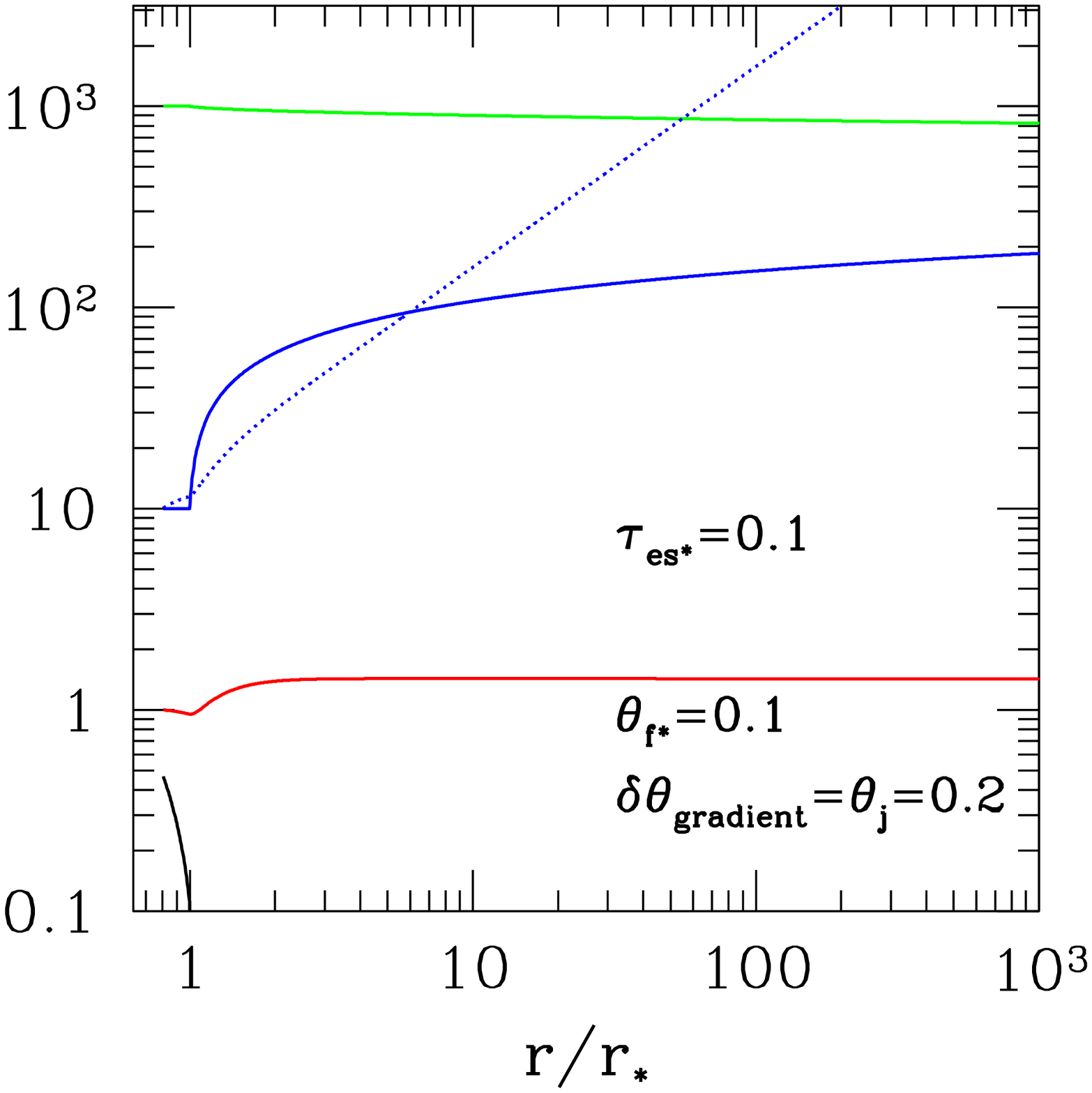}{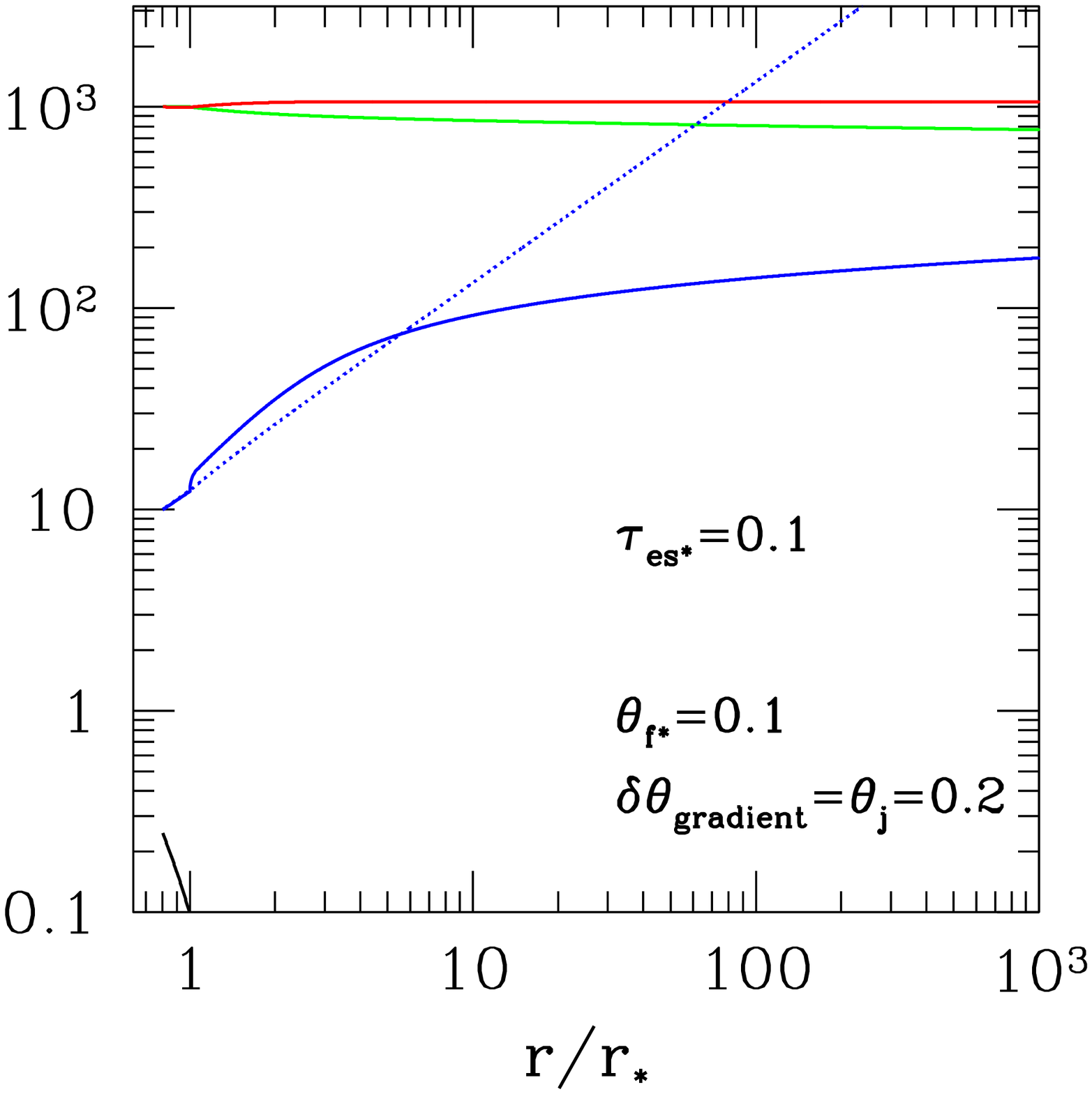}
%\plottwo{smGamCausalExtAtaust1Rs1.eps}{smGamCausalExtAtaust1Rs1000.eps}
%\plottwo{smGamCausalExtAtaust0_1Rs1.eps}{smGamCausalExtAtaust0_1Rs1000.eps}
\caption{Same as Figure \ref{fig:GamCausal1}, but for $\tau_{\rm es*} = 1$ (top plots) and $\tau_{\rm es*} = 0.1$
(bottom plots).}
\label{fig:GamCausal2}
\end{figure}
\begin{figure}[h]
\epsscale{0.85}
\plottwo{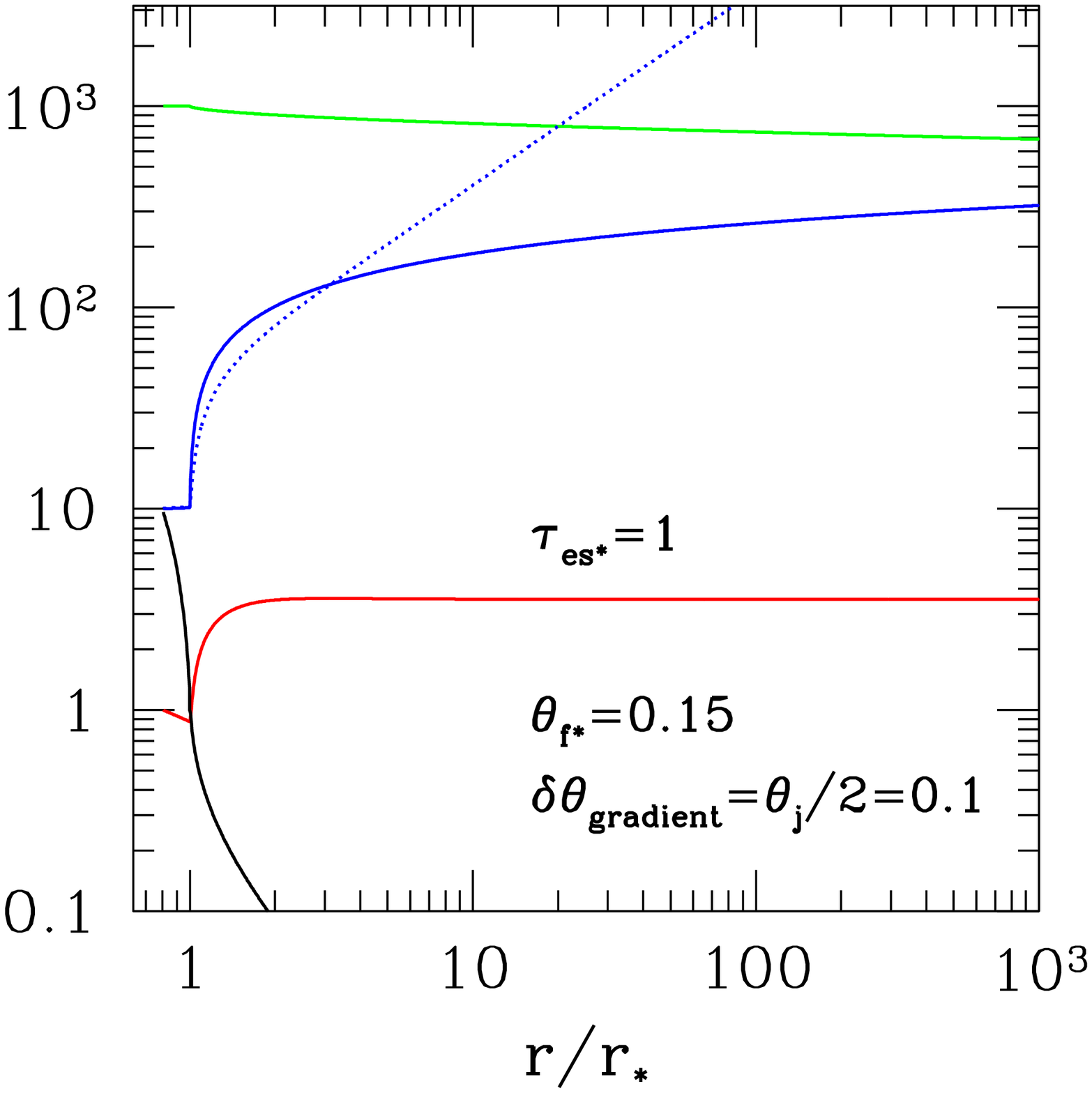}{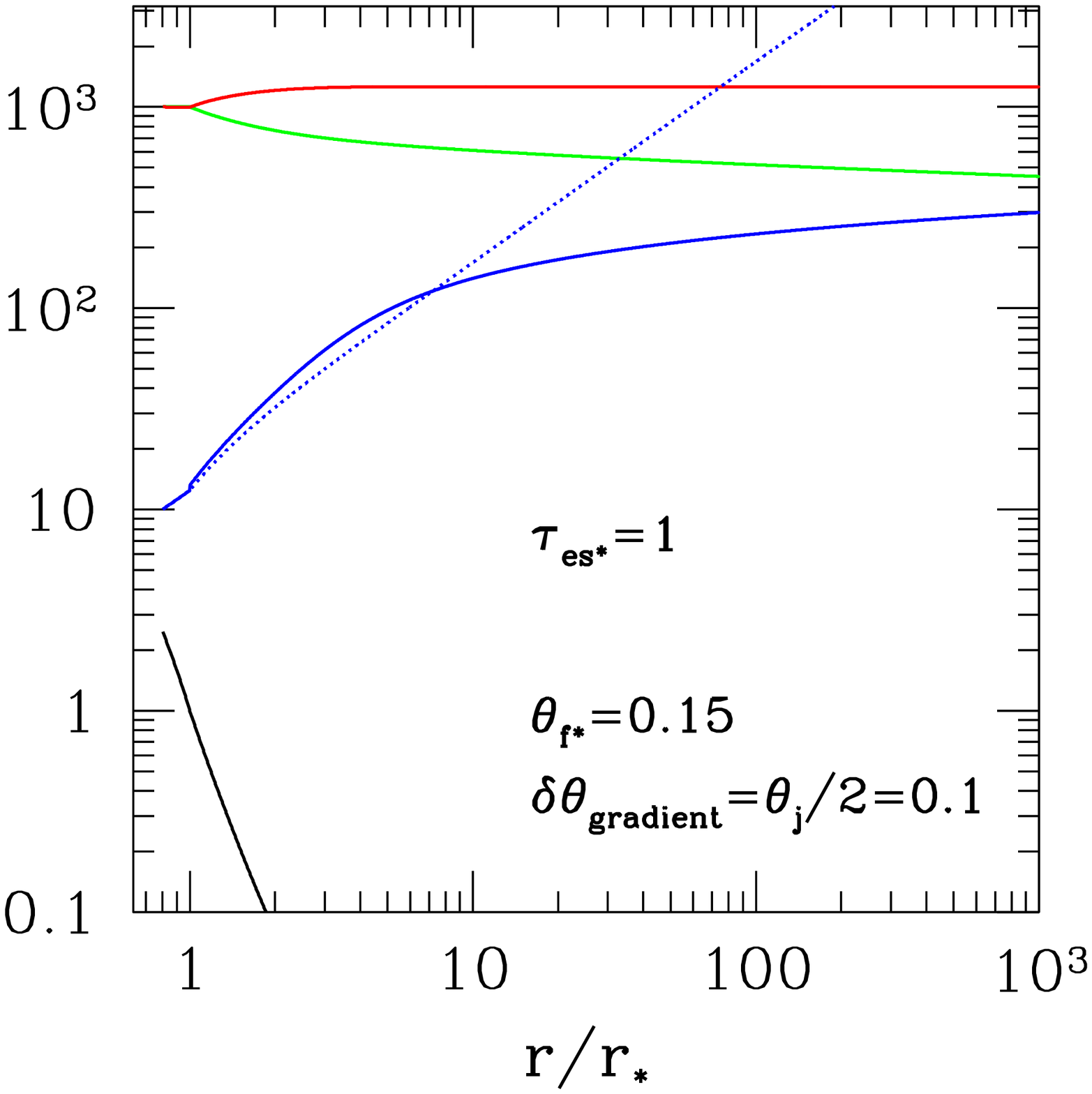}
\plottwo{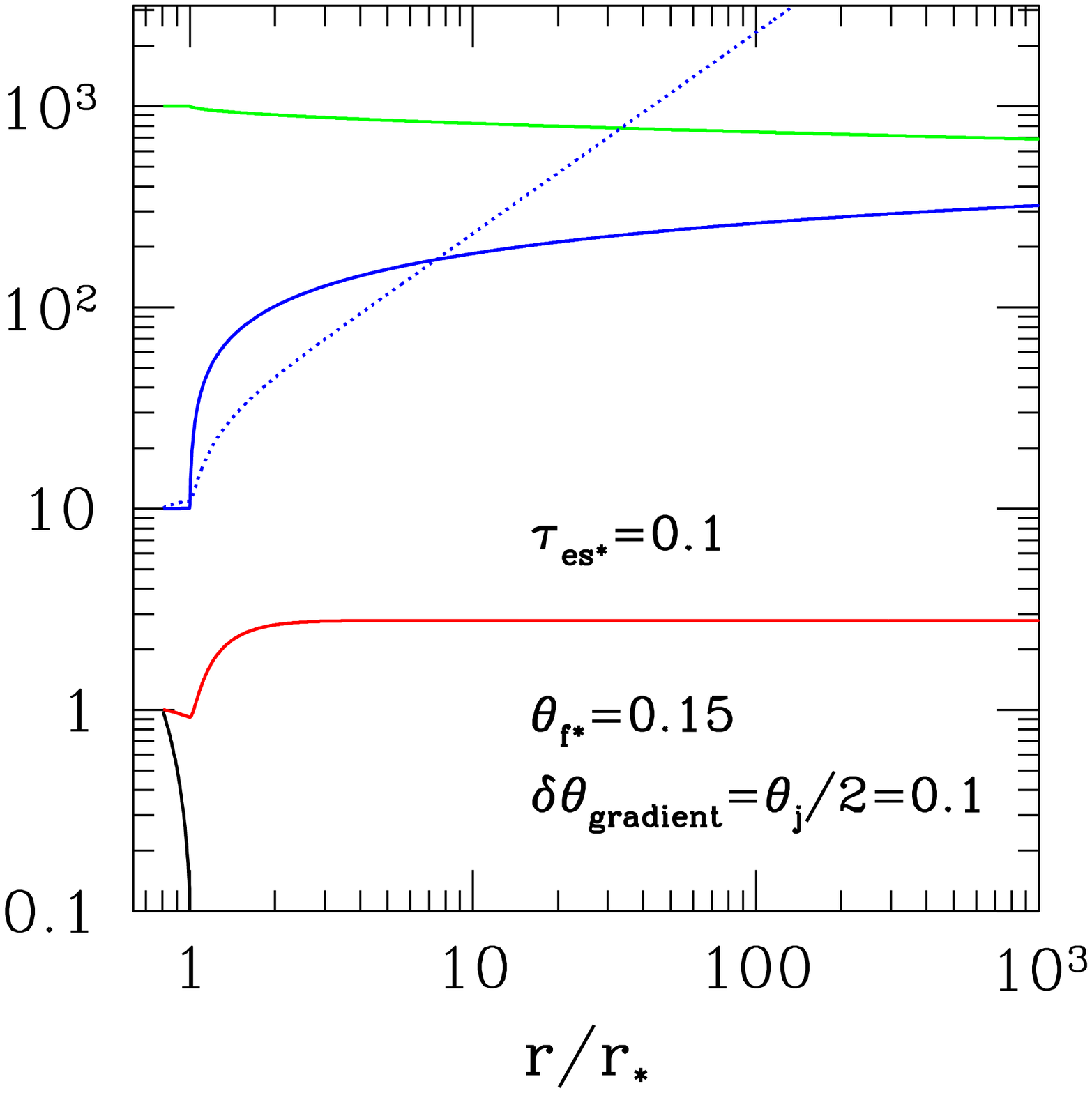}{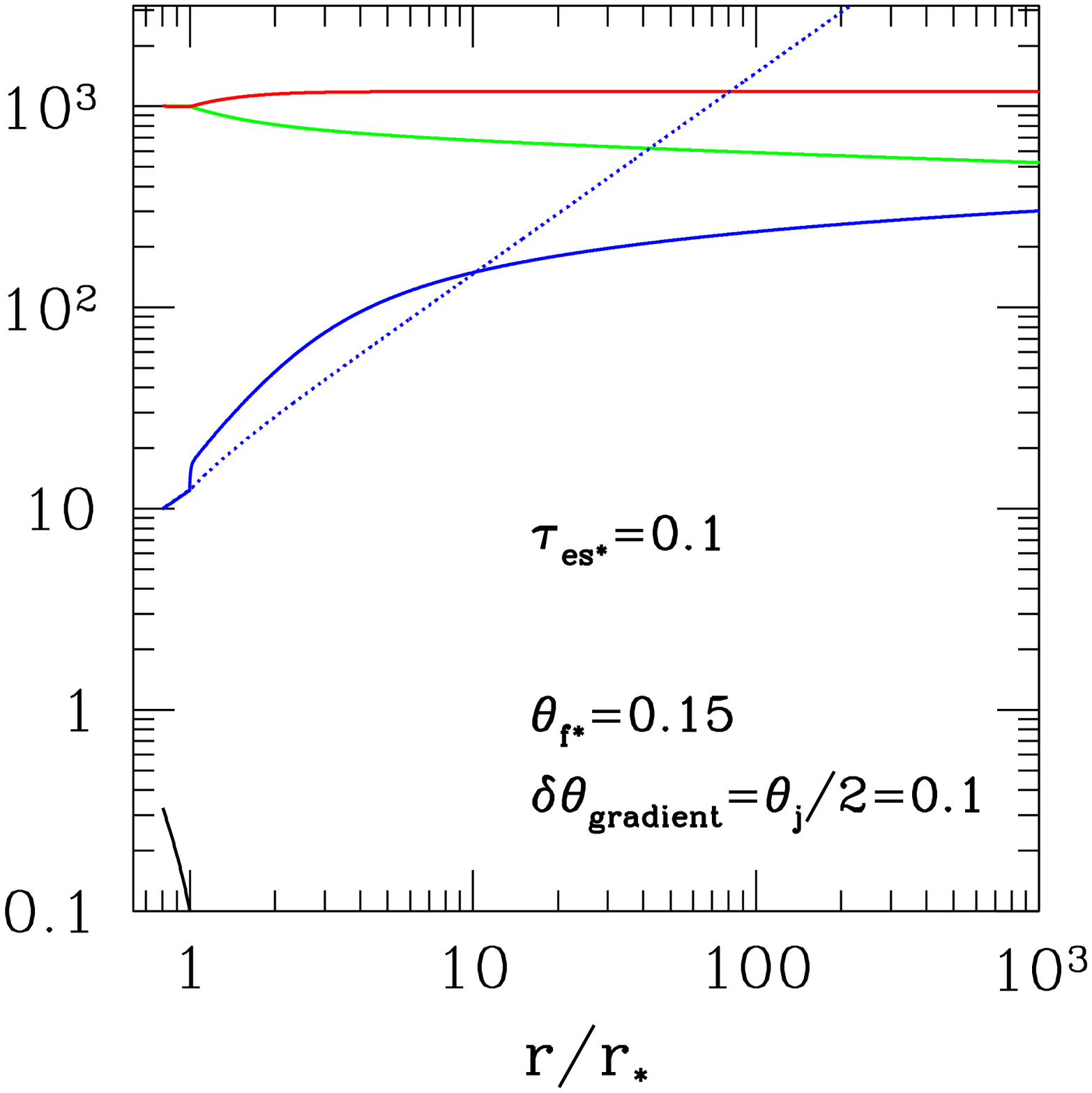}
%\plottwo{smGamCausalExtBtaust1Rs1.eps}{smGamCausalExtBtaust1Rs1000.eps}
%\plottwo{smGamCausalExtBtaust0_1Rs1.eps}{smGamCausalExtBtaust0_1Rs1000.eps}
\caption{Same as Figures \ref{fig:GamCausal1}, \ref{fig:GamCausal2}, except $\delta\theta_{\rm gradient} = \theta_j/2 = 0.1$
and $\theta_{f*} = 0.15$.  Optical depth at breakout:  $\tau_{\rm es*}=10$ in top plots and $\tau_{\rm es*} = 0.1$ in bottom plots.}
%\vskip .05in
\label{fig:GamCausal3}
\end{figure}

In our first set of integrations, the scattering optical depth is chosen to be large at the 
breakout radius.  At this point, the radiation is still tied to the matter, 
$\Gamma_{\rm eq}(r_*) = \Gamma(r_*)$, but the rapid MHD acceleration experienced by the flaring 
jet quickly forces a transition to low $\tau_{\rm es}$.  The optical depth for a photon 
propagating radially from $r=r_*$ is given by equation (\ref{eq:taustar}), 
and is thus unknown \textit{a priori}.  To impose a particular value of $\tau_{\rm es}(r_*)$, we first 
choose an approximate value of $\alpha_{\rm es *}$, evolve the equations of motion and then iterate.
The Lorentz force term in $d\Gamma/dr$ is only valid outside the fast magnetosonic critical point,
and so we take $\Gamma(r_i) \gtrsim \Gamma_c \simeq \sigma_*^{1/3}$.  Radial integrations are done using
a 5th-order Runge-Kutta algorithm with adaptive step size (see Sections 7.3, 7.5 of \citealt{kiusalaas10}). 

Results are plotted in Figure \ref{fig:GamCausal1} for an outflow with magnetization $\sigma_i = 
\sigma_* = 1000$, and both low and high radiation fluxes (${\cal R}_i = 1, 1000$) at the inner boundary.  
The action of the Lorentz force is concentrated at a small radius where the flaring is most severe, 
causing an increase in $\Gamma$ that is initially much faster than linear. 
When the radiation energy flux is weak compared 
with the magnetic Poynting flux, the outflow experiences strong but logarithmic acceleration after breakout with 
$\Gamma\propto\ln^{1/2}x$, a direct consequence of causally limited flaring, 
$\delta\theta_f \sim 1/\Gamma$ [equation (\ref{eq:gammaevol})].  While the optical depth is large, the 
photon field is advected with the plasma, remaining nearly isotropic in the comoving frame.  Once the 
optical depth falls below unity the photon field decouples and is free to self-collimate,
so that $\Gamma_{\rm eq}\propto x$.  

High radiation fluxes (${\cal R} \sim \sigma$) force the flow back to the shallower profile
$\Gamma \simeq \Gamma_{\rm eq}$, even while the magnetic flaring grows stronger
($1/\Gamma$ is larger).  The acceleration zone is therefore widened in the radial direction compared with the 
radiation-free jet.

Quite generally, we find that the terminal Lorentz factor is insensitive to the initial radiation
energy flux.  The growth in radiation energy flux outside the photosphere is therefore largely compensated 
by a further reduction in outflow magnetization.

We also consider a low scattering depth at the breakout radius. In this case the matter is only weakly coupled to 
the radiation field, and the Lorentz factor is forced well above $\Gamma_{\rm eq}$ a small distance outside the 
breakout radius (Figure \ref{fig:GamCausal2}). As in the case of higher optical depths, high radiation fluxes limit 
the growth of $\Gamma$ and force it toward $\Gamma_{\rm eq}$.  But the mismatch between $\Gamma$ and $\Gamma_{\rm eq}$
remains unless ${\cal R} \gtrsim \sigma$.  Energy is transferred from the magnetic field to the photons,
resulting in a significantly modified spectrum (Section \ref{s:spectrum}).

Faster jet flaring, corresponding to a jet edge with $\delta\theta_{\rm gradient} = 0.5 \theta_j$,
produces faster acceleration and terminal Lorentz factors closer to $\sigma_*$,  but otherwise
qualitatively similar behavior (Figure \ref{fig:GamCausal3}). 

A novel effect becomes clear when the compactness is large:  the magnetization can show a significant
reduction, dropping significantly below ${\cal R}$ and even $\Gamma$, and therefore resulting in a {\it weakly}
magnetized outflow. This is caused by the strong radiation drag at small radius, which restricts the 
growth of $\Gamma$ which allows for stronger jet flaring.

%%%%%%%%%%%%%%%%%%%%%%%%%%%%%%%%%%%%%%%%%%%%%%%%%%%%%%%%%%%%%%%%%%%%%%%%%%%%%%%%%%%
%%%%%%%%%%%%%%%%%%%%%%%%%%%%%%%%%%%%%%%%%%%%%%%%%%%%%%%%%%%%%%%%%%%%%%%%%%%%%%%%%%%
%\section{Narrow Jet with Optically Thin Radiation Field}\label{s:eom}
\vskip 5 \baselineskip
\section{Flaring, Hot Magnetized Jet:  Transparent Flow \\ across the Fast Critical Surface (Model II)}\label{s:eom}

The dynamics of the outflow can be calculated more precisely in the optically thin
regime, where the radiation field is prescribed at an emitting surface (radius $r_s$) and passively 
collimates outside that surface.  This allows us to study the critical point structure of the flow, at 
the price of neglecting the effects of multiple scattering.  In the spherical case, the emission surface 
could, if one wanted, be identified with the physical surface of a star.  But the model of a 
passively collimating photon field can also be applied to outflows that are already relativistic at the 
photosphere, including those
with a jet geometry.  In this section, we solve the wind equations for a steady, flaring jet which is 
optically thin but sub-magnetosonic at breakout.  In contrast to the model presented in Section \ref{s:causaljet}, 
here we prescribe the flaring profile in advance.

\begin{figure}[h]
\centerline{\includegraphics[width=0.8\hsize]{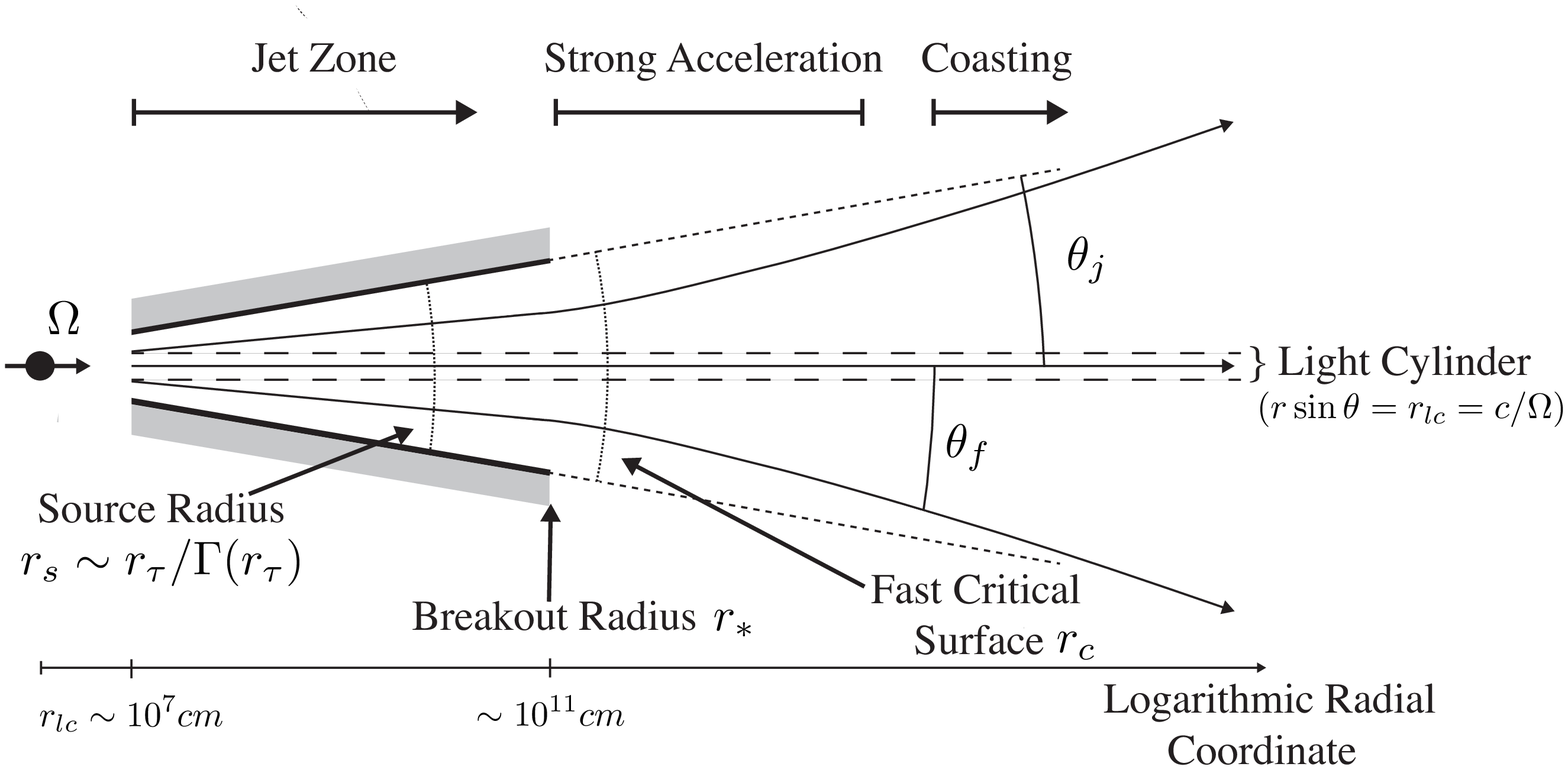}}
%\centerline{\includegraphics[width=1.2\hsize]{JetFigureSec3.eps}}
\caption{Geometry and approximate scale of the flow solutions for jet model II.}
\vskip .1in
\label{fig:JetSec3}
\end{figure}
The flow geometry is shown in Figure \ref{fig:JetSec3}.  The photon source radius sits in the confined portion 
of the jet, inside the breakout radius $r_*$.  After breakout, the optically thin flow is accelerated 
though the magnetosonic surface, whose location and shape are calculated self consistently  (analytic approximations
to the position of the critical surface can be found in Section \ref{sec:MSsurface}).  
We follow the flow along a field line $\theta_f(r)$, situated well outside the light cylinder, from just 
outside the source radius.

Our solution to the flow below the breakout point formally is in the optically thin regime.  
Because the jet has already typically attained relativistic
motion before breaking out, we can view the photon emission as arising from a virtual surface
located below the physical photosphere, at a radius $r_{s,\rm eff} \sim r_\tau/\Gamma(r_\tau)$.  At
a high radiation intensity, the matter is locked into the bulk frame defined by the photon field,
$\Gamma \simeq \Gamma_{\rm eq}\sim (r-r_s)/(\theta_j r_s)$, around breakout.
This means that the flow profile closely mimics an optically thick, radiation-dominated flow inside breakout,
and we expect that our flow solutions should adequately represent the dynamics of a jet which encounters 
a photosphere at a radius $r_* \sim r_\tau > r_s$.   

Our procedure is first to choose the poloidal field geometry 
and radiation profile, and then evolve the energy and angular momentum along the poloidal flux surfaces.  
A simple description of the photon field is possible
when the jet geometry is locally spherical - that is, when the streamlines are conical inside breakout.  This
constrains the non-radial Lorentz force to vanish at $r < r_*$, which in the small-angle limit can be written
as
\be
{1\over c}({\bf J}\times{\bf B})_\theta \simeq - {B_\phi\over 4\pi r\theta}{d(B_\phi \theta)\over d\theta} = 0
\quad \Rightarrow \quad B_\phi \propto \theta^{-1}.  
\ee
A jet with such a line current profile will {\it de-collimate} at $r > r_*$, as the external pressure is removed.  
This decollimation leads to rapid outward acceleration of the cold matter entrained in the jet, even in the
absence of radiative forcing.

Other jet profiles are easily constructed and may be more natural:  in the cold MHD jet calculation of
\cite{tchek10}, the confining surface has a parabolic structure inside breakout, transitioning to a conical
geometry outside.  Nonetheless, the radial Lorentz factor profiles that we obtain are (in the absence of radiation) 
very similar to those of \cite{tchek10}, and only depend on the magnitude of the differential flaring
between magnetic flux surfaces.  

We focus on the local dynamics within magnetic flux surfaces, taking into 
account the effect of radiation pressure.  This longitudinal dynamics is sensitive to the relative flaring
rates of neighboring flux surfaces, but not to the global profile of Poynting flux transverse to the jet axis.
Although the transverse force balance is not explicitly taken into account, we do check that i) 
the degree of magnetic flaring is consistent with causal stresses; and ii) that the transverse component of
the radiation force is weak compared with the transverse Lorentz force (so that the radiation flow is not
strong enough to comb out the field lines into a conical geometry:  see Appendix D).

%%%%%%%%%%%%%%%%%%%%%%%%%%%%%%%%%%%%%%%%%%%%%%%%%%%%%%%%%%%%%%%%%%%%%%%%%%%%%%%%%%%

\subsection{Jet Properties}\label{s:jetgeom}

To construct an optically thin radiation field, we consider the simplest case of uniform intensity $I = \int I_\nu d\nu$
at the emission radius $r_s$, as we did in Paper I, but now restrict the sampling of the radiation 
field to polar angles $\theta < \theta_j$.  The emission patch covers a small angular disk of
area $\pi(\theta_j r_s)^2$ (Figure \ref{fig:JetSec3}), and the luminosity per sterad is
$dL_{\gamma *}/d\Omega \sim \pi \theta_j^2 I$.   This allows an analytic calculation of the radiation 
force acting on a particle of arbitrary Lorentz factor and direction, which is presented in Appendix A.
This result generalizes the simpler angular moment formalism used in Section \ref{s:causaljet} and presented
in equation (\ref{eq:Flab3}). 

We showed in Paper I that if we normalize the photon intensity and the angular width of the photon beam
by fixing i) the radiation force and ii) the relativistic frame $\Gamma_{\rm eq}$ in which this force vanishes,
then other quantities, such as the mean power radiated by an electron in its bulk frame, are nearly identical to
those obtained from a radiation field that is isotropic at a relativistically moving photosphere.

The radiation streams freely outward at $r > r_s$, and its cone contracts with increasing radius.  
The size and orientation of this cone now vary with distance from the jet axis (in contrast with
the case of a spherical emission surface; Paper I).  There is generally a misalignment of the direction 
of peak radiation intensity with respect to both the radial direction, and the local flow direction.  
The alignment is strongest at a small but finite distance from the rotation axis, and produces
a peak in the radiation force there.\footnote{A tiny portion of the outer jet sits inside the light
cylinder of the engine, and formally retains strong rotation of its field lines, which reduces the overlap
between the radiation and fluid flows.  This effect will, in practice, probably be eliminated by turbulence
in the jet.}

We normalize distances to $r_s$, but measure the photon compactness (\ref{eq:chidef}) at the breakout radius,
\be\label{eq:dimvar}
x\equiv {r\over r_s}; \quad\quad  \omega \equiv {\Omega_f r_s\over c}\quad\quad
\chi_*\equiv 
%\frac{\sigma_{T}}{r_{*}\bar{m}c^{3}} \frac{dL_{*}}{d\Omega}
\frac{\sigma_{T}r_{*}\pi\theta_j^2 I}{\bar{m}c^{3}\left(x_{*}-1\right)^{2}}.
\ee
In a GRB outflow, the photosphere generally lies outside the light cylinder
of the rotating engine, so we take $\omega = \Omega_f r_s/c > 1$ in our calculations.  In this context, 
the magnetization can be most simply defined as\footnote{This definition, following \cite{michel69} and \cite{goldreich70},
differs in terms $O(v_\phi/\Omega_f r\sin\theta_f)$ to $\sigma = (c^2d\dot M/d\Omega)^{-1}dL_P/d\Omega$, the 
definition used in Section \ref{s:causaljet}; and to
$\sigma = (c^3 d\dot M/d\Omega)^{-1}\dot \Phi_\phi^2$, where $\dot\Phi_\phi = v_r r \sin\theta_f B_\phi$ is the advection rate
of toroidal flux.  We use equation (\ref{eq:sigma}) is independent of radius if the poloidal field is restricted to be purely radial, 
but the last two definitions are non-constant at $O(v_\phi/\Omega_f r\sin\theta_f)$.}
\be\label{eq:sigma}
\sigma \equiv \frac{B_r^2\Omega_f^2(r\theta_f)^2}{4\pi \Gamma\rho v_rc^3}\quad\quad (\theta_f \ll 1).
\ee 
Neglecting the radiation field, the energy and angular momentum per unit rest mass are given by
\be\label{eq:integral}
\mu c^2 = \left(\Gamma - \Omega_f{B_rB_\phi\over4\pi\Gamma\rho v_r}\right)c^2;\quad\quad
\mathcal{L} r_s c = \left(\Gamma v_\phi - {B_rB_\phi\over4\pi\Gamma\rho v_r}\right)r\theta_f,
\ee
and in a steady MHD outflow are conserved along field lines.
They can be written in a dimensionless form,
\be\label{eq:dimintegral}
\mu \;=\; \Gamma - {\sigma\over\omega x\theta_f}\frac{B_{\phi}}{B_{r}} \;\simeq\; \Gamma + \sigma;
\quad\quad
{\cal L} \;=\; \Gamma x \theta_f \beta_{\phi} - {\sigma\over \omega^2 x\theta_f}{B_\phi\over B_r} \;\equiv\; {\cal L}_m + {\cal L}_P.
\ee

%%%%%%%%%%%%%%%%%%%%%%%%%%%%%%%%%%%%%%%%%%%%%%%%%%%%%%%%%%%%%%%%%%%%%%%%%%%%%%%%%%%

\subsection{Poloidal Field Configuration}\label{sec:poloidalprofile}

To incorporate a strong radial Lorentz force into the outflow, we choose the poloidal flux surfaces by fixing the
function $\theta_f(x,\theta_{f*})$, where $\theta_{f*}$ is the polar angle at the breakout radius.   The Lorentz force is
large and positive if neighboring flux surfaces diverge from each other more rapidly than in a monopolar geometry.
The effect of this differential expansion appears in the wind equations via the function
\be\label{eq:br}
A(x,\theta_{f*}) \equiv \frac{d\ln\theta_f}{d\ln\theta_{f*}}; \quad\quad
B_{r} = {B_{r*}x_*^2\over A x^2} \left({\theta_{f*}\over \theta_f}\right)^2;\quad\quad \sigma = {\sigma_*\over A}.
\ee
We focus on the dynamics along a single flux surface, and so do not have to consider the angular dependence of
$B_{r*} = B_r(r_*)$.  The critical point structure of the longitudinal flow is insensitive to angular gradients
in the flow magnetization.

The flaring profile used in our calculations is described in Section \ref{s:flare}, followed by the numerical results.

%%%%%%%%%%%%%%%%%%%%%%%%%%%%%%%%%%%%%%%%%%%%%%%%%%%%%%%%%%%%%%%%%%%%%%%%%%%%%%%%%%%

\subsection{Longitudinal Wind Equations}

We now consider the longitudinal evolution of the outflow variables along a magnetic flux surface.  
A radiation force (\ref{eq:Flab2}) is added to the Euler equation, which becomes
\be\label{e:Euler}
\rho\,\Gamma{\bf v}\cdot\bnabla(\Gamma{\bf v}) =
{1\over 4\pi}\left[(\bnabla\cdot{\bf E}){\bf E} + (\bnabla\times{\bf B})\times{\bf B}\right] + 
             \frac{\Gamma\rho}{\bar{m}}{\bf F^{{\rm rad}}}.
\ee
Taking the dot product of equation (\ref{e:Euler}) with the unit poloidal field vector $\hat B_{\rm p}$,
defining the longitudinal derivative $\partial_l = \hat B_p\cdot\bnabla$, and taking the small-angle limit, we have
\be\label{e:Eulerdl1}
\partial_l\Gamma c^{2}-\frac{v_{\phi}}{r\theta_f}\partial_l {\cal L}_m
= -\frac{B_{\phi}}{4\pi\Gamma\rho r\theta_f}\partial_l\left(r\theta_f B_{\phi}\right) + \frac{F^{\rm rad}_p}{\bar{m}}
\ee
The $\phi$-component of equation (\ref{e:Euler}) is 
\be\label{e:Eulerdl2}
v_p\partial_l {\cal L}_m = \frac{B_p}{4\pi\Gamma\rho}\partial_l\left(r\theta_f B_{\phi}\right)+ (r\theta_f)\frac{F_\phi^{\rm rad}}{\bar{m}}.
\ee
Here ${\cal L}_m$ is the specific matter angular momentum [equation (\ref{eq:dimintegral})], and 
we have made use of the fact that the poloidal flow velocity ${\bf v}_p$ is aligned with ${\bf B}_p$. The Coulomb force 
only contributes to the transverse force balance and does not appear in equations (\ref{e:Eulerdl1}) or (\ref{e:Eulerdl2}).
Both of these features are easily derived by noting that the toroidal electric field vanishes in a steady, axisymmetric MHD wind 
(${\bf E}\cdot{\bf B} = 0$), which implies that ${\bf v}_p\times{\bf B}_p = 0$ and ${\bf E}_p\cdot {\bf B}_p = 0$.

In Appendix A we calculate the radiation force (\ref{eq:Flab2}) in a thin jet, and express the poloidal and toroidal components
in terms of dimensionless functions $R_j$, $P_j$, 
\be\label{eq:radforcedef}
F_p^{\rm rad} = \chi_* {\bar{m} c^2\over r_s} R_j(r,\Gamma,\beta_\phi);\quad\quad  
F_\phi^{\rm rad} = \chi_* {\bar{m} c^2\over r_s} P_j(r,\Gamma,\beta_\phi).
\ee
Rotation of the photon field at the emission surface tends to reduce
the azimuthal drag.  It can be incorporated by modifying the $\beta_\phi$-dependence of equations (\ref{eq:radforcedef}),
as is discussed in Appendix B, but is generally negligible when the outflow lies far outside the light cylinder ($\omega \gg 1$).

A good approximation to the poloidal force can be obtained on field lines $(x\omega)^{-1} \ll \theta_f \ll \theta_j$,
\be 
R_{j}\simeq\frac{x_{*}}{4x^{2}\Gamma^{2}}\left(1-\frac{\theta_{j}^{4}\Gamma^{4}}{3x^{4}}\right),
\ee 
in agreement with equation (\ref{eq:Flab3}).  The result for a spherical emission surface (Paper I) differs only in the 
absence of the factor $\theta^4_j$.  The vanishing of the radiation force occurs at a significantly higher Lorentz factor
when the photon beam is collimated,
\be\label{eq:gameq}
\Gamma_{\rm eq}\simeq\frac{x-1}{\theta_j}
\ee
up to a numerical factor of order unity as shown in Figure \ref{fig:Gammaeq}.

The deviation of the field lines from a purely radial direction is measured by $\Delta\theta_B = B_\theta/B_r$, which 
we take to be small, so that $B_p = (1+\Delta\theta_B^2)^{1/2}B_r \simeq B_r$, $v_p \simeq v_r$. 
As is detailed in Appendix C, the derivatives along field lines on the right hand side of equations (\ref{e:Eulerdl1}) 
and (\ref{e:Eulerdl2}) can be written in the small-angle approximation.  Ignoring the cross-field force balance then 
allows us to express (\ref{e:Eulerdl1}), (\ref{e:Eulerdl2}) as two ordinary differential equations, which can be
re-written in terms of $d\Gamma/dl$ and $d{\cal L}_m/dl$.  The various terms on the right-hand side of these 
equations can be separated into purely magnetocentrifugal pieces (which do not depend on the radiation force), 
the direct radiation force, and a cross term:
\be\label{eq:Gamp}
{d\Gamma\over dl} = \frac{\Gamma'_{\sigma}+\Gamma'_{\sigma\chi}+\Gamma'_{\chi}}{\mu_{\rm eff}}; \quad\quad
{d{\cal L}_m\over dl} = \frac{({\cal L}'_m)_{\sigma} + ({\cal L}'_m)_{\sigma\chi} + ({\cal L}'_m)_{\chi}}{\mu_{\rm eff}}
\ee
where
 \be\label{eq:Gammaprime}
\Gamma_{\sigma}^{\prime} + \Gamma_{\sigma\chi}^{\prime} + \Gamma_{\chi}^{\prime} \equiv
-\frac{\sigma}{x\theta_f\omega\beta_{r}}\Psi - 
\frac{\sigma\chi_{*}}{(x\theta_f\omega)^2\beta_r\Gamma}
\left(R_{j} + %\frac{\beta_{r}}{\beta_{p}}
\frac{B_{\phi}}{B_{r}}P_{j}\right)\Lambda +
\chi_{*}\left(R_{j}+\frac{\beta_{\phi}}{\beta_r}P_{j}\right)
\ee
\be
({\cal L}'_m)_\sigma + ({\cal L}'_m)_{\sigma\chi} + ({\cal L}'_m)_{\chi} \equiv
-\frac{\sigma}{x\theta_f\omega^2\beta_{r}}\Psi -
\frac{\sigma\chi_{*}}{x\theta_f\omega^2\beta_r\Gamma}\left(\beta_{\phi}\Lambda+{1\over\beta_r\Gamma^2}\frac{B_{\phi}}{B_{r}}\right)
\left(R_{j}+\frac{B_{\phi}}{B_{r}}P_{j}\right) + \chi_{*}\frac{x\theta_{f}}{\beta_r}P_{j}
\ee
where
\be
\Lambda \equiv 1+\frac{\beta_{\phi}}{\beta_r}\frac{B_{\phi}}{B_r}; \quad\quad
\Psi \equiv \left(1 + {\Delta\theta_{B}\over\theta_f}\right)(1 + \Lambda){\beta_\phi\over x} + 
\frac{B_{\phi}}{B_{r}}{\beta_r\over A}{dA\over dl}
\ee
and
\be\label{eq:mueff}
\mu_{\rm eff} \equiv 1-\frac{\sigma}{(\beta_r\Gamma)^{3}}\left(1+\beta_\phi^2\Gamma^2\right) - 
\frac{\sigma}{(x\theta_f\omega)^2\beta_r\Gamma}\left(1+{\beta_\phi^2\over\beta_r^2}\right) + 
\frac{2\sigma \beta_\phi}{x\omega\theta_f\,\beta_r^3\Gamma}
\ee
is the effective inertia.

%%%%%%%%%%%%%%%%%%%%%%%%%%%%%%%%%%%%%%%%%%%%%%%%%%%%%%%%%%%%%%%%%%%%%%%%%%%%%%%%%%%

\subsection{Poloidal Field Profile}\label{s:flare}

\begin{figure}[h]
\centerline{\includegraphics[width=0.4\hsize]{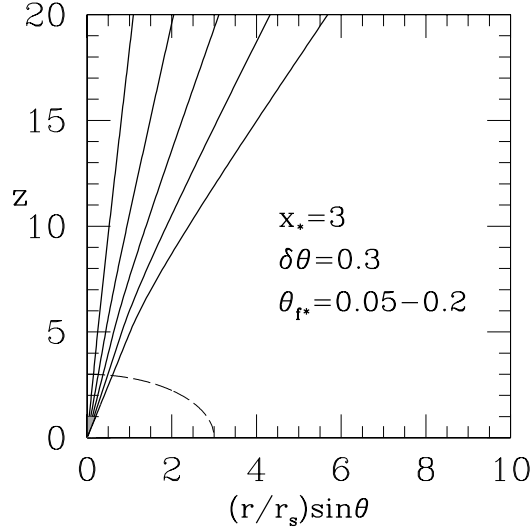}}
%\centerline{\includegraphics[width=0.4\hsize]{smSetwaXvsY.eps}}
\caption{Sample field line profiles of the type (\ref{eq:thetaprof}) with strong flaring ($\delta\theta=0.3$)
and breakout radius $x_*=r_*/r_s=3$ (dashed line). The emission radius bounds the inner grey zone.}
\label{fig:fieldline}
\vskip .2in
\end{figure}

We now prescribe the poloidal field profile outside the breakout surface, which, in a steady jet, 
also determines the poloidal streamlines.  The profile inside $r_*$ is assumed to be straight and conical,
$\theta_{f*} = \theta_{f,s}$.  A strong Lorentz force is obtained outside $r_*$  if the net change $\theta_{f,\infty}-\theta_{f*}$ 
in polar angle is itself a growing function of $\theta_{f*}$.  A simple choice, that is asymptotically conical
at a large radius, is
\be\label{eq:thetaprof}
{\theta_f(x)\over \theta_{f*}} 
= 1+\frac{\theta_{f*}}{\delta\theta}\left(1-\frac{x_{*}}{x}\right)^{\alpha}\quad\quad (x > x_*).
\ee
This connects smoothly with the inner cone if $\alpha > 1$.  
The net change in polar angle is determined by\footnote{This is the analog of the parameter
$\delta\theta_{\rm gradient}$ appearing in jet Model I, used to approximate the 
derivative in equation (\ref{eq:gammaevol}).} $\delta\theta$,
\be
\theta_{f,\infty}-\theta_{f*} = \left({\theta_{f*}\over\delta\theta}\right)\theta_{f*}.
\ee
The local change in the field line direction, relative to the total bend, is
\be
{\Delta\theta_B\over\theta_f-\theta_{f*}} = {B_\theta/B_r\over \theta_f - \theta_{f*}} = 
{2\alpha x_*\over x - x_*}.
\ee
The flux spreading factor works out to $A = 2 - \theta_{f*}/\theta_f$ for any field-line profile of the form (\ref{eq:thetaprof}). 
In the absence of radiation, the Lorentz factor can be obtained by imposing energy conservation.  It depends on the flaring profile
of the jet via
\ba\label{eq:MHDGaminf}
\Gamma &\simeq& \Gamma_* + \sigma_* - \sigma = \Gamma_* + \sigma_*\left(1-A^{-1}\right) 
    = \Gamma_* + \sigma_*{\theta_f/\theta_{f*}-1\over 2\theta_f/\theta_{f*}-1};\nn
\Gamma_{\infty} &\simeq& \Gamma_* + \sigma_* {\theta_{f*}/\delta\theta \over 1 + 2\theta_{f*}/\delta\theta}.
\ea
The acceleration tends to be more concentrated in radius for smaller values of the parameter $\alpha$;
in what follows $\alpha=2$. In Figure \ref{fig:fieldline} we show sample field lines given by (\ref{eq:thetaprof}) with strong flaring ($\delta\theta=0.3$) for several values of $\theta_{f*}$.
%%%%%%%%%%%%%%%%%%%%%%%%%%%%%%%%%%%%%%%%%%%%%%%%%%%%%%%%%%%%%%

\subsection{Position of Fast Magnetosonic Surface}\label{sec:MSsurface}

When the flow speed surpasses the fast magnetosonic speed, radial magnetic disturbances are swept downstream
and cannot interact with the part of the jet interior to the fast critical surface.  The inertia of the electromagnetic field
also becomes insignificant in the radial force balance, so that radiation pressure is relatively more important.
We first consider how the position $x_c=r_c/r_s$ of the critical surface is modified by 
field-line flaring, and then consider the effects of radiation.  The critical surface sits at infinite
radius only if the poloidal magnetic field is constrained to be radial and radiation is absent \citep{goldreich70}.  

The critical surface is obtained by setting $\mu_{\rm eff} = 0$ in equations (\ref{eq:Gamp}).   
Retaining $\chi_* = 0$, and assuming $\sigma \gg 1$, this corresponds to $\Gamma \simeq \sigma^{1/3}$ 
[a vanishing coefficient of $d\Gamma/dl$ in equation (\ref{eq:Eulerdl1b})], and $\Psi = 0$.  When 
$\delta\theta_f = O(\theta_j)$, the magnetofluid rapidly accelerates outside radius $r_*$, and so we can 
expand $A \simeq 1$ near this radius:
\be
{dA\over d\ln l} \simeq \alpha{\theta_{f*}\over\delta\theta}\left(1 - {x_*\over x}\right)^{\alpha-1}{x_*\over x};\quad\quad
\Psi \simeq {1\over x^2\theta_f\omega} - (x\theta_f\omega){dA\over dl}.
\ee
Here we have approximated $\beta_\phi \simeq 1/x\theta_f\omega \ll 1$.  Then, for $\alpha = 2$,
\be\label{eq:xclowchi}
{x_c\over x_*} \simeq 1 + {\delta\theta/\theta_{f*} \over 2 (x_*\theta_{f*}\omega)^2}.
\ee

In the absence of magnetic-field flaring, the radiation stress forces the fast surface in from infinity (Paper I).  Taking 
instead $\delta\theta = \infty$ but allowing for finite $\chi_*$, the fast surface corresponds to $\Gamma \simeq \sigma^{1/3}$
and $\Gamma^{\prime}_{\sigma}\simeq\Gamma^{\prime}_{\chi}$.   Then
\be \label{eq:xchighchi}
{x_c\over x_*} \simeq \frac{4\sigma^{5/3}}{\chi_*\left(x_{*}-1\right)^{2}\omega^{2}\theta_{f*}^{2}}.
\ee
This differs from the spherical case (Paper I) mainly by the factor $1/\theta_{f*}^2$.  At a very high compactness,
the flow is tied to the collimating radiation field, and thus the critical surface is pulled in to where
$\Gamma_{\rm eq}(x,\theta_{f})\simeq\Gamma_c\simeq\sigma^{1/3}$. 
\begin{figure}[h]
\epsscale{1.1}
\plottwo{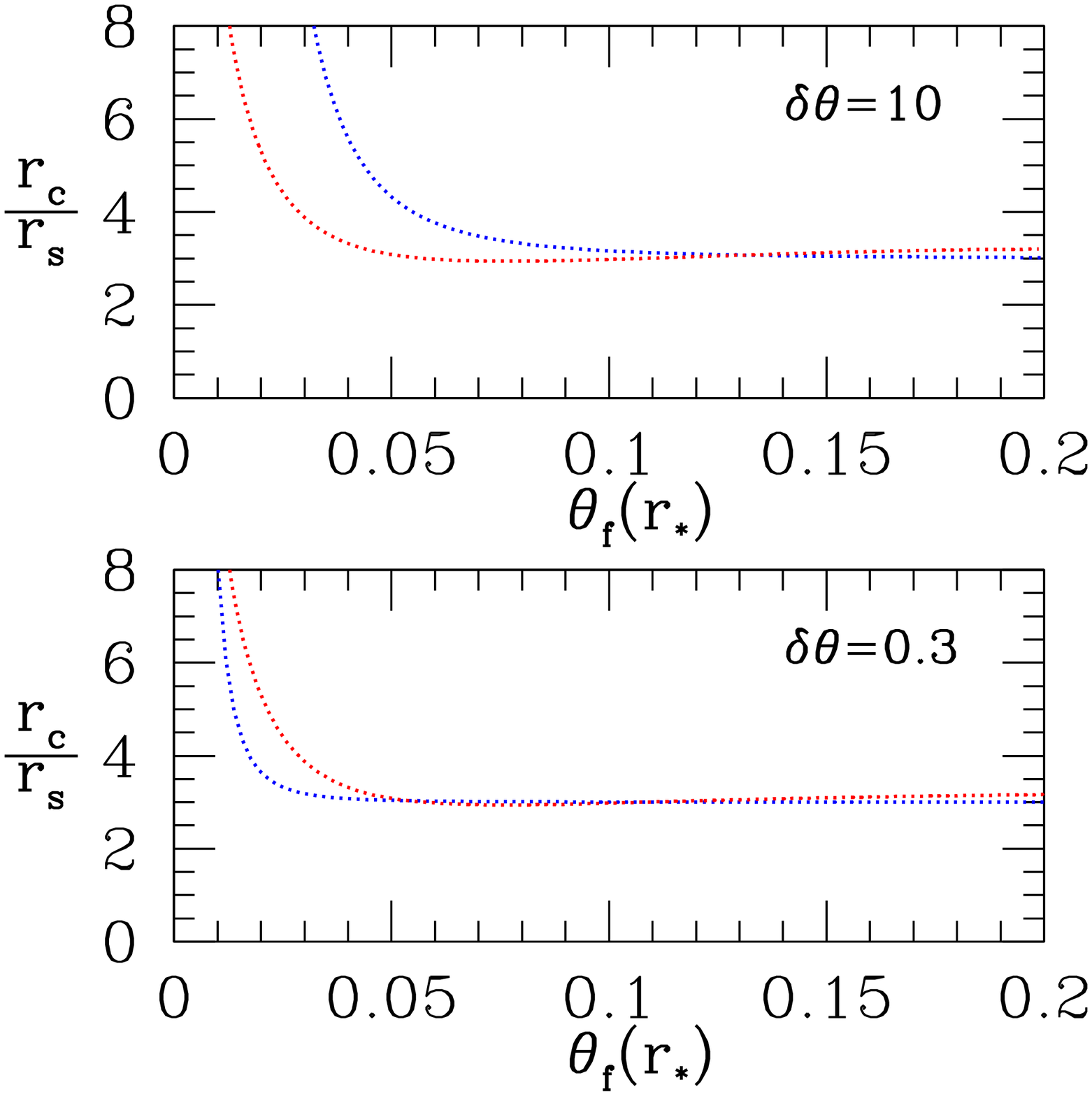}{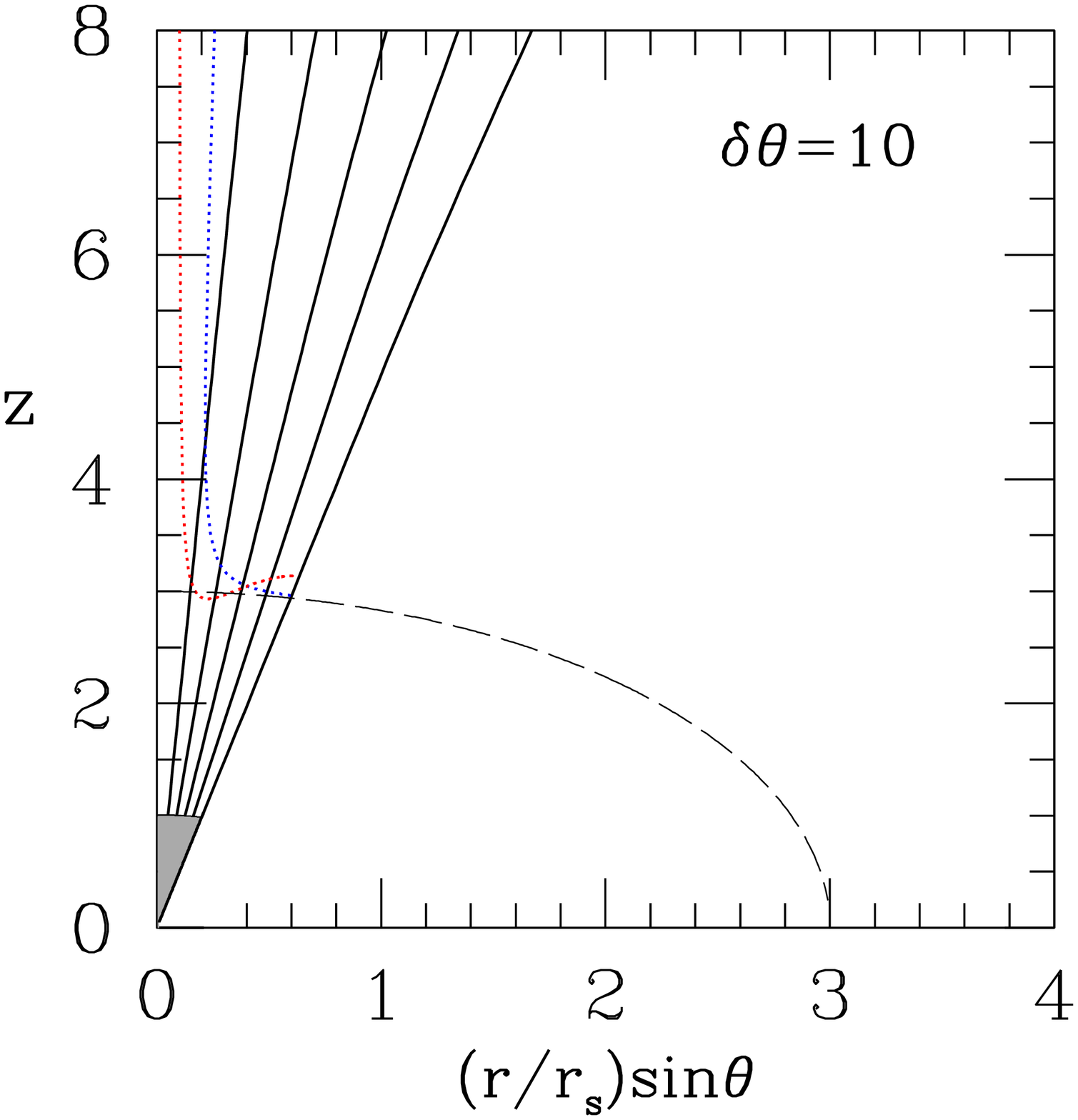}
%smCriticalLocuswasaxcthca.eps\plottwo{smCriticalLocuswasaxcthca.eps}{smCriticalLocusSetsaXvsY.eps}
\caption{Magnetosonic surface inside a flaring jet at low radiation compactness (dotted blue, $\chi \rightarrow 0$) 
and high compactness [dotted red, $\chi/\sigma \ll 1$ but still satisfying the bound given in equation (\ref{eq:maxchi})].
\textit{Left panels:}  Radius $x_c=r_c/r_s$ of the magnetosonic point as a function of field line footprint angle,
for weak flaring (top) and strong flaring (bottom). \textit{Right panel:} Two dimensional depiction of the 
magnetosonic surface (weak flaring).}
\label{fig:MSsurface}
\end{figure}

The fast surface is shown as a function of angle in Figure \ref{fig:MSsurface}, for a breakout radius $x_* = 3$. 
At low radiation compactness, this surface typically lies just outside breakout, $x_c \gtrsim x_*$, in agreement with 
equation (\ref{eq:xclowchi}).  As the compactness is increased, the critical surface can either move inward or outward, depending
on the location where $\Gamma_{\rm eq} = \sigma^{1/3}$. The critical surface is typically 
pulled inward near the rotation axis if the magnetic field is weakly flared, and also at larger polar angles if $x_* \gg \theta_j \sigma^{1/3}$. 
Then its position follows equation (\ref{eq:xchighchi}) until reaching the high-compactness limit at
\be 
\chi_*\simeq \frac{4\sigma^{4/3}}{x_*\omega^2\theta_{f*}^2\theta_j}. 
\ee
If alternatively the breakout radius is small, then the critical surface is pushed out by radiation drag,
\be 
\frac{x_{c}}{x_{*}}\simeq\left(\frac{\chi_{*}}{24\sigma^{1/3}}\frac{\delta\theta}{\theta_{f*}}\frac{\theta_{j}^{4}}{x_{*}^{4}}\right)^{1/7},
\ee
reaching its high-compactness limit at
\be 
\chi_{*}\simeq 24{\theta_{f*}\over\delta\theta}\left(\frac{\theta_j}{x_j}\right)^3\sigma^{8/3}.
\ee

The deviation of $x_c$ toward large radius that is seen close to the rotation axis is due to a combination of effects:
a reduction in the outward Lorentz force due to the weaker field-line flaring; and a mis-match between the radiation
and matter flows driven by strong rotation.  The first effect dominates at low $\chi_*$.  The change in critical radius
at high $\chi_*$ can be estimated using equation (\ref{eq:Gammaeqlimit}) for $\Gamma_{\rm eq}$ near the axis:
\be 
x_c\simeq\frac{\sigma^{1/3}}{\omega\theta_f}, \quad\quad (\theta_f \ll \omega^{-1}\theta_j^{-1}).
\ee

%%%%%%%%%%%%%%%%%%%%%%%%%%%%%%%%%%%%%%%%%%%%%%%%%%%%%%%%%%%%%%
%%%%%%%%%%%%%%%%%%%%%%%%%%%%%%%%%%%%%%%%%%%%%%%%%%%%%%%%%%%%%%
\subsection{Numerical Results}\label{s:results}
We now examine the solutions to the wind equations (\ref{eq:Gamp})-(\ref{eq:mueff}) that we have derived
for a geometrically and optically thin jet. The singularity at the fast 
magnetosonic critical point, and the stiffness of the equations associated with large values of $\sigma$
and $\chi$, means that simple integration techniques such as Runge-Kutta are inadequate.
To determine the position of the critical point and the flow solution inside it, we use the 
relaxation method described in Paper I (see also \citealt{london82}).  
\begin{figure}[h]
\epsscale{1.02}
\plottwo{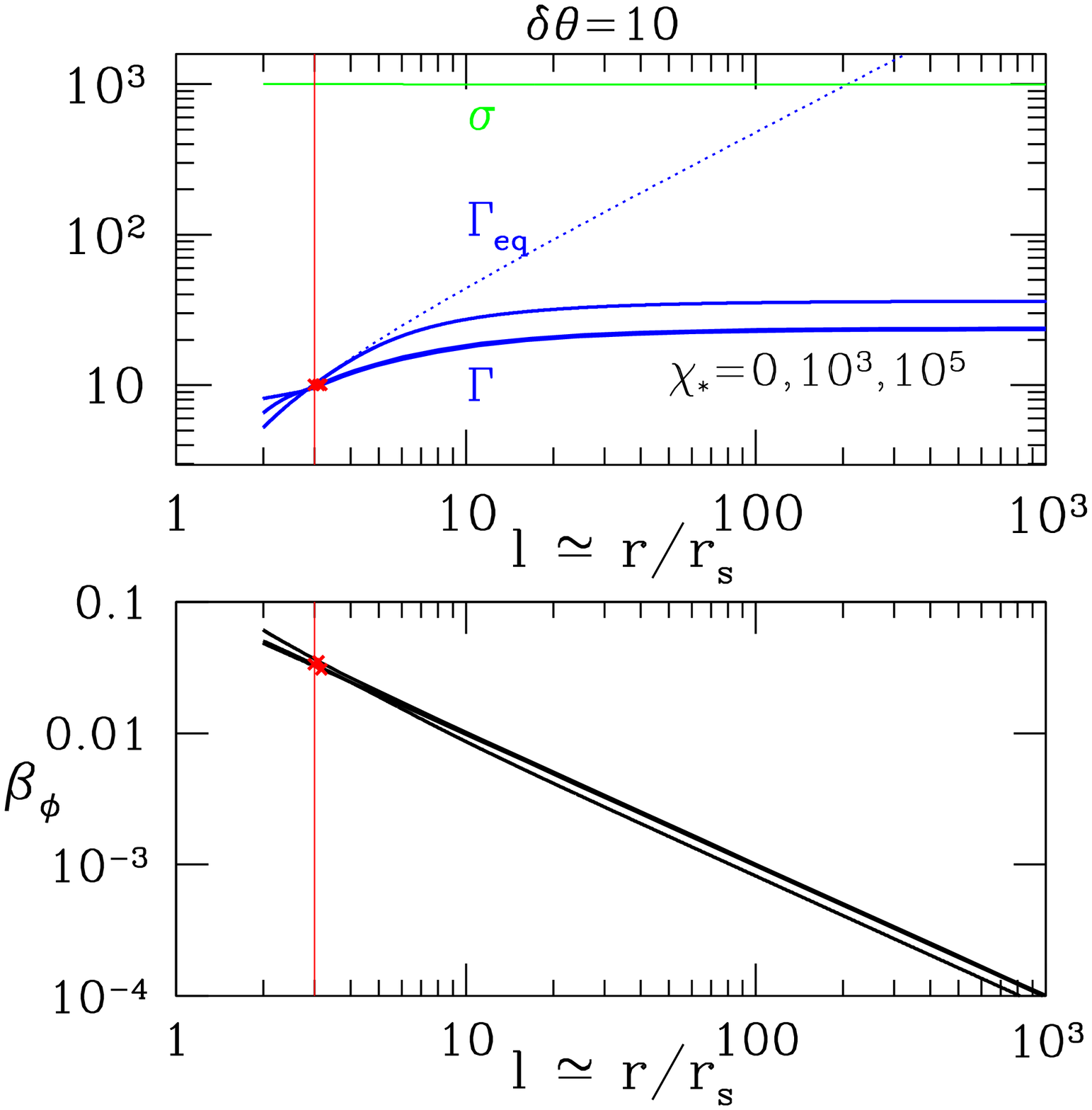}{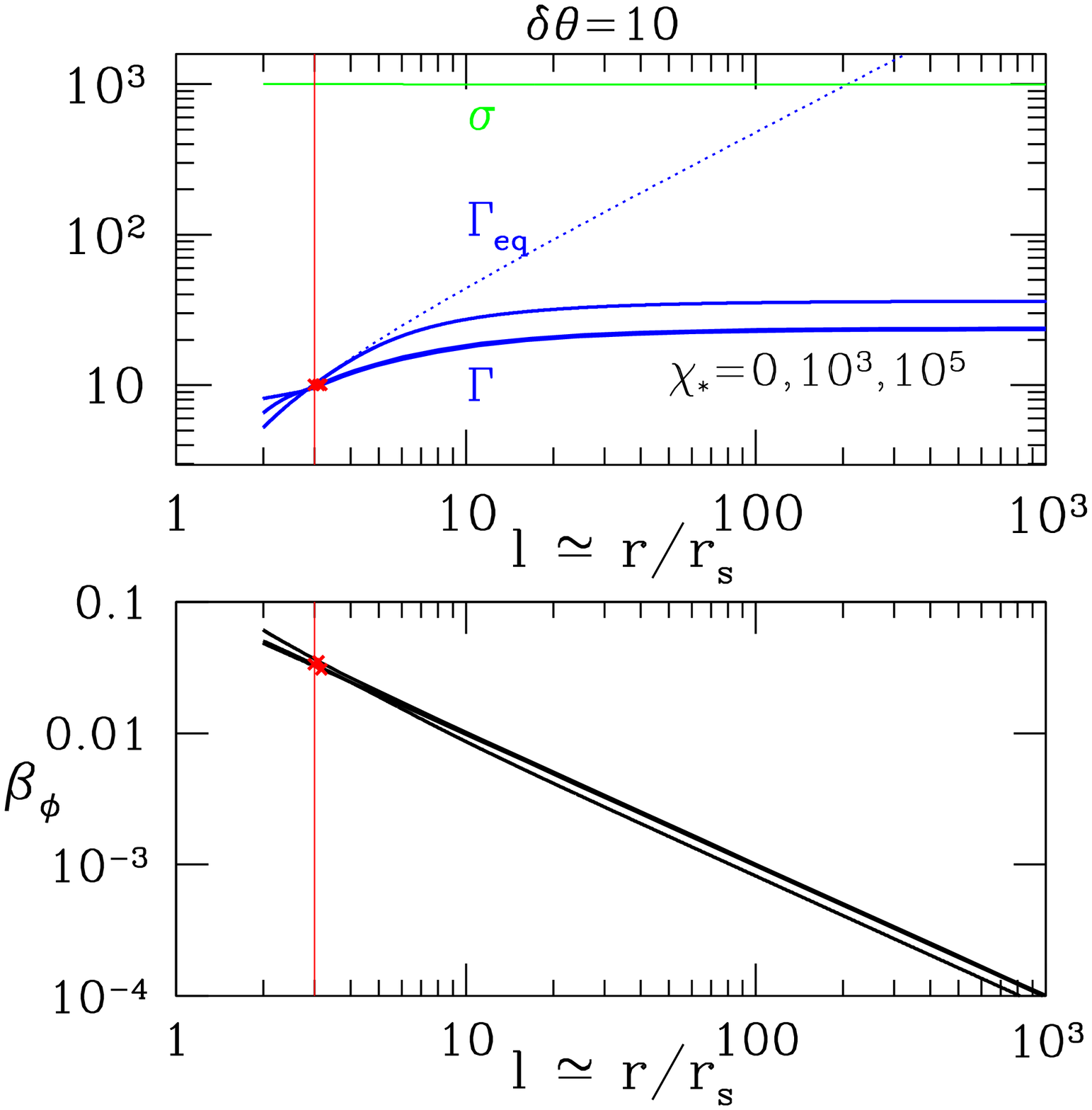}
\centerline{\includegraphics[width=0.45\hsize]{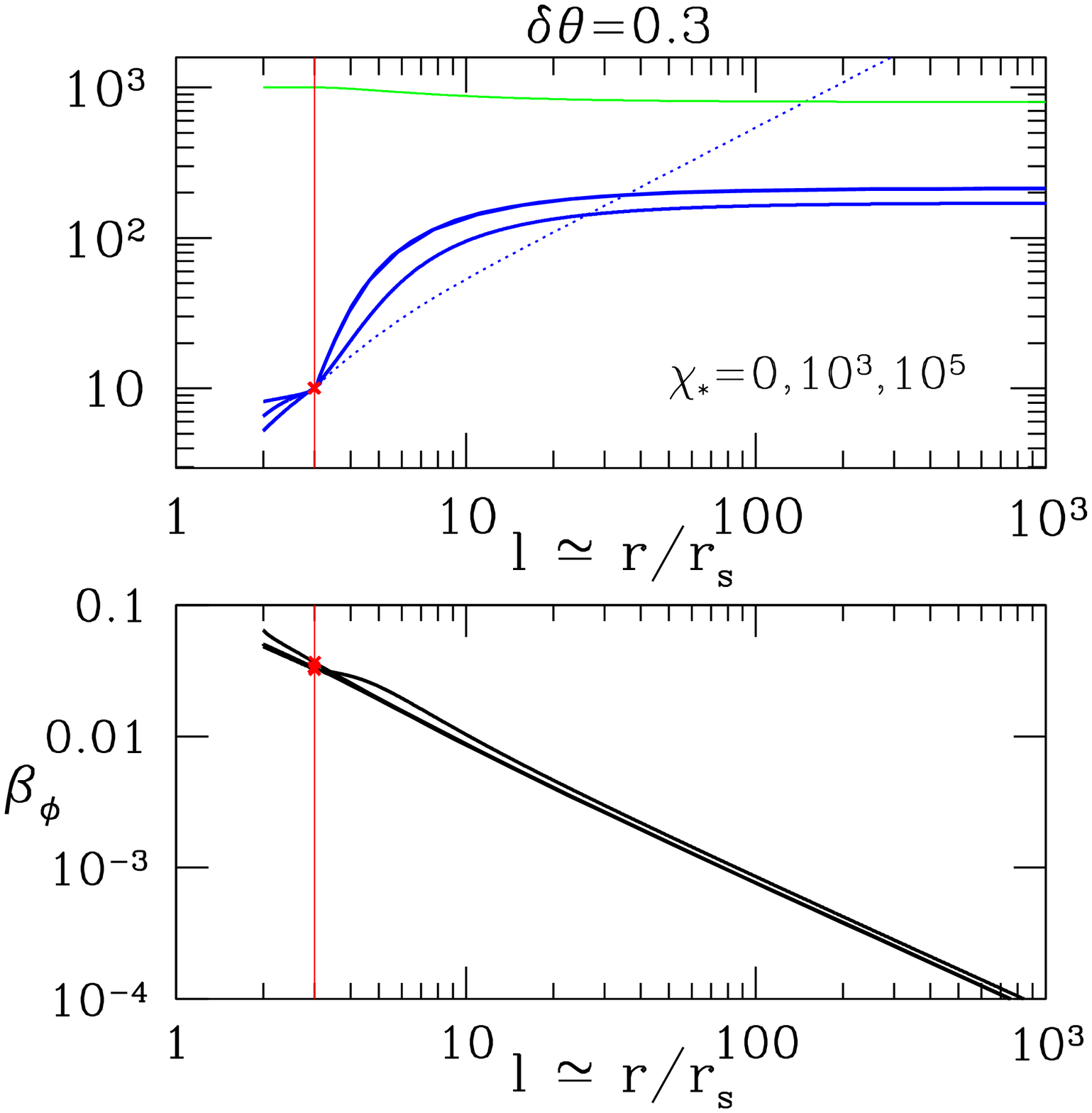}}
%\plottwo{smGamBetaMultipleChisa.eps}{smGamBetaMultipleChira.eps}
%\centerline{\includegraphics[width=0.5\hsize]{smGamBetaMultipleChiwa.eps}}
\caption{Acceleration in a thin, strongly magnetized jet ($\sigma_*=1000$, $\omega=100$, $\theta_j=0.2$),
along a field line anchored at $\theta_{f*}=0.1$. The radiation compactness at the breakout 
radius $x_*=r_*/r_s=3$ is varied: $\chi_*=1,10^3,10^5$, corresponding to the top to bottom curves on the left side. 
\textit{Left panel:} $\delta\theta=10$ (weak flaring); \textit{right panel:} 
$\delta\theta=1$; {\it bottom panel:} $\delta\theta = 0.3$ (strong flaring). (The distance along the field line $l$ differs little from the radial coordinate since the degree of flaring is small).} 
\vskip .3in
\label{fig:jetprofiles}
\end{figure}

The inner boundary radius $r_i$
of the integration is chosen somewhat differently than in Paper I:  we set it to twice
the photon emission radius ($x_i=2$) because we only evaluate the radiation force
where photons propagate at small angles with respect to the jet axis (requiring that $x_i - 1 \gg \theta_j$).
The solutions for $\Gamma$ and ${\cal L}_m$ obtained by an integration inside the critical point are required to be smooth near $x_i$;
avoiding sharp gradients restricts the boundary values at $x_i$ to a narrow range.  We also make a first guess for 
the critical point radius $x_c$.  The regularity of the solution at $x_c$ then allows us to determine
the flow variables at the critical point from the equations
\be\label{eq:reg}
\left(\Gamma'_{\sigma}+\Gamma'_{\sigma\chi}+\Gamma'_{\chi}\right)_{x_c}\;=\;0\;=\;\mu_{\rm eff}(x_c).
\ee 
An approximate solution is chosen which connects the inner boundary values to the critical point.\footnote{This 
procedure is followed for a discrete set of values of $\chi_*$, after which the inner boundary condition on 
$\Gamma$, $\mathcal{L}_m$ is chosen from an interpolating function.}  This solution,
along with the position of the critical point, is then relaxed to within a desired tolerance 
using a Newton-Raphson method, all the while satisfying the regularity condition (\ref{eq:reg}).
As a last step, the flow outside the critical point is obtained by shooting outward using a fifth-order Runge-Kutta algorithm.

Solutions are obtained for a range of photon compactness and a high magnetization ($\sigma_*=1000$).  The part 
of the jet studied sits well outside the light cylinder, $\omega=100$ in equation (\ref{eq:dimvar}).
Choosing the flaring profile (\ref{eq:thetaprof}), we follow the flow along
a field line with initial footprint $\theta_{f*}=0.1$, in a jet of half-opening angle 
$\theta_j=0.2$ and a breakout radius $x_*=3 = 1.5x_i$.   The magnitude of the jet flaring
is adjusted by choosing the parameter $\delta\theta$, with values $10,1,0.3$ corresponding to a
net angular shift $\theta_{f,\infty}/\theta_{f*} - 1 = 0.1\theta_{f*}$, $\theta_{f*}$, $3.3\,\theta_{f*}$
between breakout and infinity.  The maximal flaring chosen ($\delta\theta = 0.3$) still satisfies 
equation (\ref{eq:causalcons}), and so the divergence of neighboring magnetic field lines is consistent
with causal stresses.

The results are show in Figure \ref{fig:jetprofiles}.  At low radiation compactness, they resemble 
those obtained by \cite{tchek10} for a cold MHD jet.  A slow, nearly linear, increase in $\Gamma$ within the star 
is followed by rapid (but logarithmic) growth beyond the breakout point, where the field lines begin to diverge.
As $\chi_*$ increases above $\sim \sigma$, photon drag begins to dominate the weak Lorentz force
inside the breakout radius, and $\Gamma$ tends to $\Gamma_{\rm eq}\simeq x/\theta_j$
[equation (\ref{eq:gameq})].  After breaking out, the fluid is 
quickly accelerated through the fast point.    A strong radiation field forces the position of
the critical point to a radius where $\Gamma_{\rm eq} \simeq \sigma_*^{1/3}$ (in this case,
the displacement is outward).  We do not search for solutions with the
fast point inside the star, corresponding to $x_* \ga 3$.  
\begin{figure}[h]
\centerline{\includegraphics[width=0.5\hsize]{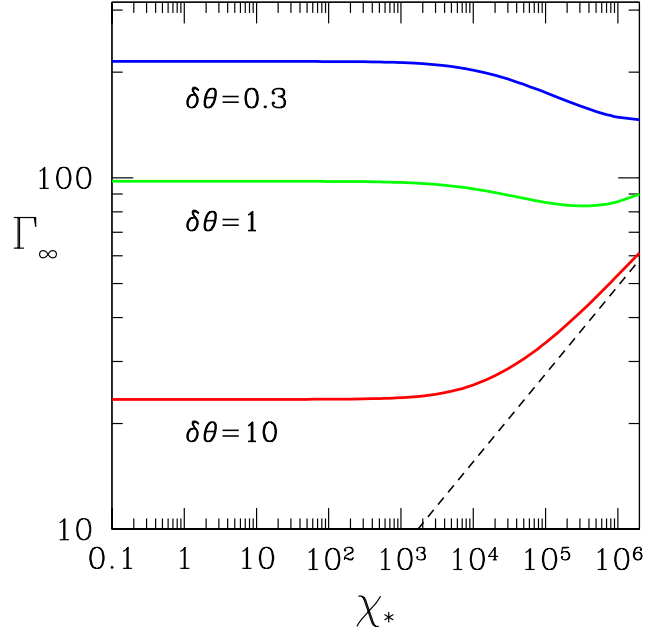}}
%\centerline{\includegraphics[width=0.8\hsize]{smGamInfwrsxs3TH0_1.eps}}
\caption{Asymptotic Lorentz factor of flaring magnetized jet ($\sigma_*=1000$, $\omega=100$, $\theta_j = 0.2$),
as a function of compactness $\chi_*$ at the breakout radius.  Curves correspond to a field line anchored at $\theta_{f*}=0.1$
and represent varying degrees of flaring (weak: $\delta\theta=10$; intermediate:  $\delta\theta = 1$; 
strong: $\delta\theta = 0.3$).  Dashed black line:  asymptotic Lorentz factor $0.8(\chi_*x_*/4\theta_j)^{1/4}$ 
of an unmagnetized, radial, monopolar outflow.}
\label{fig:gaminfty}
\end{figure}

The influence of the radiation field can be clearly seen in the dependence of asymptotic
Lorentz factor $\Gamma_\infty$ on $\chi_*$ (Figure \ref{fig:gaminfty}).  When flaring is strong 
($\delta\theta \lesssim 1$), the matter is pushed rapidly to $\Gamma > \Gamma_{\rm eq}$ and
it feels a net drag outside the magnetosonic point.  The asymptotic Lorentz factor is reduced
from (\ref{eq:MHDGaminf}).  At very high $\chi_*$, the radiation
drag is able to suppress the acceleration and, therefore, the asymptotic Lorentz factor. 

The minimal compactness needed to significantly affect the post-breakout flow can be estimated by
equating the leading terms in (\ref{eq:Gammaprime}) at $x=3x_*/2$, the point where the 
relative flaring of poloidal field lines is maximal.  This gives
\be 
\chi_*\gtrsim\sigma_*^{5/3}{\theta_{f*}\over \delta\theta}.
\ee
In this case, the sign of the effect of radiation pressure outside the critical point still
depends on the degree of magnetic field line flaring.  
If flaring is weaker ($\delta\theta \gtrsim 1$) then the details of the flow profile near breakout differ
from the unmagnetized flow, but $\Gamma_\infty$ is still well approximated by an unmagnetized,
radiatively driven flow: 
\be 
\Gamma_\infty\simeq 0.8(\chi_*x_*/\theta_j)^{1/4}\quad\quad(\sigma=0,~\delta\theta\rightarrow\infty). 
\ee

Our solutions, in the part of parameter space that we have explored, satisfy two basic constraints.
First, the outflow is optically thin at (or near) breakout if the compactness sits below the bound
(\ref{eq:chimax}).  Second, the component of the radiation force transverse to the poloidal flux
surfaces must remain small compared with the transverse Lorentz force that is implied by the chosen
flaring profile.  The corresponding upper bound (\ref{eq:maxchi}) on the compactness is derived in Appendix \ref{s:transrad}.

%%%%%%%%%%%%%%%%%%%%%%%%%%%%%%%%%%%%%%%%%%%%%%%%%%%%%%%%%%%%%
%%%%%%%%%%%%%%%%%%%%%%%%%%%%%%%%%%%%%%%%%%%%%%%%%%%%%%%%%%%%%%

\section{Spectrum of Scattered Photons}\label{s:spectrum}

We now consider the self-consistent spectrum of photons that scatter off a
hot electromagnetic outflow near its photosphere.  Our focus is solely on
the signature of the differential flow of matter and photons -- that is, we
neglect any internal processes that would heat particles or induce small-scale
deviations from a uniform flow.

We first consider the
jet model of Section \ref{s:causaljet}, in which the optical depth 
is finite but the flow is considered only outside its fast magnetosonic surface.
Then we turn to the optically thin jet model of Sections \ref{s:eom} and \ref{s:results},
in which the entire flow is solved inside and outside the critical surface,
but the region interior to the photosphere is ignored.  As in Paper I, we neglect
any internal dissipation in the outflow, which can contribute to the high-energy
tail of the spectrum \citep{thompson94,giannios06,beloborodov10}.

%%%%%%%%%%%%%%%%%%%%%%%%%%%%%%%%%%%%%%%%%%%%%%%%%%%%%%%%%%%%%%

\subsection{Spectrum in Jet Model I (Super-magnetosonic): Monte Carlo Method}\label{s:smethod}

Here we follow the photon field self-consistently across the jet photosphere,
which is assumed to sit outside the fast magnetosonic point (Figure \ref{fig:JetSec2}).
The exchange of energy between photons and magnetic field was calculated 
in Section \ref{s:causaljet} in parallel with the flaring rate of the poloidal field lines 
outside a fixed breakout radius.  The radiation force on the matter, and the evolution of
the equilibrium Lorentz factor $\Gamma_{\rm eq}$ of the radiation field, defined in equations
({\ref{eq:moment}) and ({\ref{eq:Flab3}), are both calculated by taking angular moments 
(\ref{eq:fn}) of the intensity.

To calculate the emergent spectrum, we i) take the flow velocity profile
as a given background, and then ii) inject photons from the inner radius $r_i = 0.8r_*$ with 
an isotropic distribution in a frame moving with Lorentz factor $\Gamma_{{\rm eq},i}$.
Photon parameters in the rest frame of the `star' at $r > r_i$ are obtained by a 
simple Lorentz boost.  Defining a radial direction cosine by $\mu = \hat k\cdot\hat r$,
one has $\mu = (\mu' + \beta)/(1 + \beta \mu')$, 
$\omega = \Gamma(1 + \beta\mu') \omega'$, where the prime labels the matter rest frame.
Deviations from radial flow are assumed small compared with the width of the photon beam.

Electron scatterings are handled in the Thomson approximation (the outflow moves relativistically
in the case of a GRB), by drawing a random number $1 - e^{-\Delta\tau_{\rm es}}$.
The position of the next scatter point is calculated by integrating
\be
\Delta\tau_{\rm es}[r_1,r_2,\mu(r_1)] = \alpha_{\rm es*}r_*
\int_{r_1}^{r_2}\left[1-\mu(r)\beta(r)\right] {r_* dr\over \mu(r)\beta(r){r}^2}
\ee
along the photon ray.  Note that the outflow solution of
Section \ref{s:causaljet} has been iterated so that the coefficient $\alpha_{\rm es*}$
corresponds to a prescribed value of the radial optical depth $\tau_{\rm es}(r_*,\infty,1)$ at the breakout radius.  Other initial flow parameters are defined
at $r_i$.
The direction cosine evolves from a scattering radius $r$ to $r_2 > r$ according to
\be
1 - \mu(r_2)^2 =  \left({r\over r_2}\right)^2\left[1-\mu(r)^2\right]
\quad\quad(r_2 > r).
\ee
The photon escapes if $\Delta\tau_{\rm es}$ exceeds the total optical depth along the ray.

The frequency distribution of the outgoing photons is first obtained with a monochromatic
photon source, $I_\nu = I_0\nu_0\delta(\nu-\nu_0)$.  This output spectrum
is then convolved a source spectrum that is either a pure blackbody, 
or a function\footnote{This corresponds to the lower-frequency half of the Band function 
\citep{band93}, extended to all frequencies.} that mimics the low-frequency slope of
a GRB, $F_\nu = {\rm const}\times e^{-h\nu/kT_0}$.  For both types of seed spectrum,
the temperature $T_0$ is normalized by requiring $F_\nu$ to peak at the fixed
reference frequency $\nu_0$.

Scatterings are taken to be elastic in the bulk frame, where the matter is assumed
cold, so that the outgoing and ingoing frequencies satisfy the usual Doppler relation,
\be\label{eq:doppler}
{\nu_{\rm em}\over\nu}  \;=\;  {1-\beta\mu\over 1-\beta\mu_{\rm em}}
\;=\; {1+\beta\mu_{\rm em}'\over 1+\beta\mu'}.
\ee
After transforming $\mu$ to the local bulk frame, we pick scattering angles $\theta_s'$,
$\phi_s'$ with respect to the flow direction. The direction cosine
of the outgoing photon is determined via $\mu_{\rm em}' = \mu'\cos\theta_s' + (1-\mu'^2)^{1/2}\sin\theta_s'\cos\phi_s'$, followed by a boost to the stellar frame.

The peak of the seed photon distribution is stretched to higher frequencies
when the outflow Lorentz factor $\Gamma \ga \Gamma_{\rm eq}$ [equation (\ref{eq:gameq})].
A scattered photon has a frequency in the range $\nu_{\rm min} < \nu < \nu_{\rm max}$, where
\be\label{eq:numax}
\nu_{\rm max} = {1-\beta\mu_{\rm min}\over 1-\beta}\nu_0 \simeq \left(1+{\Gamma^2\over x^2}\right)\nu_0
\quad (x,\Gamma\gg 1)
\ee
and $\nu_{\rm min} = [(1-\beta)/(1+\beta)]\nu_0 \sim \nu_0/4\Gamma^2$.  

%%%%%%%%%%%%%%%%%%%%%%%%%%%%%%%%%%%%%%%%%%%%%%%%%%%%%%%%%%%%%%

\subsection{Spectrum in Jet Model I:  Results}\label{s:sresults}

Figure \ref{fig:Spectra1} shows spectra for the case where matter and radiation field are 
initially locked together at $\Gamma_i = \Gamma_{{\rm eq},i} = 10$.   The curves correspond to 
a variety of optical depths, as well as low and high initial photon fluxes, ${\cal R}_i = 1, 10^3$.
The peak of the spectrum is somewhat broadened compared
with a pure blackbody, and the segment shortward in frequency of the peak has a flattened
spectrum, although not as flat as is seen in GRBs.  A similar effect was seen in Paper I
in the case of hot electromagnetic outflows accelerating along a radial, monopolar monopolar
magnetic field.   
\begin{figure}[h]
\epsscale{1.}
\plottwo{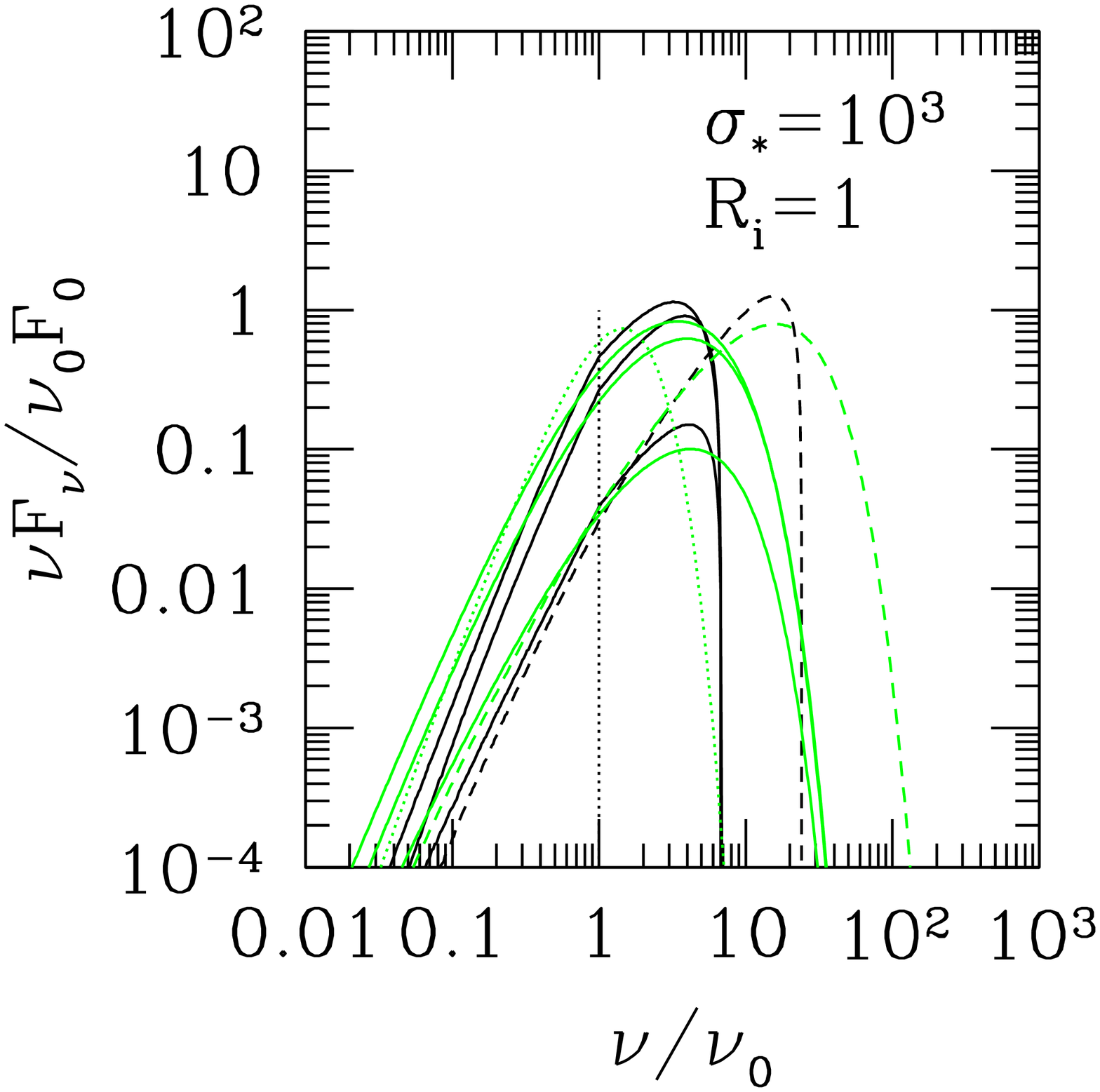}{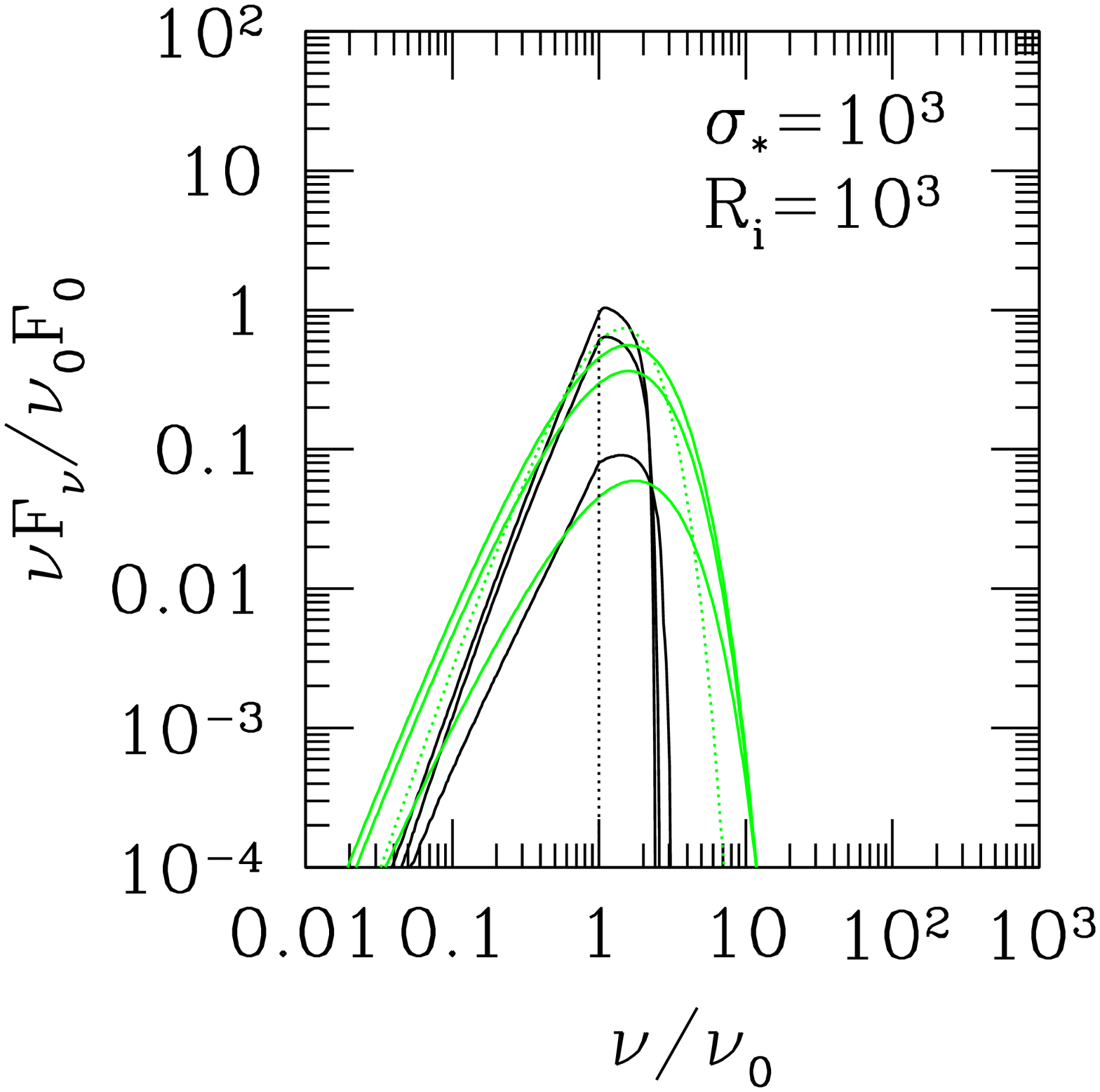}
\plottwo{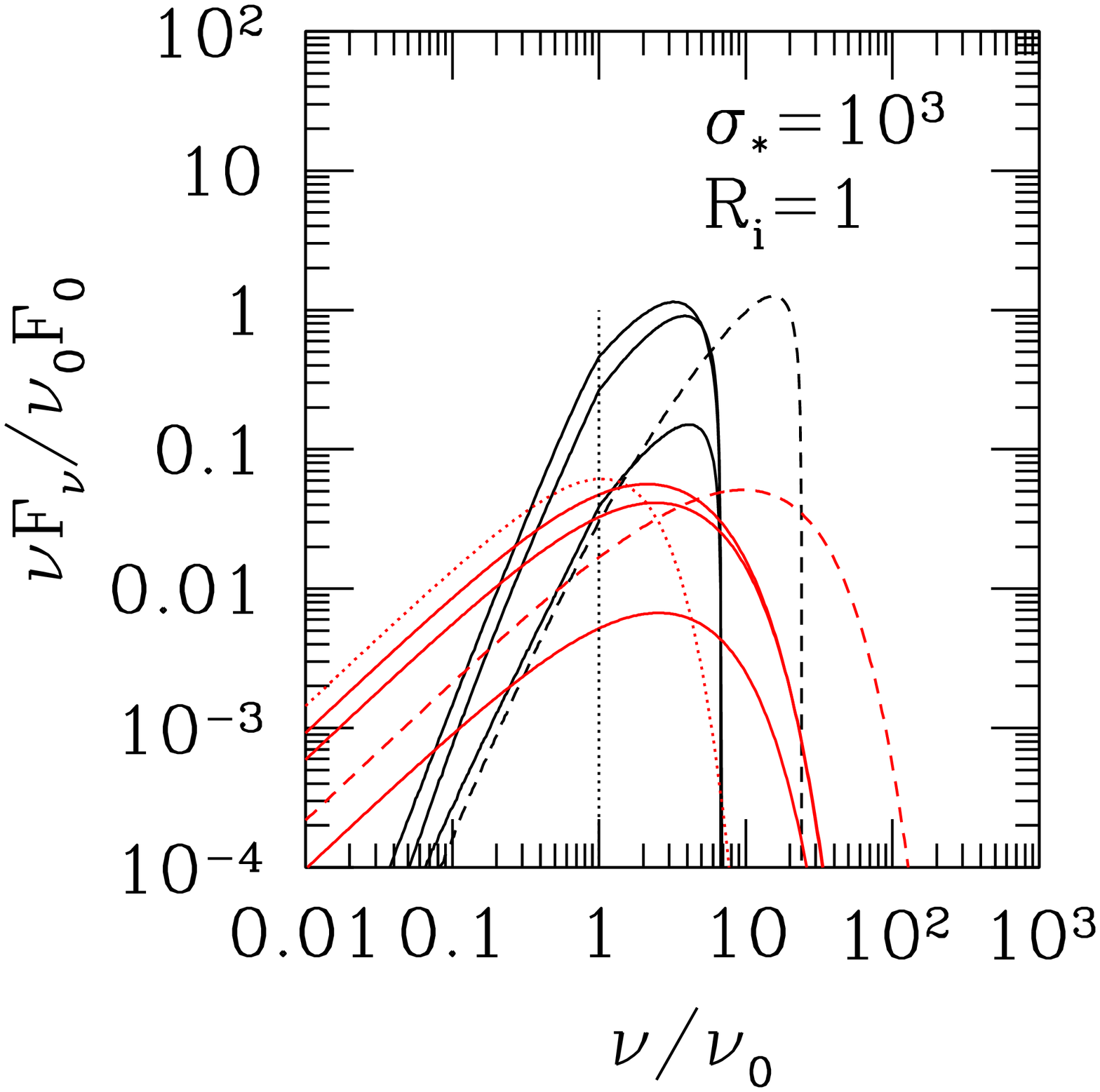}{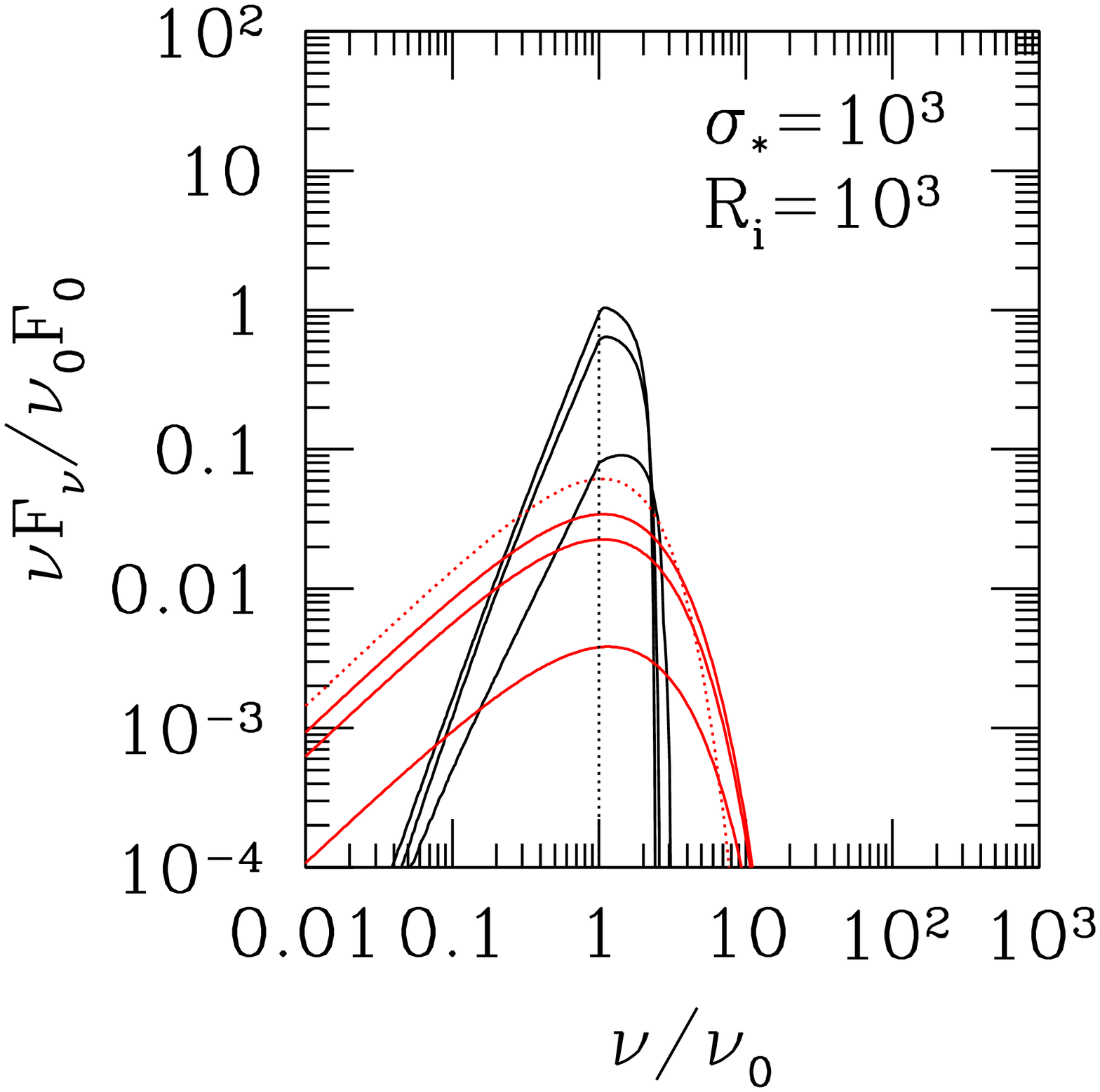}
%\plottwo{spectrum_bb_R1.ps}{spectrum_bb_R1000.ps}
%\plottwo{spectrum_hybrid_R1.ps}{spectrum_hybrid_R1000.ps}
\caption{Spectra of photons emerging from a relativistic jet with a radial profile calculated by the method
of Section \ref{s:causaljet}, for various values of the optical depth at breakout
$\tau_{\rm es*} = 0.1, 1, 10$ [equation (\ref{eq:taustar})].  Black curves correspond to a monochromatic 
seed, green curves to a blackbody seed, red curves to a GRB-like seed spectrum,
$F_\nu = {\rm const}\times e^{-h\nu/kT_0}$ [the lower-frequency half of the Band (1993) function
extended to all frequencies].  Normalized photon energy flux at the inner radius 
$r_i = 0.8 r_*$:  ${\cal R}_i = 1$ (left), ${\cal R}_i = 10^3$ (right).   Initial Lorentz factor 
$\Gamma_i = 10$.  Bulk frame of the radiation field $\Gamma_{{\rm eq},i} = 10$, except for the dashed 
curves which correspond to $\Gamma_{{\rm eq},i} = \Gamma_i/2 = 5$, $\tau_{\rm es*} = 0.1$.}
\label{fig:Spectra1}
\vskip .2in
\end{figure}
The spectrum below the peak is flattened even more if the radiation field emerging at the
jet photosphere is broader than the matter Lorentz cone:  the dashed curve in the
${\cal R}_i = 1$ panels is the result for $\Gamma_{{\rm eq},i} = \Gamma_i/2 = 5$ and a 
low optical depth ($\tau_{\rm es *} = 0.1$) at the breakout radius .  As expected from the above
argument, the peak of the spectrum is stretched upward in frequency above the peak of
the seed spectrum.  

%%%%%%%%%%%%%%%%%%%%%%%%%%%%%%%%%%%%%%%%%%%%%%%%%%%%%%%%%%%%%%

\subsection{Spectrum in Jet Model II (Trans-magnetosonic)}\label{s:sresults2}

The output spectrum is calculated by a similar method to that described in Section \ref{s:smethod}.
The background flow is prescribed, in this case by the solutions obtained in Section \ref{s:results}, 
and the flow is approximated as radial.
Since we are not, now, following the outflow across its photosphere, and the entire simulation
volume is assumed optically thin, we take a similar input photon distribution as was used 
to calculate the flow acceleration:  the intensity is constant, $I = I_0$ for $\theta < \theta_j = 0.2$ radian.
The jet breakout radius is taken to be $x_* = 3$, as in Figure
\ref{fig:jetprofiles}, and the radial optical depth $\Delta\tau_{\rm res}(x_i =2,\infty,\mu=1) = 1$. Results are shown in Figures \ref{fig:Spectra3}-\ref{fig:Spectra4} for three different degrees
of magnetic field flaring:   strong [corresponding to $\delta\theta=0.3$ in equation
(\ref{eq:thetaprof})],  intermediate ($\delta\theta = 1$), and weak ($\delta\theta = 10$). 
\begin{figure}[h]
\epsscale{.91}
\plottwo{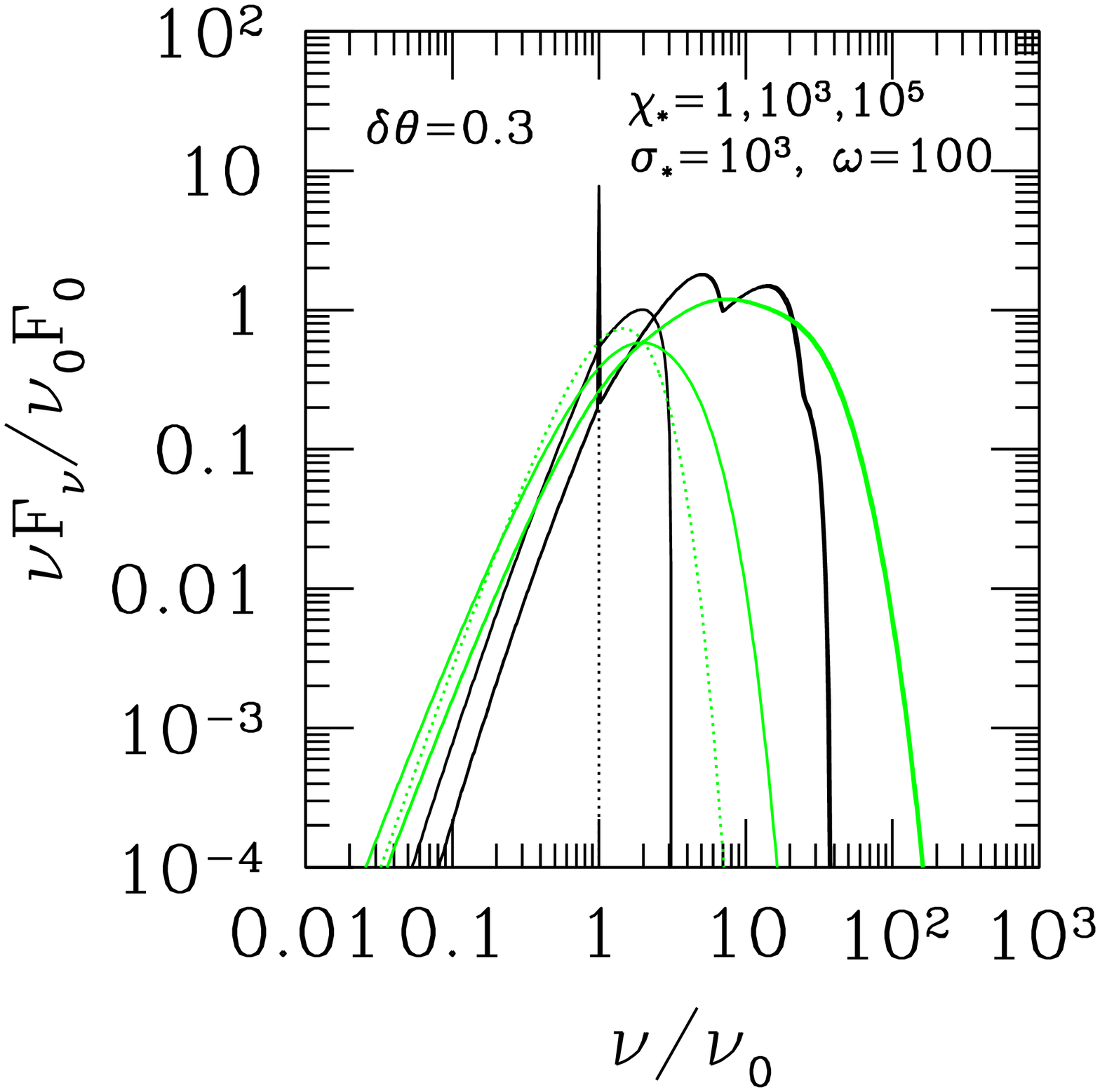}{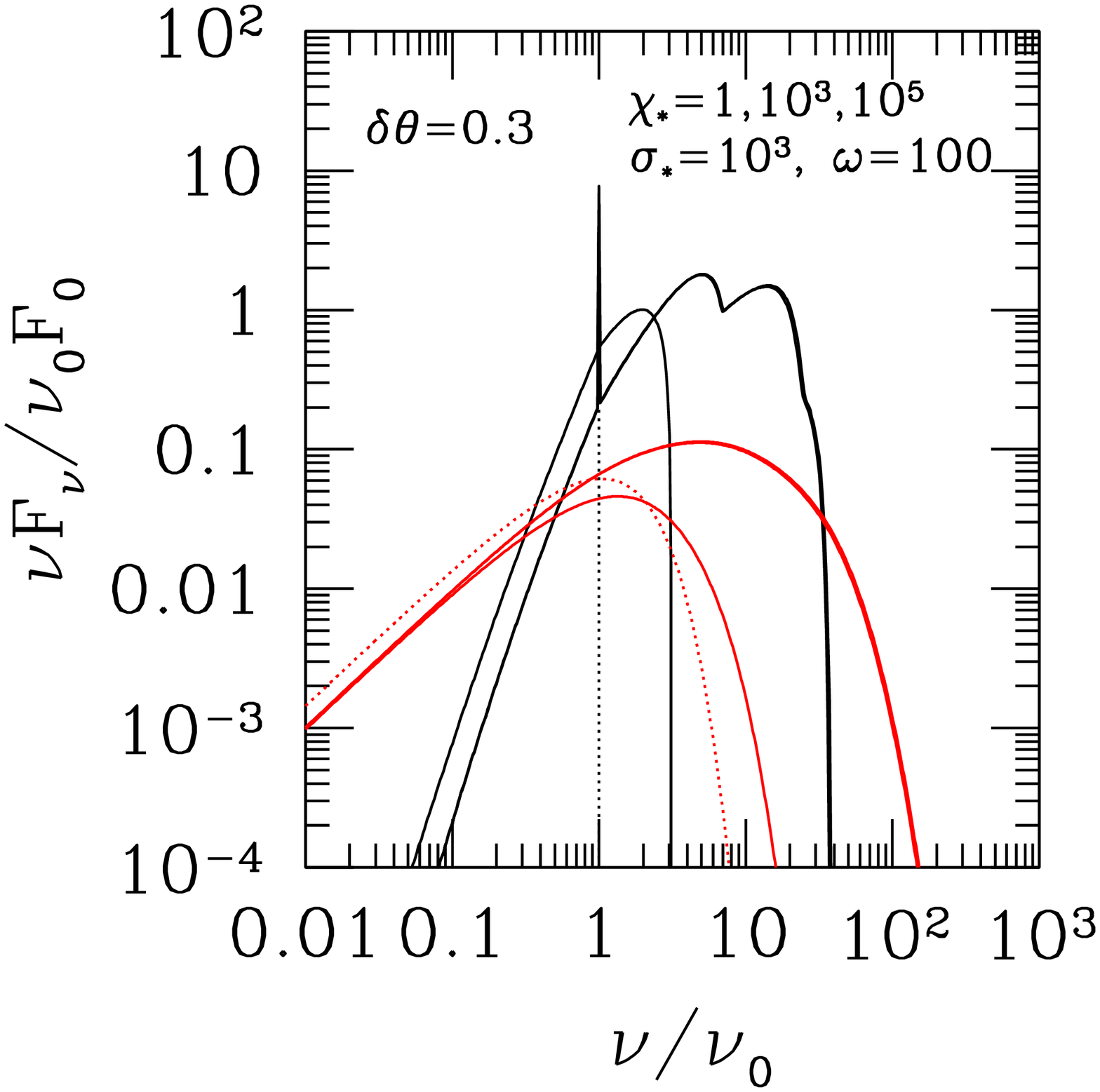}
%\plottwo{spectrum_bb_delta0.3.ps}{spectrum_hybrid_delta0.3.ps}
\caption{Photon spectrum emerging from a highly magnetized jet with radial profile given
in Figure \ref{fig:jetprofiles}, corresponding to strong magnetic flaring
[$\delta\theta = 0.3$ in equation (\ref{eq:thetaprof})].  The  
radiation compactness $\chi_* = 1, 10^3, 10^5$ at the breakout
radius $x_* = 3$.  (The spectra extend to higher frequency at a lower compactness.)
The optical depth to radially moving photons is unity at the inner boundary $x=2$.  
Black lines: monochromatic seed spectrum. \textit{Left Panel:} black-body photon source 
(green lines). \textit{Right Panel:} GRB-like seed spectrum, $F_\nu = {\rm const}\times
e^{-h\nu/kT_0}$.  Dotted curves: source spectrum.}
\vskip .1in
\label{fig:Spectra3}
\end{figure}
\begin{figure}[h]
\epsscale{.91}
\plottwo{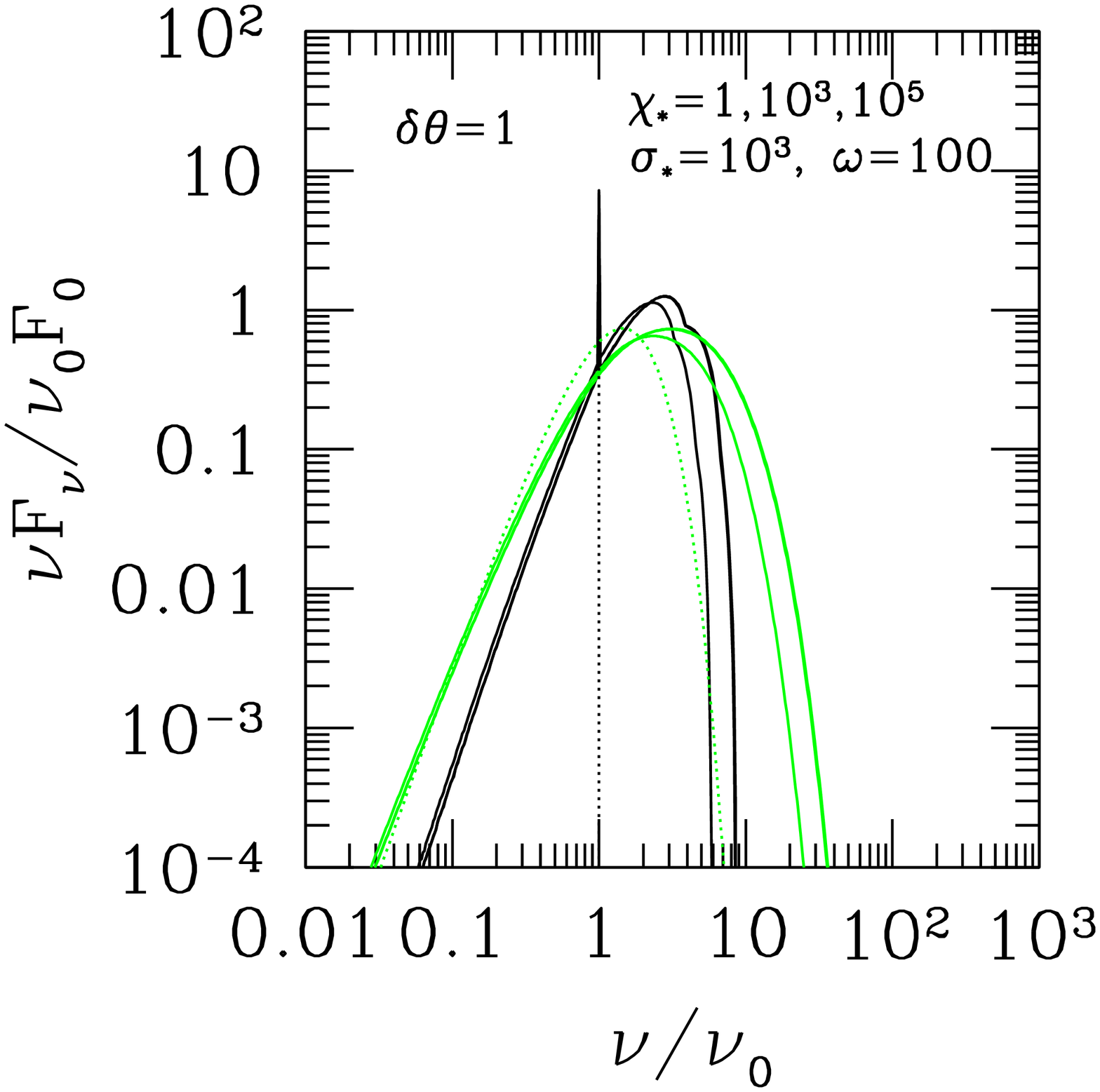}{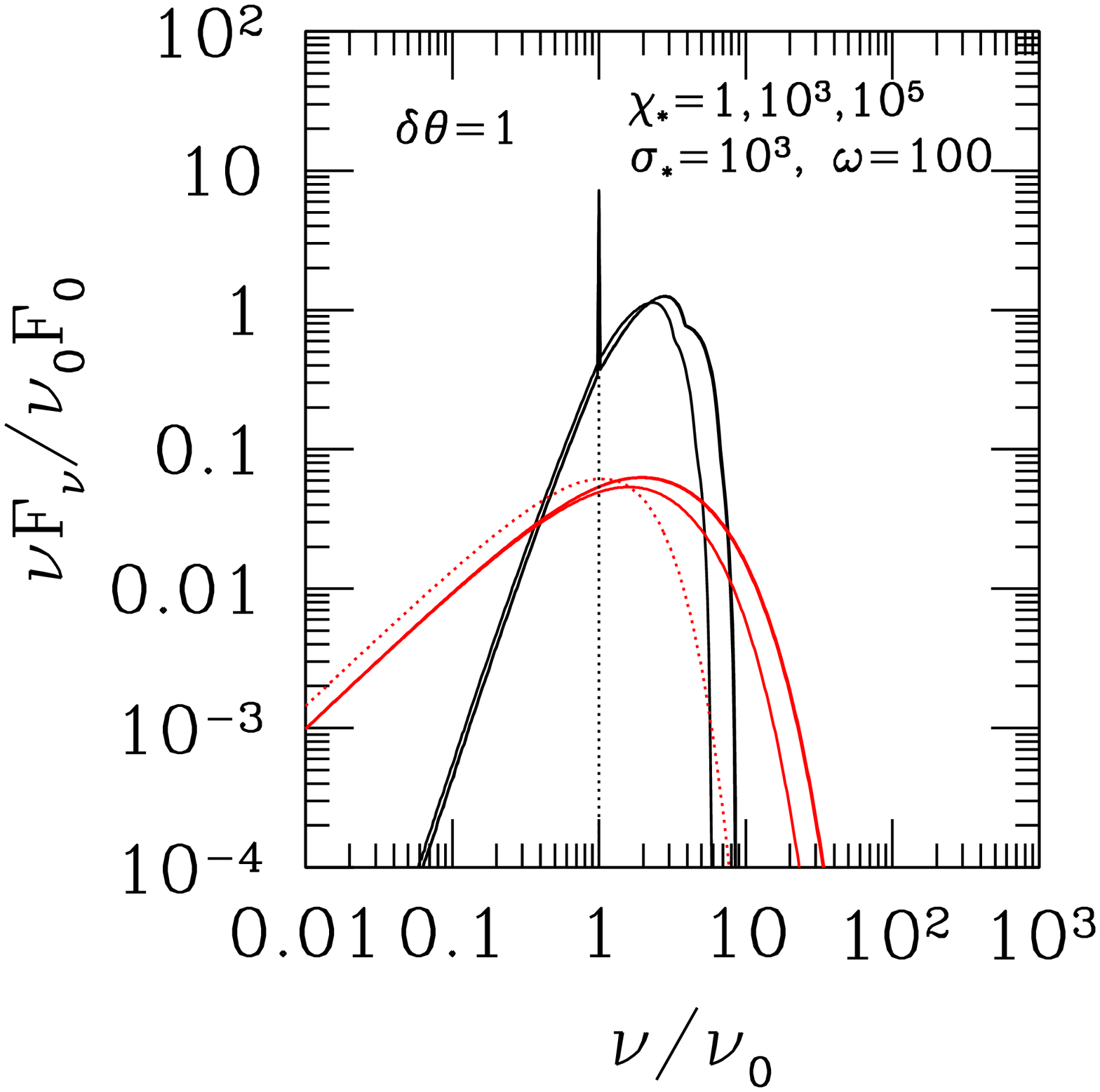}
\plottwo{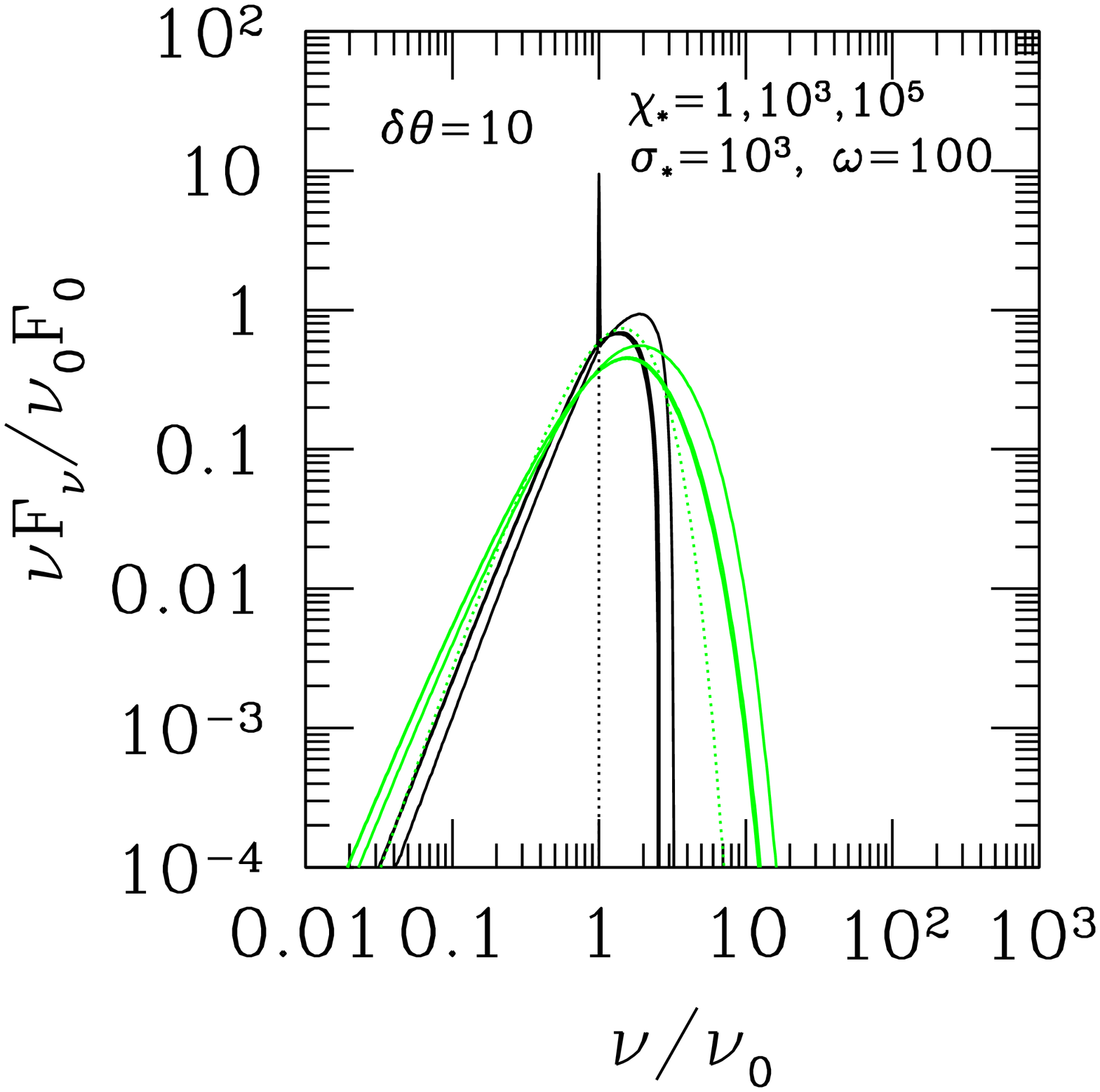}{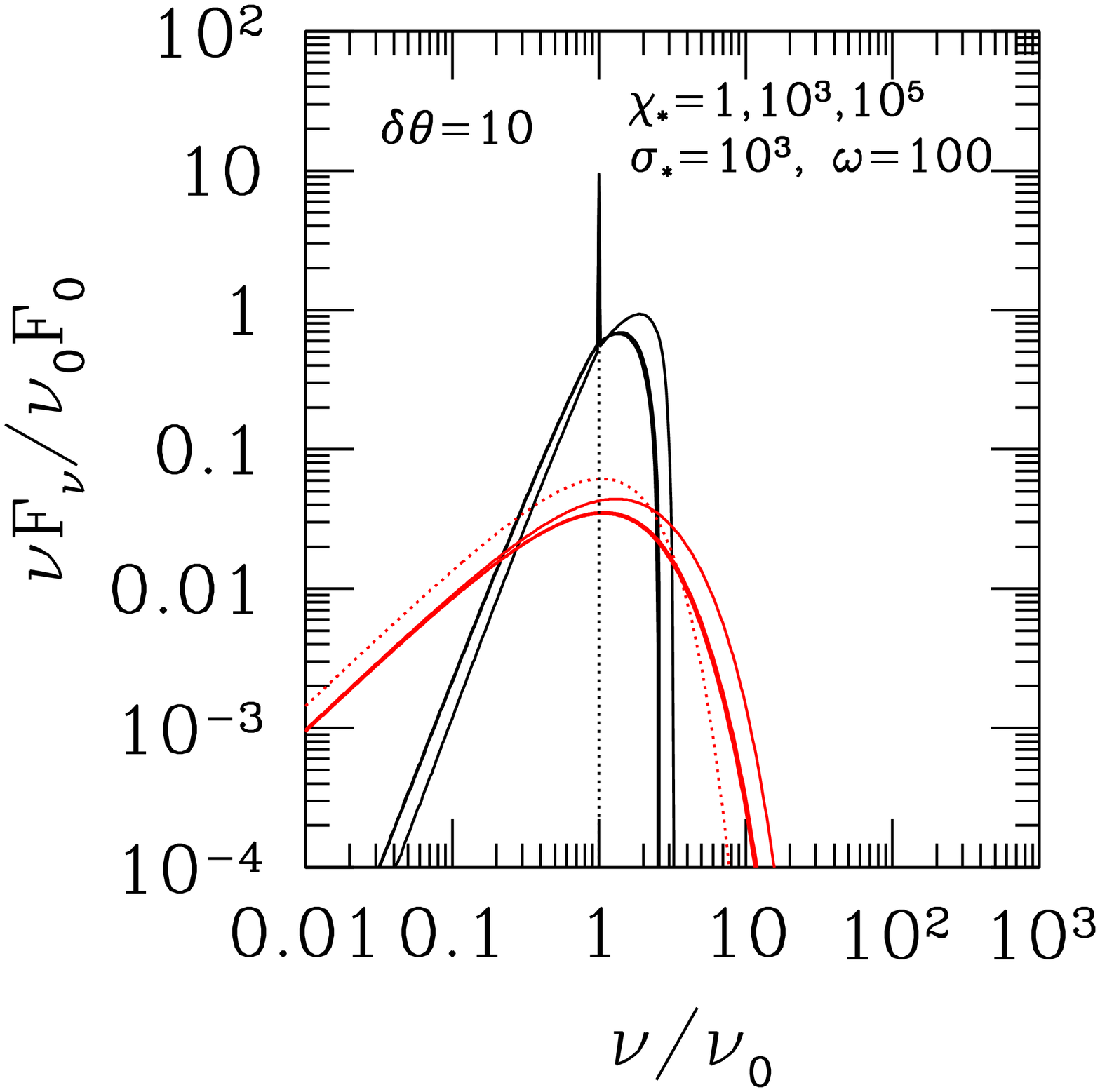}
%\plottwo{spectrum_bb_delta1.ps}{spectrum_hybrid_delta1.ps}
\caption{Same as \ref{fig:Spectra3}, but with $\delta\theta = 1 = 5\theta_j$
(top), and $\delta\theta = 10 = 50\theta_j$ (bottom).}
\label{fig:Spectra4}
\vskip .05in
\end{figure}
%\begin{figure}[h]
%\epsscale{.85}
%\plottwo{spectrum_bb_delta10.ps}{spectrum_hybrid_delta10.ps}
%\caption{Same as \ref{fig:Spectra3}, but with $\delta\theta = 10 = 50\theta_j$.}
%\label{fig:Spectra5}
%\end{figure}
The high-energy extension of the spectrum becomes broader as the radiation compactness is reduced:  
stronger radiation drag limits the increase in $\Gamma$ above $\Gamma_{\rm eq}$.  In the case of a 
monochromatic input spectrum, one notices the appearance of a few distinct orders of 
Compton scattering.  The bumps in the spectrum are smoothed out when convolved with a 
blackbody source.

%%%%%%%%%%%%%%%%%%%%%%%%%%%%%%%%%%%%%%%%%%%%%%%%%%%%%%%%%%%%%%
\section{Summary}

We have examined the effect of intense radiation pressure on a cold, magnetized
outflow with a jet geometry.  The poloidal magnetic field lines are allowed
to deviate from spherical symmetry, e.g. due to breakout from a confining medium.  
The outflow experiences a strong outward Lorentz force 
as a result, so that the magnetofluid and radiation field have a tendency to flow
differentially outside the transparency surface.  We have considered the combined
dynamics of the magnetofluid and radiation, and as well as the modification to the
radiation spectrum by multiple scattering.

We first considered the transition zone straddling the scattering photosphere.
While the jet is still optically thick, its magnetization
is suppressed by the inertia of the advected radiation.  Outside the 
breakout point, the jet experiences a strong outward Lorentz force, which
forces a rapid reduction in optical depth.  This approach assumes that the 
fast critical surface lies deep in the jet, but calculates the radial flaring
of the field lines self-consistently with a simple causal prescription, and
calculates the interaction between the radiation and matter for arbitrary
scattering depth.

If the jet is still optically thick at breakout, then the emergent spectrum is modestly
broadened and hardened below the peak.  On the other hand, breakout outside
the transparency surface results in a photon beam that is significantly
broader than the Lorentz cone of the accelerating jet, and therefore results
in a more extended high-energy component to the spectrum.  A stronger radiation
field suppresses the accelerating effect of jet flaring and brings the
spectrum closer to the original thermal input.  Broadening of the photon beam
could also be due to scattering by a shell of slower material entrained at the jet
head \citep{thompson06}.

Our second approach to the problem focuses on the zone outside the jet photosphere,
but allows for large enough magnetization that the flow passes through the fast critical
surface just outside the breakout radius.  We then solve for the flow profile along
magnetic flux surfaces, both inside and outside the critical point.  In doing this,
we choose a realistic angular distribution for the radiation field but prescribe
a flaring profile for the poloidal field lines.   The cross-field force balance is
not solved self-consistently, but we check that in all cases the transverse component
of the radiation force is small compared with the transverse Lorentz force that is
implied by the chosen field profiles.  

As regards the longitudinal motion along magnetic flux surfaces,
we define the critical compactness $\chi$ of the radiation field
above which the matter and radiation are locked, and the Lorentz force
is subdominant.  For small jet flaring, the radiation force leads to an increase in
terminal Lorentz factor at high values of $\chi$, but can somewhat  suppress the acceleration
if the flaring is strong.  The extent of the high-energy component of the spectrum is
shown to depend in an interesting way on the degree of flaring and the position of 
the photosphere relative to the breakout radius.  

Issues not addressed in this paper include the effects of multiple scattering
at the magnetosonic critical surface, an ambient radiation field generated far outside
the engine (e.g. \citealt{li92a,beskin04}), or the feedback of an intense radiation flow
on the poloidal structure of the magnetic field.  An effect specific to gamma-ray bursts
involves the sidescattering of gamma-rays outside the forward shock, combined with the
radiative acceleration of the pair-enriched material up to a Lorentz factor comparable 
to that of the relativistic ejecta \citep{tm00,beloborodov02}.  This delays
the deceleration of the ejecta, and makes the medium ahead of the shock 
optically thick to scattering \citep{thompson06}.  Photons side-scattered through
large angles would continue to interact with jet material at a smaller radius,
creating pairs downstream of the forward shock, delaying the decoupling of the photons from
the jet fluid, and generating a high-energy tail to the photon spectrum by bulk Comptonization.
This means that the outermost shell of jet material (of a thickness $\sim \theta_j^2 r_*$)
may avoid strong outward acceleration during jet breakout.  However, jet material flowing at 
much greater distances back of the jet head sees weaker Compton drag and rates of 
pair creation during breakout.  The slow forward shell becomes geometrically thin as it
is pushed outward and, eventually, subject to a corrugation instability \citep{thompson06}.  

\acknowledgments  We thank the NSERC of Canada for support.

%%%%%%%%%%%%%%%%%%%%%%%%%%%%%%%%%%%%%%%%%%%%%%%%%%%%%%%%%%%%%%
\begin{appendix}

\section{Geometry of Scattering in a Narrow Jet}
\label{s:tests}
We now calculate the radiation force on plasma moving on a general trajectory within a thin jet 
of opening angle $\theta_j\ll1$, following the setup of Section \ref{s:jetgeom}.   Our goal is to
obtain an analytic expression for this force, which is possible by assuming
a uniform intensity $I = \int I_\nu d\nu$ at the `emission' surface (radius $r_s$), and taking
this surface to be locally spherical.   When considering the interaction with matter, this intensity 
distribution gives similar results to a radiation field that is locally isotropic in the 
relativistic frame of the emitting medium.  (See the discussion in Section \ref{s:jetgeom} and Section 
3.3 of Paper I.)  The result generalizes the simpler angular moment formalism used in Section 
\ref{s:causaljet} and presented in equation (\ref{eq:Flab3}). 

Photons are emitted from coordinates $\{\theta_\gamma,\phi_\gamma\}$ within a patch of angular
radius $\theta_\gamma \leq \theta_j$, and scatter in the jet the position \{$x  = r/r_s > 1$, 
$\theta_f$, $\phi_f = 0$\}.  
The presence of an absorbing surface at the edge of the jet would change the radiation force at
angles $\theta_f > \theta_j$.  Given the uncertain nature of the medium outside
the jet, we restrict the calculation of the force to angles $< \theta_j$.
The photon trajectory is tilted with respect to the radial line passing through the scattering point, 
by an angle (Figure \ref{fig:3DJet})
\be
\theta_{\gamma,r}=\frac{(\theta_{\gamma,r})_s}{x-1}.
\ee
Here
\be
(\theta_{\gamma,r})_s = \left(\theta_f^2+\theta_\gamma^2-2\theta_f \theta_\gamma\cos\phi_\gamma\right)^{1/2}
\ee
is the corresponding angle measured on the `emission' surface, and we have assumed that $x-1 \gg \theta_j$ in 
making the expansion in $x$.
%%%%%%%%%%%%NEW%%%%%%%%%%%%%%%%%%
The intensity at the point of scattering can then be expressed as 
\be\label{eq:thetas}
I = \int d\nu I_\nu = I_0 \quad{\rm for}\quad \theta_{\gamma,r} < \sqrt{\theta_f^2+\theta^2_j-2\theta_f \theta_j\cos\phi_\gamma}
\left({r_s\over r-r_s}\right).
\ee
\begin{figure}[h]
\centerline{\includegraphics[width=0.4\hsize]{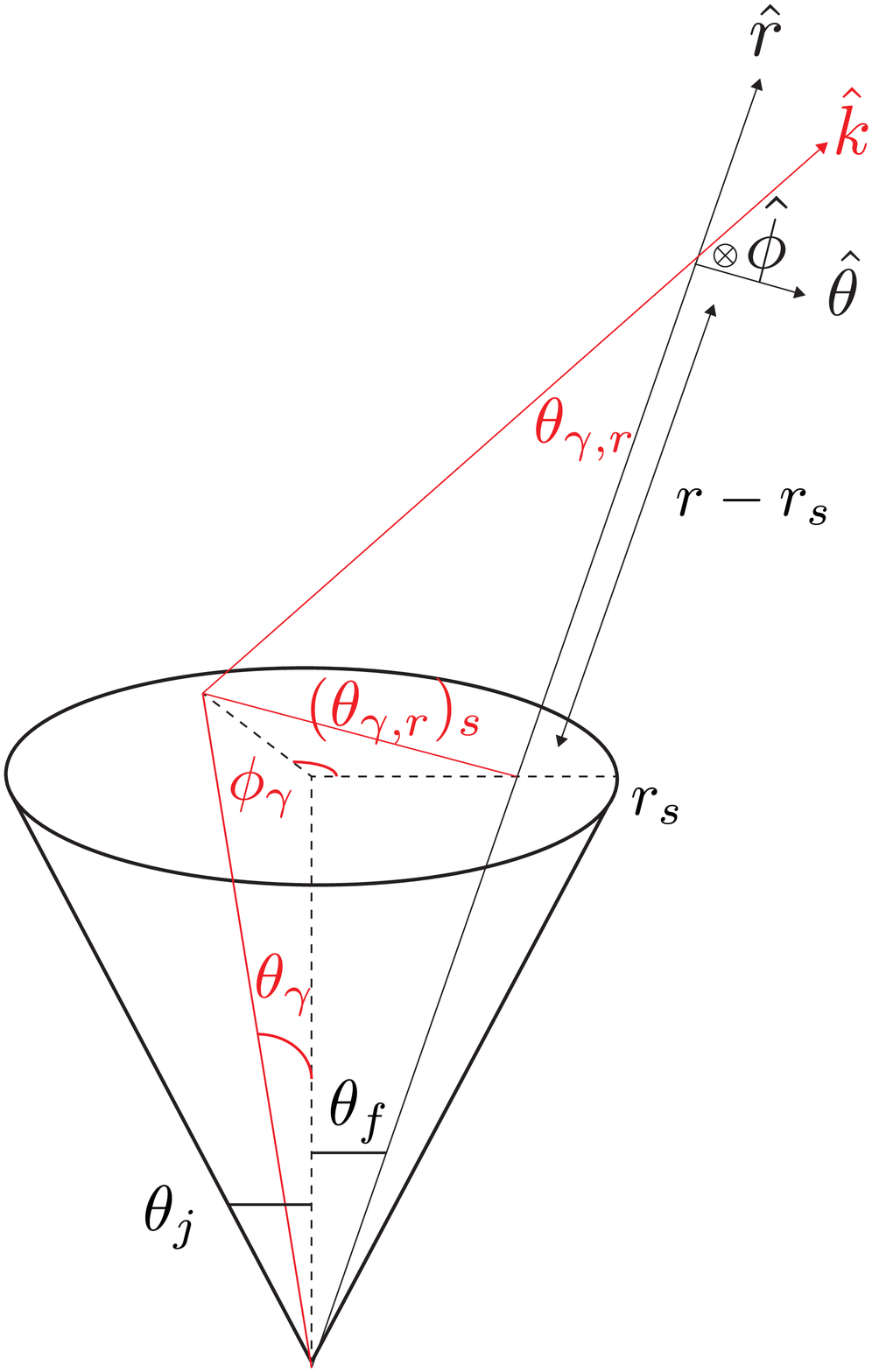}}
%\centerline{\includegraphics[width=0.4\hsize]{3DJet.ps}}
\caption{Geometry of photon emission and scattering in a thin jet.  The emission surface is a circular
patch of radius $\theta_j r_s$.  We take the photon intensity to be independent of angle and uniform on 
this surface.  Coordinates $\{\theta_\gamma,\phi_\gamma\}$ label the point of emission, and
$\theta_{\gamma,r}$ is the angle $\cos^{-1}(\hat k\cdot \hat r)$ at the point of scattering on
a field line of polar angle $\{\theta_f, \phi_f = 0\}$.}
\vskip .1in
\label{fig:3DJet}
\end{figure}
%%%%%%%%%%%%NEW%%%%%%%%%%%%%%%%%%

The unit wave vector in the local $(\hat{r},\hat{\theta},\hat{\phi}) $ coordinate system is
\be
\hat{k} 
= \left(k_{r},k_{\theta},k_{\phi}\right)
= \left(1-{1\over 2}\theta_{\gamma,r}^2,\;\frac{\theta_f-\theta_\gamma\cos\phi_\gamma}{x-1},\;-\frac{\theta_\gamma\sin\phi_\gamma}{x-1}\right) 
\ee
with the poloidal component
\be
k_{p}
= \frac{\beta_{r}}{\beta_{p}}k_{r}+\frac{\beta_\theta}{\beta_{p}}k_{\theta}
= \frac{k_{r}+\Delta\theta_{B}k_{\theta}}{\sqrt{1+\Delta\theta_{B}^{2}}}\simeq1-\frac{1}{2}\left(\frac{\theta_f}{x-1}-\Delta\theta_{B}\right)^{2}-\frac{\theta_\gamma^{2}}{2\left(x-1\right)^{2}}+\frac{\theta_\gamma}{x-1}\left(\frac{\theta_f}{x-1}-\Delta\theta_{B}\right)\cos\phi_\gamma\ee
where $\Delta\theta_B = B_\theta/B_r$ is the angle that a bending field line makes with the local radial vector. 
To evaluate the lab-frame radiation force (\ref{eq:Flab2})
we relate the solid angle of incoming photons to the emission coordinates via
\be
d\Omega=\frac{\theta_{\gamma}}{\left(x-1\right)^{2}}d\theta_{\gamma}d\phi_{\gamma}.
\ee
The poloidal and toroidal radiation force is then evaluated as follows. We begin by writing 
\be 
1-\bbeta\cdot\hat{k} = A + B\cos\phi_\gamma + C\sin\phi_\gamma
\ee
and express the components of the wave vector as
\be 
\hat{\beta}_{p,\phi}\cdot\hat{k} = D_{p,\phi} + E_{p,\phi}\cos\phi_\gamma + F_{p,\phi}\sin\phi_\gamma. 
\ee
Integrating first over $\phi_\gamma$ and then $\theta_\gamma$ at the `emission'
surface gives
\be 
F^{\rm rad}_{p,\phi}=\frac{\sigma_{T}I}{c\left(x-1\right)^{2}}\int_{0}^{\theta_{j}}d\theta_{\gamma}\theta_{\gamma}\left[2\pi AD_{p,\phi}+\pi BE_{p,\phi}+\pi CF_{p,\phi}-\beta_{p,\phi}\Gamma^{2}\left(2\pi A^{2}+\pi B^{2}+\pi C^{2}\right)\right]
\ee

\be
=\frac{\bar{m}c^{2}}{r_{s}}\chi_{*}\frac{\left(x_{*}-1\right)^{2}}{x_{*}\left(x-1\right)^{2}}\left[\frac{1}{2}G_{p,\phi}+\frac{1}{4}\theta_{j}^{2}H_{p,\phi}+\frac{1}{6}\theta_{j}^{4}K_{p,\phi}\right]
\ee
with\be 
G_{p}\simeq2\left(1-\beta_{p}\right)\left[1-\beta_{p}\Gamma^{2}\left(1-\beta_{p}\right)\right]-\frac{1}{2}\beta_{p}\left(1+\beta_{p}^{2}\Gamma^{2}\right)\left(\frac{\theta_{f}}{x-1}-\Delta\theta_{B}\right)^{4}
\ee
\be 
H_{p}\simeq-\frac{\beta_{p}^{3}\Gamma^{2}}{\left(x-1\right)^{2}}\left[2\left(\frac{\theta_{f}}{x-1}-\Delta\theta_{B}\right)^{2}+\frac{\beta_{\phi}^{2}}{\beta_{p}^{2}}\right]
\ee
\be 
K_p=-\frac{\beta_{p}}{2\left(x-1\right)^{4}}\left(1+\beta_{p}^{2}\Gamma^{2}\right). 
\ee
\be 
G_{\phi}=-2\beta_{\phi}\Gamma^{2}\left[\left(1-\beta_{p}\right)^{2}+\beta_{p}\left(1-\beta_{p}\right)\left(\frac{\theta_{f}}{x-1}-\Delta\theta_{B}\right)^{2}+\frac{1}{4}\beta_{p}^{2}\left(\frac{\theta_{f}}{x-1}-\Delta\theta_{B}\right)^{4}\right]
\ee
\be 
H_{\phi}=-\frac{\beta_{\phi}}{\left(x-1\right)^{2}}\left[1+\Gamma^{2}\left(2\beta_{p}(1-\beta_{p})+\beta_{\phi}^{2}\right)+2\beta_{p}^{2}\Gamma^{2}\left(\frac{\theta_{f}}{x-1}-\Delta\theta_{B}\right)^{2}\right]
\ee
\be 
K_{\phi}=-\frac{\beta_{\phi}\beta_{p}^{2}\Gamma^{2}}{2\left(x-1\right)^{4}}
\ee
where $G_p$ and $H_p$ are accurate to first order in $\Gamma^{-2}$ and $(\theta/x)^2$.
The equilibrium Lorentz factor of the photon field, the frame in which ${\bf F}^{\rm rad}$ vanishes, is 
found by solving $\Gamma^\prime_\chi=0$ in (\ref{eq:Gammaprime}). The results are shown in Figure 
\ref{fig:Gammaeq} for the poloidal field line profile described in Sec. \ref{sec:poloidalprofile}. At a radius $x-1 \gg \theta_j$ one finds
$\Gamma_{\rm eq}\simeq x/\theta_j$, with a coefficient of order unity that depends on the 
footprint angle and flaring profile.  A thin jet defines a relativistic frame at relatively small distances from the `emission' surface, as compared
with a spherically symmetric outflow (for which $\Gamma_{\rm eq}\simeq 3^{1/4}x$).
Photons arriving at a scattering point from large angles provide relatively
strong drag.

The jet fluid maintains rapid rotation around the light cylinder, $\theta_f \sim 1/x\omega$,
where the fluid flow is less aligned with the radiation field and $\Gamma_{\rm eq}$ is reduced.
Estimating $\beta_\phi\simeq1/x\omega\theta_f$ (just outside the light cylinder), one finds
\be \label{eq:Gammaeqlimit}
\Gamma_{{\rm eq }}\simeq
\left(\frac{\theta_{j}^{4}}{3(x-1)^{4}} 
+ \frac{1}{x^{4}\omega^{4}\theta_{f}^{4}}\right)^{-\frac{1}{4}},
\ee
valid for all $x$.

\begin{figure}[h]
\centerline{\includegraphics[width=0.5\hsize]{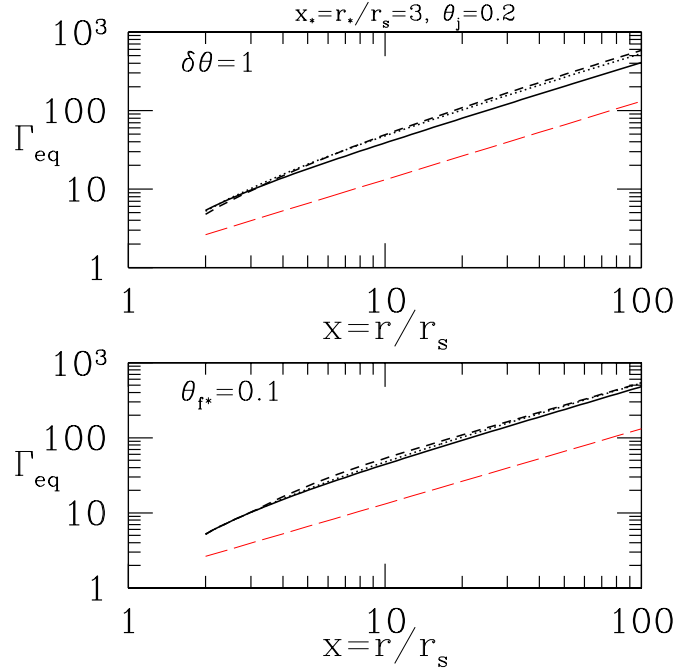}}
%\centerline{\includegraphics[width=0.8\hsize]{smGamEqsmalltheta.eps}}
\caption{Lorentz factor at which material in a thin jet ($\theta_j = 0.2$) feels a vanishing radiation force ${\bf F}^{\rm rad}$.
\textit{Upper panel:} Moderately flared magnetic field [$\delta\theta = 1$ in equation (\ref{eq:thetaprof})], with black lines
corresponding to different footprints at the breakout radius:  $\theta_{f*} = 0.05, 0.1, 0.15$ (solid, dotted and dashed).
\textit{Lower panel:} Different degrees of magnetic flaring, $\delta\theta=10,1,0.3$ (weak, medium, strong flaring)
for a field line anchored at $\theta_{f*} = 0.1$.  Long-dashed red line:  $\Gamma_{\rm eq}$ for a spherical emission 
surface and monopolar poloidal field.}\label{fig:Gammaeq}
\vskip .2in
\end{figure}

%%%%%%%%%%%%%%%%%%%%%%%%%%%%%%%%%%%%%%%%%%%%%%%%%%%%%%%%%%%%%%
%%%%%%%%%%%%%%%%%%%%%%%%%%%%%%%%%%%%%%%%%%%%%%%%%%%%%%%%%%%%%%

\section{Accounting for Rotation of the Photon Field}
The photon source rotates rapidly in some cases, e.g. a rapidly rotating
star such as a millisecond magnetar, or the merged remnant of a white dwarf binary.  
We can approximate the effect of a rotating emission surface by setting
\be\label{eq:betarad}
\beta_{\phi}\rightarrow
\beta_{\phi}-\frac{\beta_{\phi,R}}{x\theta_f}
\ee  
in equations (\ref{eq:radforcedef}).  Here $\beta_{\phi,R}$ is
a constant representing the aberration of the outflowing photons at
$r = r_s$ ($x = 1$).
In this situation, plasma near the emission surface can more 
easily co-rotate with the radiation field while still being accelerated outward.

The value of $\beta_{\phi,R}$ depends on the type of source.  One
has $\beta_{\phi,R} \sim \Omega r_s/c \equiv \omega $ when the photons flow from
the surface of a star of radius $r_s$ through a transparent wind.
On the other hand, if the outflow is optically thick in a narrow radial zone
close to the engine, then one expects $\beta_{\phi,R} \sim (\Omega r_s/c)^{-1}
\sim \omega^{-1}$ based on the conservation of angular momentum from the
light cylinder out to the transparency surface ($x = 1$). 

%%%%%%%%%%%%%%%%%%%%%%%%%%%%%%%%%%%%%%%%%%%%%%%%%%%%%%%%%%%%%%

\section{Wind Equations for Jet Model II}
Here we derive the equations (\ref{eq:Gamp})-(\ref{eq:mueff}) for the longitudinal development of Lorentz factor and particle 
angular momentum along magnetic flux surfaces.   Beginning with the poloidal and toroidal components of the Euler equation, 
(\ref{e:Eulerdl1}) and (\ref{e:Eulerdl2}), we expand the derivatives on the right hand side as
\be\label{e:EulerRHS}
 \partial_{l}\left(r\theta B_{\phi}\right) = B_{\phi}\partial_{l}(r\theta_f) + 
 r\theta\frac{B_{\phi}}{B_{r}}\partial_{l}B_{r}+r\theta_f B_{r}\partial_{l}\left(\frac{B_{\phi}}{B_{r}}\right).
\ee
This can be evaluated using  $\partial_{l}(r\theta_{f}) \simeq \theta_f + B_\theta/B_r = \theta_{f}+\Delta\theta_{B}$,
equation (\ref{eq:br}) for $\partial_l B_r$, and
\be
v_{r}\partial_{l}\left(\frac{B_{\phi}}{B_{r}}\right) = 
\partial_{l}v_{\phi}\left(1+\frac{v_{\phi}}{v_{r}}\frac{B_{\phi}}{B_{r}}\right) - 
\frac{B_{\phi}}{B_{r}}\frac{\partial_{l}\Gamma}{v_{r}\Gamma^{3}}-\Omega_f\left(\theta_f + \Delta\theta_B\right),
\ee
where 
\be
{\partial_{l}v_{\phi}\over v_\phi} = \frac{\partial_{l}{\cal L}_m}{{\cal L}_m} - 
{1\over r}\left(1+\frac{\Delta\theta_{B}}{\theta_{f}}\right)-{\partial_l\Gamma\over \Gamma}.
\ee

The wind equations (\ref{e:Eulerdl1}), (\ref{e:Eulerdl2}) are now transformed into ordinary differential equations
by ignoring the cross-field force balance:
\be\label{eq:Eulerdl1b}
\left[1 - {\sigma\over (x\theta_f\omega)^2\Gamma} \left(\beta_\phi\Lambda + {1\over \beta_r\Gamma^2}{B_\phi\over B_r}\right) 
          {B_\phi\over B_r}\right]{d\Gamma\over dl} - 
{1\over x\theta_f}\left[\beta_{\phi}-\frac{\sigma}{(x\theta_f\omega)^2\Gamma}\frac{B_{\phi}}{B_{r}}\Lambda\right]{d{\cal L}_m\over dl}
=\frac{\sigma \Psi}{(x\theta_f\omega)^2}\frac{B_{\phi}}{B_{r}} + \chi_{*}R_{j};
\ee
\be\label{eq:Eulerdl2b}
\frac{\sigma}{(x\theta_{f}\omega)^{2}\Gamma}\left(\beta_{\phi}\Lambda + \frac{1}{\beta_r\Gamma^2}\frac{B_{\phi}}{B_{r}}\right){d\Gamma\over dl} +
{1\over x\theta_f}\left[\beta_r-\frac{\sigma}{(x\theta_f\omega)^2\Gamma}\Lambda\right]{d{\cal L}_m\over dl} = 
-\frac{\sigma \Psi}{(x\theta_f\omega)^2} + \chi_{*}P_{j},
\ee
where
\be
\Lambda \equiv 1+\frac{\beta_{\phi}}{\beta_r}\frac{B_{\phi}}{B_r}; \quad\quad
\Psi \equiv \left(1 + {\Delta\theta_{B}\over\theta_f}\right)(1 + \Lambda){\beta_\phi\over x} + 
\frac{B_{\phi}}{B_{r}}{\beta_r\over A}{dA\over dl}.
\ee

Two simple tests of these equations are made possible by neglecting the radiation force.  The energy and angular momentum 
integrals (\ref{eq:dimintegral}) are now related by 
\be
{d\Gamma\over dl} - \omega{d{\cal L}_m\over dl} = 0.
\ee
This equation is recovered by summing (\ref{eq:Eulerdl1b}) and $B_\phi/B_r$ times (\ref{eq:Eulerdl2b}), and making use
of Ferraro's law (\ref{eq:induction}).  Second, outside the fast point the inertia of the magnetofluid
is dominated by the matter:  the coefficient of $d\Gamma/dl$ in equation (\ref{eq:Eulerdl1b}) 
is $\simeq 1 - \sigma/\Gamma^3$ and approaches unity.  Since in addition $\beta_\phi \rightarrow 0$, the term involving 
${\cal L}_m$ can be neglected and one finds 
\be
{d\Gamma\over dl} \simeq {\sigma_*\over A^2}{dA\over dl}\quad\quad (\Gamma^3 \gg \sigma).
\ee
The same result can be obtained from the integral equations (\ref{eq:sigmaf}), (\ref{eq:dimintegral}) and (\ref{eq:br}),
\be\label{eq:gammamhd}
\Gamma = \Gamma_* + \sigma_*\left(1-A^{-1}\right).
\ee
Equation (\ref{eq:Gamp}) is then obtained by solving (\ref{eq:Eulerdl1b}) and (\ref{eq:Eulerdl2b}) for $d\Gamma/dl$ and $d\mathcal{L}_m/dl$.

\section{Cross-field Forces}\label{s:transrad}
Though we ignore the cross-field force balance, it is useful to estimate the
transverse radiation force and compare it with the Lorentz force that is 
implied by a given field-line profile.  Our procedure becomes inconsistent
if the transverse radiation force dominates, because the radiation field
will then comb out the field lines in the radial direction.

The polar component of equation (\ref{eq:Flab2}) gives the estimate,
\be \label{eq:radcross}
F_{\theta}^{\rm rad} \simeq \frac{\bar{m}c^{2}}{r_{s}}\chi_{*}
\frac{\left(x_{*}-1\right)^{2}}{2x_{*}\left(x-1\right)^{2}\Gamma^{2}}
\left(\frac{\theta_{f}}{x-1}-\frac{1}{2}\Delta\theta_{B}\right).
\ee
The first term on the right-hand side represents the force imparted by photons streaming 
from a finite polar cap toward particles on off-axis field lines.  The second represents 
the drag imparted as the poloidal particle flow bends across the radiation field.
The cross-field Lorentz force is given by 
\be
\frac{\bar{m}}{4\pi\rho}\left[(\bnabla\times{\bf B})\times{\bf B}\right]_{\theta} \simeq
\frac{\bar{m}B_{\phi}^{2}}{4\pi\rho r\theta_{f}}=\bar{m}c^{2}\frac{\sigma}{r\theta_{f}}.
\ee
Requiring this to be greater than (\ref{eq:radcross}) gives 
an upper bound on the radiation compactness at jet breakout ($x > x_* \gg 1$),
\be \label{eq:maxchi}
\frac{\chi_{*}}{\sigma}<\frac{2x\Gamma^{2}}{x_{*}\theta_f}
\left(\frac{\theta_{f}}{x}-\frac{\Delta\theta_{B}}{2}\right)^{-1}.
\ee
Our calculations can admit values of $\chi_*$ as large as $\sim 10^4 \sigma$ without any inconsistency,
given the typical jet parameters $\theta_j \sim 0.2$, $x_* \sim 3$, $\Gamma > \Gamma_* \sim 10$.

\end{appendix}

%%%%%%%%%%%%%%%%%%%%%%%%%%%%%%%%%%%%%%%%%%%%%%%%%%%%%%%%%%%%%%

%%%%%%%%%%%%%%%%%%%%%%%%%%%%%%%%%%%%%%%%%%%%%%%%%%%%%%%%%%%%%%


\begin{thebibliography}{} 
\bibitem[Amati et al.(2002)]{amati02} Amati, L., Frontera, F., Tavani, M., et al.\ 2002, \aap, 390, 81 
\bibitem[Band et al.(1993)]{band93} Band, D., Matteson, J., Ford, L., et al.\ 1993, \apj, 413, 281 
\bibitem[Begelman \& Li(1994)]{begelman94} Begelman, M.~C., \& Li, Z.-Y.\ 1994, \apj, 426, 269
\bibitem[Beloborodov(2002)]{beloborodov02} Beloborodov, A.~M.\ 2002, \apj, 565, 808
\bibitem[Beloborodov(2010)]{beloborodov10} Beloborodov, A.~M.\ 2010, \mnras, 407, 1033 
\bibitem[Beloborodov(2011)]{beloborodov11} Beloborodov, A.~M.\ 2011, \apj, 737, 68  
\bibitem[Beskin et al.(2004)]{beskin04} Beskin, V.~S., Zakamska, N.~L., \& Sol, H.\ 2004, \mnras, 347, 587
\bibitem[Beskin \& Nokhrina(2006)]{beskin06} Beskin, V.~S., \& Nokhrina, E.~E.\ 2006, \mnras, 367, 375 
%\bibitem[Blandford \& Znajek(1977)]{bz77} Blandford, R.~D., \& Znajek, R.~L.\ 1977, \mnras, 179, 433 
%\bibitem[Blandford \& Payne(1982)]{bp82} Blandford, R.~D., \& Payne, D.~G.\ 1982, \mnras, 199, 883 
%\bibitem[Bucciantini et al.(2012)]{bucciantini12} Bucciantini, N., 
%Metzger, B.~D., Thompson, T.~A., \& Quataert, E.\ 2012, \mnras, 419, 1537
%\bibitem[Contopoulos(2005)]{contopolous05} Contopoulos, I.\ 2005, \aap, 442, 579 
\bibitem[Camenzind(1987)]{camenzind87} Camenzind, M.\ 1987, \aap, 184, 341
\bibitem[Dessart et al.(2009)]{dessart09} Dessart, L., Ott, 
C.~D., Burrows, A., Rosswog, S., \& Livne, E.\ 2009, \apj, 690, 1681 
\bibitem[Drenkhahn \& Spruit(2002)]{drenkhahn02} Drenkhahn, G.,
\& Spruit, H.~C.\ 2002, \aap, 391, 1141 
%\bibitem[Duncan \& Thompson(1992)]{dt92} Duncan, R.~C., 
%\& Thompson, C.\ 1992, \apjl, 392, L9 
\bibitem[Ferraro(1937)]{ferraro37} Ferraro, V.~C.~A.\ 1937, \mnras, 97, 458 
%\bibitem[Ghisellini et al.(2000)]{ghisellini00} Ghisellini, G., Lazzati, D., Celotti, A., \& 
%Rees, M.~J.\ 2000, \mnras, 316, L45 
\bibitem[Giannios(2006)]{giannios06} Giannios, D.\ 2006, \aap, 457, 763 
\bibitem[Giannios \& Spruit(2007)]{giannios07} Giannios, D., \& Spruit, H.~C.\ 2007, \aap, 469, 1 
\bibitem[Goldreich \& Julian(1970)]{goldreich70} Goldreich, P., \& Julian, 
W.~H.\ 1970, \apj, 160, 971 
%\bibitem[Granot et al.(2011)]{granot11} Granot, J., Komissarov, S.~S., \& Spitkovsky, A.\ 2011, \mnras, 411, 1323 
%\bibitem[Guess(1962)]{guess62} Guess, A.~W.\ 1962, \apj, 135, 855 
%\bibitem[Goodman(1986)]{goodman86} Goodman, J.\ 1986, \apjl, 308, L47 
%\bibitem[Kippenhahn et al. (1967)]{kippenhahn67}{Kippenhahn et al. (1967)}
%Kippenhahn,\, R., Weigert,\, A., Hofmeister,\, E. (1967): Computing Stellar 
%evolution. Meth.\,Comp.\,Phys., 7, 129
\bibitem[Kiusalaas(2010)]{kiusalaas10} Kiusalaas, J. 2010, Numerical 
Methods in Engineering with Python (2nd ed.; New York: Cambridge University Press)
\bibitem[Kulkarni et al.(1999)]{kulkarni99} Kulkarni, S.~R., 
Djorgovski, S.~G., Odewahn, S.~C., et al.\ 1999, \nat, 398, 389
%\bibitem[Lazzati et al.(2011)]{lazzati11} Lazzati, D., Morsony, B.~J., \& Begelman, M.~C.\ 2011, \apj, 732, 34
\bibitem[Li et al.(1992a)]{li92a} Li, Z.-Y., Begelman, M.~C., \& Chiueh, T.\ 1992, \apj, 384, 567 
\bibitem[Li et al.(1992b)]{li92b} Li, Z.-Y., Chiueh, T., \& Begelman, M.~C.\ 1992, \apj, 394, 459L 
\bibitem[London \& Flannery(1982)]{london82} London, R.~A., \& 
Flannery, B.~P.\ 1982, \apj, 258, 260 
%\bibitem[Lyutikov \& Blandford(2003)]{lyutikov03} Lyutikov, M., 
%\& Blandford, R.\ 2003, arXiv:astro-ph/0312347 
\bibitem[MacFadyen \& Woosley(1999)]{macfadyen99} MacFadyen, A.~I., \& 
Woosley, S.~E.\ 1999, \apj, 524, 262 
\bibitem[Meszaros \& Rees(1997)]{meszaros97} Meszaros, P., \& 
Rees, M.~J.\ 1997, \apjl, 482, L29 
%\bibitem[McKinney(2006)]{mckinney06} McKinney, J.~C.\ 2006, \mnras, 368, L30 
\bibitem[Michel(1969)]{michel69} Michel, F.~C.\ 1969, \apj, 158, 727 
%\bibitem[Nakar \& Piran(2005)]{nakar05} Nakar, E., \& Piran, T.\ 2005, \mnras, 360, L73 
\bibitem[Paczynski(1998)]{paczynski98} Paczynski, B.\ 1998, \apjl, 494, L45 
%\bibitem[Pe'er(2008)]{peer08} Pe'er, A.\ 2008, \apj, 682, 463
%\bibitem[Press et al. (2007)]{press07} Press, William H. and Teukolsky, Saul A. 
%and Vetterling, William T. and Flannery, Brian P. 2007, Numerical Recipes 3rd 
%Edition: The Art of Scientific Computing (3rd ed.; New York: Cambridge University Press)
\bibitem[Rees \& M{\'e}sz{\'a}ros(2005)]{rees05} Rees, M.~J., \& M{\'e}sz{\'a}ros, P.\ 2005, \apj, 628, 847 
\bibitem[Russo \& Thompson(2012)]{russo12} Russo, M. \& Thompson, C., \apj, submitted
%\bibitem[Shemi \& Piran(1990)]{shemi90} Shemi, A., \& Piran, T.\ 1990, \apjl, 365, L55 
%\bibitem[Sikora et al.(1994)]{sikora94} Sikora, M., Begelman, M.~C., \& Rees, M.~J.\ 1994, \apj, 421, 153 
%\bibitem[Spitkovsky(2006)]{spitkovsky06} Spitkovsky, A.\ 2006, \apjl, 648, L51 
\bibitem[Tchekhovskoy et al.(2009)]{tchek09} Tchekhovskoy, A., 
McKinney, J.~C., \& Narayan, R.\ 2009, \apj, 699, 1789 
\bibitem[Tchekhovskoy et al.(2010)]{tchek10} Tchekhovskoy, A., 
Narayan, R., \& McKinney, J.~C.\ 2010, 15, 749 
\bibitem[Thompson(1994)]{thompson94} Thompson, C.\ 1994, \mnras, 270, 480 
\bibitem[Thompson(2006)]{thompson06} Thompson, C.\ 2006, \apj, 651, 333 
\bibitem[Thompson \& Madau(2000)]{tm00} Thompson, C., \& Madau, P.\ 2000, \apj, 538, 105 
\bibitem[Vlahakis \& K{\"o}nigl(2003a)]{vlahakis03a} Vlahakis, N., \& K{\"o}nigl, A.\ 2003, \apj, 596, 1080 
\bibitem[Vlahakis \& K{\"o}nigl(2003b)]{vlahakis03b} Vlahakis, N., \& K{\"o}nigl, A.\ 2003, \apj, 596, 1104 
\bibitem[Woosley(1993)]{woosley93} Woosley, S.~E.\ 1993, \apj, 405, 273 
\bibitem[Zhang \& Yan(2011)]{zhang11} Zhang, B., \& Yan, H.\ 2011, \apj, 726, 90
\end{thebibliography}
\end{document}